\documentclass[showpacs,preprintnumbers,amsmath,amssymb]{revtex4}

\usepackage{epsfig}% Include figure files
\usepackage{graphicx}% Include figure files
\usepackage{dcolumn}% Align table columns on decimal point
\usepackage{bm}% bold math

\begin{document}

%%%%%%%%%%%%%%%%%%%%%%%%%%%%%%%%%%%%%%%%%%%%%%%%%%%%%%%%%%%%%%%%%%%%%%%%%%
%                               Title                                    %
%%%%%%%%%%%%%%%%%%%%%%%%%%%%%%%%%%%%%%%%%%%%%%%%%%%%%%%%%%%%%%%%%%%%%%%%%%

\title{Normal-Ordered Wave-Function Factorization of the 1D Hubbard Model
for Finite Values of the On-site Repulsion $U$}
\author{J. M. P. Carmelo}
\affiliation{GCEP-Center of Physics, University of Minho, Campus
Gualtar, P-4710-057 Braga, Portugal}
\date{23 May 2003}
%\date{\today}

%%%%%%%%%%%%%%%%%%%%%%%%%%%%%%%%%%%%%%%%%%%%%%%%%%%%%%%%%%%%%%%%%%%%%%%%%%
%                              abstract                                  %
%%%%%%%%%%%%%%%%%%%%%%%%%%%%%%%%%%%%%%%%%%%%%%%%%%%%%%%%%%%%%%%%%%%%%%%%%%

\begin{abstract}
In this paper we find that in the thermodynamic limit and for the
the ground-state normal-ordered 1D Hubbard model the wave function
of excited states contained in few-electron excitations factorizes
for all values of the on-site Coulombian repulsion $U$. This
factorization results from the non-interacting character of the
pseudofermions whose occupancy configurations describe these
excited states. Our study includes the introduction of the
pseudoparticle - pseudofermion unitary transformation and of an
operator algebra for both the pseudoparticles and the
pseudofermions. The pseudofermion description takes into account
the relationship between the rotated electrons and the holons,
spinons, and $c0$ pseudoparticles. [Rotated electrons are related
to the electrons by a canonical transformation.] As the
corresponding pseudoparticles, the $c\nu$ pseudofermions (and
$s\nu$ pseudofermions) are $\eta$-spin zero $2\nu$-holon composite
quantum objects (and spin zero $2\nu$-spinon composite quantum
objects) where $\nu=1,\,2,...$. The pseudofermions are non
interacting and thus have no residual interactions, in contrast to
the corresponding pseudoparticles, whose statistics we classify
according to the generalized Pauli principle. The physics behind
the invariance of the pseudoparticles under the above
transformations for specific values of the bare momentum is also
studied and discussed.
\end{abstract}


\maketitle
%%%%%%%%%%%%%%%%%%%%%%%%%%%%%%%%%%%%%%%%%%%%%%%%%%%%%%%%%%%%%%%%%%%%%%%%%%
%                              body of paper                             %
%%%%%%%%%%%%%%%%%%%%%%%%%%%%%%%%%%%%%%%%%%%%%%%%%%%%%%%%%%%%%%%%%%%%%%%%%%
\section{INTRODUCTION}

Recently there has been a renewed experimental interest in the
exotic one-electron and two-electron spectral properties of
quasi-1D materials
\cite{spectral0,Hussey,Menzel,Fuji02,Hasan,Ralph,Gweon,Vescoli,Mori}.
Some of these experimental studies observed unusual
finite-energy/frequency spectral properties, which are far from
being well understood. For low values of the energy, the
microscopic electronic properties of these materials are usually
described by systems of coupled chains. For finite values of the
energy larger than the transfer integrals for electron hopping
between the chains, the one-electron (1D) Hubbard model is
expected to provide a good description of the physics of these
materials \cite{Menzel,Hasan,Ralph,Gweon,Vescoli,Mori}. This is
confirmed by the recent quantitative studies of Refs.
\cite{spectral0,spectral}. Similar unusual spectral properties
observed in two-dimensional (2D) high-$T_c$ superconductors could
result from effective quasi-1D charge and spin transport
\cite{Menzel,Granath,Orgard,Carlson,Zaanen,Antonio}. However, the
non-perturbative nature of the 1D Hubbard model implies that the
electronic creation and annihilation operators do not provide a
suitable operational description for the study of the
finite-energy spectral properties. Thus, the first step for the
study of these properties is the introduction of a suitable
operational description. Except in the limit of infinite on-site
Coulombian repulsion $U\rightarrow\infty$
\cite{Ogata,Ricardo,Penc95,Penc96}, the introduction of such a
description is an open problem of great physical interest. For low
values of energy useful information about the effects of the
non-perturbative electronic correlations is provided by
two-component conformal-field theory
\cite{Belavin,Frahm,Carmelo91,Carmelo92,Carmelo97pp}.
Unfortunately, that method does not apply for finite values of
energy.

In view of the above-mentioned unusual finite-energy/frequency
spectral properties observed in real experiments, which are far
from being well understood, efforts towards the introduction of a
suitable operational description to deal with the finite-energy
problem are welcome. In this paper we introduce an operational
representation for the 1D Hubbard model
\cite{Lieb,Takahashi,Martins98,Hubbard} in terms of
non-interacting {\it pseudofermions}. We find that in the
thermodynamic limit the wave function of excited states contained
in the few-electron excitations factorizes into separated
contributions corresponding to different pseudofermion branches.
(A few-electron excitation is generated by application onto the
ground state of operators whose expression involves the product of
a few electronic creation and/or annihilation operators.) Such
factorization occurs for all values of the on-site Coulombian
repulsion $U$ of the ground-state normal-ordered 1D Hubbard model.
The pseudofermion operational description is closely related to
the pseudoparticle representation previously considered in the
literature \cite{Carmelo97,II}, and is the natural starting point
for studies of the finite-energy/frequency few-electron spectral
properties. As a result of the wave-function factorization, the
pseudofermion description is more suitable for the study of the
overlap between few-electron excitations and the energy
eigenstates than the pseudoparticle representation.

Our starting point is a holon, spinon, and $c0$ pseudoparticle
representation, which refers to the whole Hilbert space of the
model \cite{I}. The relation between the original electrons and
these elementary quantum objects involves the concept of {\it
rotated electron}. The rotated electrons are related to the
electrons by a unitary transformation first introduced in Refs.
\cite{Harris,Mac}. For such rotated electrons double occupation is
a good quantum number for all values of $U$. Except for the $c0$
pseudoparticles, all pseudoparticle branches introduced in Ref.
\cite{Carmelo97} have a composite character in terms of holons or
spinons \cite{I}. The pseudofermions are related to the
pseudoparticles by a second unitary transformation. The charge or
spin carried by the pseudoparticle and its holon or spinon
contents remains invariant under such a transformation. It follows
that except for the $c0$ pseudofermions, all remaining
pseudofermion branches are composite objects of holons or spinons.
The concepts of local pseudoparticle and effective lattice widely
used in this paper are introduced in Ref. \cite{IIIb}.

The pseudofermion operational description introduced in this paper
and the associated factorization of the wave function of energy
eigenstates contained in few-electron excitations are used
elsewhere in the study of the finite-energy spectral properties
\cite{spectral0,spectral,V}. The theoretical predictions of Refs.
\cite{spectral0,spectral} seem to describe both qualitatively and
quantitatively the one-electron removal spectral lines observed by
photoemission experiments for finite values of the energy in real
quasi-1D materials. We thus expect that our pseudofermion
operational description is useful for the further understanding of
the exotic properties displayed by low-dimensional materials.

The paper is organized as follows: In Sec. II we introduce the 1D
Hubbard model and the rotated electrons. The elementary holon,
spinon, and $c0$ pseudoparticle description and associated
$\alpha\nu$ pseudoparticle representation are summarized in Sec.
III. Moreover, in that section we introduce the operator algebra
for the pseudoparticles and the statistics of these quantum
objects according to the generalized Pauli principle
\cite{Haldane91}. In Sec. IV we consider the ground-state
normal-ordered pseudoparticle operator description and introduce
useful ground-state quantities. The pseudofermion description and
the relationship between pseudoparticle and pseudofermion
operators are introduced and discussed in Sec. V. This includes
the introduction of the pseudofermion anticommutator algebra. In
Sec. VI we study the pseudofermion energy and momentum spectra and
introduce and discuss the factorization of the few-electron
Hilbert subspace of the ground-state normal-ordered 1D Hubbard
model. The investigation of the laws under the pseudoparticle -
pseudofermion transformation of several quantum objects and
quantities is the subject of Sec. VII. Finally, in Sec. VIII we
present the discussion and the concluding remarks.

%%%%%%%%%%%%%%%%%%%%%%%%%%%%%%%%%%%%%%%%%%%%%%%%%%%%%%%%%%%%%%%%%%%%%%%%%%
\section{THE 1D HUBBARD MODEL AND ROTATED ELECTRONS}

In a chemical potential $\mu $ and magnetic field $H$ the 1D
Hubbard Hamiltonian can be written as,

\begin{equation}
\hat{H}={\hat{H}}_{SO(4)} + \sum_{\alpha =c,\,s}\mu_{\alpha}\,
2{\hat{S}}_{\alpha}^z \, , \label{H}
\end{equation}
where the Hamiltonian

\begin{equation}
{\hat{H}}_{SO(4)} = {\hat{H}}_{H} - (U/2)\,\hat{N} + (U/4)\,N_a \,
; \hspace{1cm}{\hat{H}}_{H} = \hat{T}+U\,\hat{D} \, , \label{HH}
\end{equation}
has $SO(4)$ symmetry. Here ${\hat{H}}_{H}$ is the ``simple"
Hubbard model,

\begin{equation}
\hat{T}=-t\sum_{\sigma=\uparrow ,\,\downarrow
}\sum_{j=1}^{N_a}\Bigl[c_{j,\,\sigma}^{\dag}\,c_{j+1,\,\sigma} +
h. c.\Bigr] \, , \label{opT}
\end{equation}
is the {\it kinetic-energy} operator, and

\begin{equation}
\hat{D} =
\sum_{j=1}^{N_a}c_{j,\,\uparrow}^{\dag}\,c_{j,\,\uparrow}\,
c_{j,\,\downarrow}^{\dag}\,c_{j,\,\downarrow} =
\sum_{j=1}^{N_a}\hat{n}_{j,\,\uparrow}\,\hat{n}_{j,\,\downarrow}
\, , \label{opD}
\end{equation}
is the electron double-occupation operator. On the right-hand side
of Eq. (\ref{H}) we have that $\mu_c=\mu$, $\mu_s=\mu_0 H$,
$\mu_0$ is the Bohr magneton, and the number operators,

\begin{equation}
{\hat{S }}_c^z=-{1\over 2}[N_a-\hat{N}] \, ; \hspace{1cm} {\hat{S
}}_s^z= -{1\over 2}[{\hat{N}}_{\uparrow}- {\hat{N}}_{\downarrow}]
\, , \label{Sz}
\end{equation}
are the diagonal generators of the $\eta$-spin and spin $SU(2)$
algebras \cite{HL,Yang89,Essler}, respectively. We consider that
the number of lattice sites $N_a$ is large and even and that
$N_a/2$ is odd. The electronic number operators on the right-hand
side of Eq. (\ref{Sz}) read ${\hat{N}}=\sum_{\sigma=\uparrow
,\,\downarrow }\,\hat{N}_{\sigma}$ and
${\hat{N}}_{\sigma}=\sum_{j=1}^{N_a}\hat{N}_{j,\,\sigma}$, where
the operator $\hat{N}_{j,\,\sigma} = c_{j,\,\sigma }^{\dagger
}\,c_{j,\,\sigma }$ counts the number of spin $\sigma$ electrons
at real-space lattice site $j$. On the right-hand side of Eqs.
(\ref{opT})-(\ref{Sz}) the operator $c_{j,\,\sigma}^{\dagger}$
(and $c_{j,\,\sigma}$) creates (and annihilates) a spin $\sigma $
electron at lattice site $j=1,2,...,N_a$. We denote the lattice
constant by $a$ and the lattice length by $L=N_a\,a$.

The momentum operator is given by,

\begin{equation}
\hat{P} = \sum_{\sigma=\uparrow ,\,\downarrow }\sum_{k}\,
\hat{N}_{\sigma} (k)\, k = {L\over 2\pi}\sum_{\sigma=\uparrow
,\,\downarrow }\,\int_{-\pi/a}^{+\pi/a} dk\, \hat{N}_{\sigma}
(k)\, k \, . \label{Popel}
\end{equation}
Here the spin $\sigma$ momentum distribution operator reads
$\hat{N}_{\sigma} (k) = c_{k,\,\sigma }^{\dagger }\,c_{k,\,\sigma
}$, where the operator $c_{k,\,\sigma}^{\dagger}$ (and
$c_{k,\,\sigma}$) creates (and annihilates) a spin $\sigma $
electron at momentum $k$. The operators $c_{k,\,\sigma}^{\dagger}$
and $c_{k,\,\sigma}$ are related to the above operators
$c_{j,\,\sigma}^{\dagger}$ and $c_{j,\,\sigma}$ by the following
Fourier transforms,

\begin{equation}
c_{k,\,\sigma}^{\dagger} =
{1\over\sqrt{L}}\sum_{j=1}^{N_a}e^{ik\,aj}\,
c_{j,\,\sigma}^{\dagger} \, ; \hspace{1cm} c_{k,\,\sigma} =
{1\over\sqrt{L}}\sum_{j=1}^{N_a}e^{-ik\,aj}\, c_{j,\,\sigma} \, .
\label{relkj}
\end{equation}

The Hamiltonian $\hat{H}_{SO(4)}$ given in Eq. (\ref{HH}) commutes
with the six generators of the $\eta$-spin and spin $SU(2)$
algebras and has $SO(4)$ symmetry \cite{HL,Yang89,Essler}. While
the expressions of the two corresponding diagonal generators are
given in Eq. (\ref{Sz}), the off-diagonal generators of these two
$SU(2)$ algebras read

\begin{equation}
{\hat{S}}_c^{\dagger}=\sum_{j=1}^{N_a}(-1)^j
c_{j,\,\downarrow}^{\dagger}\,c_{j,\,\uparrow}^{\dagger} \, ;
\hspace{1cm} {\hat{S}}_c =\sum_{j=1}^{N_a}(-1)^j
c_{j,\,\uparrow}\,c_{j,\,\downarrow} \, , \label{Sc}
\end{equation}
and

\begin{equation}
{\hat{S}}_s^{\dagger}=
\sum_{j=1}^{N_a}c_{j,\,\downarrow}^{\dagger}\,c_{j,\,\uparrow} \,
; \hspace{1cm}
{\hat{S}}_s=\sum_{j=1}^{N_a}c_{j,\,\uparrow}^{\dagger}\,
c_{j,\,\downarrow} \, , \label{Ss}
\end{equation}
respectively.

Throughout this paper we use units of Planck constant one and
denote the electronic charge by $-e$. The Bethe-ansatz solvability
of the 1D Hubbard model (\ref{H}) is restricted to the Hilbert
subspace spanned by the lowest-weight states (LWSs)
\cite{Lieb,Takahashi} or highest-weight states (HWSs)
\cite{Martins98} of the $\eta$-spin and spin algebras, that is by
the states whose $S_{\alpha}$ and $S_{\alpha}^z$ numbers are such
that $S_{\alpha}= -S_{\alpha}^z$ or $S_{\alpha}=S_{\alpha}^z$,
respectively, where $\alpha =c$ for charge and $\alpha =s$ for
spin. In this paper we choose the $\eta$-spin and spin LWSs
description of the Bethe-ansatz solution. In this case, that
solution describes energy eigenstates with electronic densities
$n=N/L$ and spin densities $m=[N_{\uparrow}-N_{\downarrow}]/L$ in
the domains $0\leq n \leq 1/a$ and $0\leq m \leq n$, respectively.
Some of our results correspond to the ranges $0< n < 1/a$ and $0<
m < n$. The description of the states corresponding to the
extended domains $0\leq n \leq 1/a$\, ; $1/a\leq n \leq 2/a$ and
$-n\leq m \leq n$\, ; $-(2/a-n)\leq m \leq (2/a-n)$, respectively,
is achieved by application onto the latter states of off-diagonal
generators of the $\eta$-spin and spin $SU(2)$ algebras
\cite{I,Essler}.

Each lattice site $j=1,2,...,N_a$ of the model (\ref{H}) can
either be doubly occupied, empty, or singly occupied by a
spin-down or spin-up electron. The maximum number of electrons is
$2N_a$ and corresponds to density $n=2/a$. Besides the $N$
electrons, it is useful to consider $[2N_a-N]$ {\it electronic
holes}. (Here we use the designation {\it electronic hole} instead
of {\it hole}, in order to distinguish this type of hole from the
pseudoparticle hole and pseudofermion hole.) Our definition of
electronic hole is such that when a lattice site is empty, we say
that it is occupied by two electronic holes. If a lattice site is
singly occupied, we say that it is occupied by an electron and an
electronic hole. If a lattice site is doubly occupied, it is
unoccupied by electronic holes. The same definition holds for the
rotated-electronic holes. We note that the lattice occupied by
rotated electrons is identical to the original electronic lattice.

The electron - rotated-electron unitary transformation maps the
electrons onto rotated electrons such that rotated-electron double
occupation, no occupation, and spin-up and spin-down single
occupation are good quantum numbers for all values of $U/t$. We
call $c_{j,\,\sigma}^{\dag}$ the electrons that occur in the 1D
Hubbard model (\ref{H}) and (\ref{HH}), while the operator
${\tilde{c}}_{j,\,\sigma}^{\dag}$ such that
${\tilde{c}}_{j,\,\sigma}^{\dag} =
{\hat{V}}^{\dag}(U/t)\,c_{j,\,\sigma}^{\dag}\,{\hat{V}}(U/t)$
represents the rotated electrons, where the electron -
rotated-electron unitary operator ${\hat{V}}(U/t)$ is defined
below. Similarly, $c_{j,\,\sigma}^{\dag} =
{\hat{V}}(U/t)\,{\tilde{c}}_{j,\,\sigma}^{\dag}\,{\hat{V}}^{\dag}(U/t)$.
Note that $c_{j,\,\sigma}^{\dag}$ and
${\tilde{c}}_{j,\,\sigma}^{\dag}$ are only identical in the
$U/t\rightarrow\infty$ limit where electron double occupation
becomes a good quantum number.

The operators ${\hat{V}}^{\dag}(U/t)$ and ${\hat{V}}(U/t)$
associated with the electron - rotated-electron unitary
transformation can be written as,

\begin{equation}
{\hat{V}}^{\dag}(U/t) = e^{-\hat{S}} \, ; \hspace{0.5cm}
{\hat{V}}(U/t) = e^{\hat{S}} \, . \label{SV}
\end{equation}
The operator $\hat{S}$ of Eq. (\ref{SV}) is uniquely defined by
the following two equations,

\begin{equation}
{\tilde{H}}_H = {\hat{V}}^{\dag}(U/t)\,{\hat{H}}_H\,{\hat{V}}(U/t)
= {\hat{H}}_H + [{\hat{H}}_H,\,{\hat{S}}\,] + {1\over
2}\,[[{\hat{H}}_H,\,{\hat{S}}\,],\,{\hat{S}}\,] + ... \,
,\label{HHtil}
\end{equation}
and

\begin{equation}
[{\hat{H}}_H,\,{\hat{V}}^{\dag}(U/t)\,\hat{D}\,{\hat{V}}(U/t)] =
[{\hat{H}}_H,\,\tilde{D}] = 0 \, , \label{HHDtil}
\end{equation}
where the Hamiltonian ${\hat{H}}_H$ is given in Eq. (\ref{HH}) and
the rotated-electron double occupation operator $\tilde{D}$ reads,

\begin{equation}
\tilde{D} \equiv {\hat{V}}^{\dag}(U/t)\,\hat{D}\,{\hat{V}}(U/t) =
\sum_{j}\, {\tilde{c}}_{j,\,\uparrow }^{\dagger
}\,{\tilde{c}}_{j,\,\uparrow }\, {\tilde{c}}_{j,\,\downarrow
}^{\dagger }\,{\tilde{c}}_{j,\,\downarrow } \, . \label{Doptil}
\end{equation}
Here $\hat{D}$ is the electron double occupation operator given in
Eq. (\ref{opD}). The operator (\ref{Doptil}) commutes with the 1D
Hubbard model. We denote the rotated-electron double occupation by
$D_r$. It is a good quantum number for all values of $U/t$.

The transformation associated with the electron - rotated-electron
unitary operator ${\hat{V}}(U/t)$ was introduced in Ref.
\cite{Harris}. The studies of that reference referred to large
values of $U/t$ and did not clarify for arbitrary values of $U/t$
the relation of rotated-electron double occupation to the quantum
numbers provided by the Bethe-ansatz solution. However, this
transformation is uniquely defined for all values of $U/t$ by Eqs.
(\ref{SV})-(\ref{HHDtil}). Equations (\ref{HHtil}) and
(\ref{HHDtil}) can be used to derive an expression for the unitary
operator order by order in $t/U$. The authors of Ref. \cite{Mac}
carried out this expansion up to eighth order (see foot note
[12]).

%%%%%%%%%%%%%%%%%%%%%%%%%%%%%%%%%%%%%%%%%%%%%%%%%%%%%%%%%%%%%%%%
\section{THE PSEUDOPARTICLE OPERATORS AND THE HOLON, SPINON, AND
$c0$ PSEUDOPARTICLE BASIC DESCRIPTION}

According to the studies of Ref. \cite{Carmelo97}, there is an
infinite number of pseudoparticle branches: the $c0$
pseudoparticles and the $\alpha\nu$ pseudoparticles such that
$\alpha =c,\,s$ and $\nu=1,2,...$. The $\alpha\nu$ pseudoparticle
notation considered in this paper is related to the $c$
pseudoparticle and $\alpha,\gamma$ pseudoparticle notation of Ref.
\cite{Carmelo97} as follows: $c0\equiv c$ for $\nu =0$,
$\nu=\gamma$ and $c\nu\equiv c,\gamma$ for $\nu =1,2,...$, and
$\nu=\gamma +1$ and $s\nu\equiv s,\gamma +1$ for $\nu=1,2,...$.
Moreover, we denote by {\it bare momentum} $q$ the momentum
carried by the pseudoparticles, while in Ref. \cite{Carmelo97} it
was called {\it band momentum}. Our designation is justified by
the form of the pseudofermion momentum, as further discussed
below. Note that within our notation, the general designation of
$\alpha\nu$ pseudoparticle refers to the $\alpha =c$ branches such
that $\nu =0,1,2,...$ and $\alpha =s$ branches such that $\nu
=1,2,...$. Elsewhere it is shown that the $c\nu$ pseudoparticles
and $s\nu$ pseudoparticles are $2\nu$-holon and $2\nu$-spinon
composite objects, respectively, where $\nu =1,2,...$ \cite{I}.

The introduction of the pseudofermion operational description
studied in this paper requires the use of an operator
representation for the pseudoparticles. Thus, in this section we
introduce an operational description for the $\alpha\nu$
pseudoparticles and the holons and spinons which are not part of
composite pseudoparticles. An operational description for the
pseudoparticles of bare-momentum $q$ was introduced in Ref.
\cite{Carmelo97}. However, such a description did not take into
account the holon (and spinon) composite character of the $c\nu$
pseudoparticles (and $s\nu$ pseudoparticles). Moreover, the spinon
description used in the studies of that reference is not valid for
the whole Hilbert space. This affected the values of the entries
of the pseudoparticle statistical-interaction matrix
\cite{Haldane91}. In this section we provide the correct values
for the entries of such a matrix. Another limitation was the lack
of a representation for the pseudoparticle operators in terms of
spatial coordinates. The concepts of local $\alpha\nu$
pseudoparticle and effective $\alpha\nu$ lattice are introduced in
Ref. \cite{IIIb}. The studies of that reference include a
description of the local pseudoparticles in terms of the
rotated-electron site distribution configurations. The absence of
such a local operational description is one of the reasons why the
studies of Ref. \cite{Carmelo97} did not provide useful
information about the main issue of the relation between the
pseudoparticle and electronic operators. Also the concept of
rotated electron is important for the study of that issue and was
not considered in the studies of such a reference. All these
concepts are valuable and necessary for the application of the
operational pseudofermion representation introduced in this paper
to the evaluation of few-electron spectral functions \cite{V}.

In Appendix A we summarize the basic properties of the $\alpha\nu$
pseudoparticles which are needed for our studies. This includes
introduction to the $\alpha\nu$ pseudoparticle bare momentum,
effective $\alpha\nu$ lattice, and useful ground-state quantities.

%%%%%%%%%%%%%%%%%%%%%%%%%%%%%%%%%%%%%%%%%%%%%%%%%%%%%%%%%%%%%%%%
\subsection{THE PSEUDOPARTICLE OPERATORS AND PSEUDOPARTICLE STATISTICS
ACCORDING TO THE GENERALIZED PAULI PRINCIPLE}

Generation and removal of pseudoparticles is in general associated
with creation and/or annihilation of electrons. Yet there are also
transitions which change the numbers of these quantum objects at
constant spin $\sigma$ electron numbers. One can introduce
elementary operators for creation or annihilation of $\alpha\nu$
pseudoparticles. In this subsection, we introduce two alternative
representations corresponding to pseudoparticle operators, both in
terms of the bare-momentum $q$ and spatial coordinates. These
representations refer to the bare-momentum pseudoparticles and
local pseudoparticles \cite{IIIb}, respectively.

Let us introduce the bare-momentum $\alpha\nu$ pseudoparticle
creation (and annihilation) operator $b^{\dag }_{q,\,\alpha\nu}$
(and $b_{q,\,\alpha\nu}$) which creates (and annihilates) a
$\alpha\nu$ pseudoparticle of bare momentum $q$. In addition, we
introduce the local $\alpha\nu$ pseudoparticle creation operator
$b^{\dag }_{x_j,\,\alpha\nu}$ and annihilation operator
$b_{x_j,\,\alpha\nu}$. These bare-momentum and local
pseudoparticle operators are related as follows,

\begin{equation}
b^{\dag }_{q,\,\alpha\nu} =
{1\over\sqrt{L}}\sum_{j=1}^{N^*_{\alpha\nu}}e^{iq\,x_j}\, b^{\dag
}_{x_j,\,\alpha\nu} \, ; \hspace{1cm} b_{q,\,\alpha\nu} =
{1\over\sqrt{L}}\sum_{j=1}^{N^*_{\alpha\nu}}e^{-iq\,x_j}\,
b_{x_j,\,\alpha\nu} \, . \label{elop}
\end{equation}
The local $\alpha\nu$ pseudoparticle creation (and annihilation)
operator $b^{\dag }_{x_j,\,\alpha\nu}$ (and $b_{x_j,\,\alpha\nu}$)
creates (and annihilates) a local $\alpha\nu$ pseudoparticle at
the effective $\alpha\nu$ lattice site of spatial coordinate $x_j
=a_{\alpha\nu}\,j$, where $j=1,2,...,N^*_{\alpha\nu}$. The
effective $\alpha\nu$ lattice constant $a_{\alpha\nu}$ is defined
in Eq. (\ref{aan}) of Appendix A, where the concept of an
effective $\alpha\nu$ lattice, introduced in Ref. \cite{IIIb} is
described. The conjugate variable of the bare-momentum $q_j$ of
the $\alpha\nu$ pseudoparticle branch is the space coordinate
$x_j$ of the corresponding effective $\alpha\nu$ lattice. This is
different to the electronic operators of Eq. (\ref{relkj}), where
the conjugate variable of the momentum $k_j$ is the space variable
of the original electronic lattice. In reference \cite{IIIb}, the
pseudoparticle site distribution configurations in the effective
$\alpha\nu$ lattices are related to the corresponding
rotated-electron site distribution configurations.

The pseudoparticles obey a Pauli principle relative to the
bare-momentum occupancy configurations, {\it i.e.} a discrete
bare-momentum value $q_j$ can either be unoccupied or singly
occupied by a pseudoparticle. For $\nu>0$ composite $\alpha\nu$
pseudoparticles, the number of discrete momentum values
$N^*_{\alpha\nu}$ is not the same for all energy eigenstates.
Thus, these objects cannot be classified as fermions or bosons. In
order to classify the $\alpha\nu$ pseudoparticle according to the
generalized Pauli principle introduced in Ref. \cite{Haldane91},
we consider the $\alpha\nu$ dimensions,

\begin{equation}
d_{\alpha\nu}\equiv 1+N^*_{\alpha\nu} -N_{\alpha\nu} = 1 +
N^h_{\alpha\nu} \, , \label{dan}
\end{equation}
where according to Eqs. (\ref{N*})-(\ref{Nhcsn}) of Appendix A,
$N_{\alpha\nu}$ and $N^h_{\alpha\nu}$ are the number of
$\alpha\nu$ pseudoparticles and $\alpha\nu$ pseudoparticle holes,
respectively, and $N^*_{\alpha\nu}=N_{\alpha\nu}+
N^h_{\alpha\nu}$. A transition to an excited energy eigenstate
producing deviations $\Delta N_{\alpha\nu}$ in the $\alpha\nu$
pseudoparticle numbers, leads to the following deviations in the
corresponding $\alpha\nu$ dimension,

\begin{equation}
\Delta
d_{\alpha\nu}=-\sum_{\alpha'=c,\,s}\,\sum_{\nu'=\delta_{\alpha',\,s}}^{\infty}
g_{\alpha\nu,\,\alpha'\nu'}\,\Delta N_{\alpha'\nu'} \, .
\label{Ddang}
\end{equation}
According to the generalized Pauli principle \cite{Haldane91}, the
parameters $g_{\alpha\nu,\,\alpha'\nu'}$ on the right-hand side of
this equation are the entries of the statistical-interaction
matrix. From the use of Eqs. (\ref{N*})-(\ref{Nhcsn}) of Appendix
A, we find that for the $\alpha\nu$ pseudoparticles such a
statistical-interaction matrix has infinite dimension and its
entries are given by,

\begin{equation}
g_{c0,\,\alpha\nu}=\delta_{\alpha,\,c}\,\delta_{\nu,\,0} \, ;
\hspace{1cm} g_{\alpha\nu,\,c0} = \delta_{\alpha,\,c}
-\delta_{\alpha,\,s} \, ; \hspace{1cm}
g_{\alpha\nu,\,\alpha'\nu'}= \delta_{\alpha,\,\alpha'}\, \Bigl(\nu
+ \nu' - \vert\nu - \nu'\vert\Bigl) \, ; \hspace{0.5cm} \nu
,\,\nu'
>0 \, . \label{gan}
\end{equation}
This fully defines the statistics of the $\alpha\nu$
pseudoparticles. We emphasize that the $c0$ pseudoparticle entry
$g_{c0,\,\alpha\nu}$ given in Eq. (\ref{gan}) has Fermionic
character. This is related to the fact that the number of sites of
the effective $c0$ lattice is constant and given by $N_a$, and
that the corresponding effective lattice constant $a_{c0}$ equals
the electronic lattice constant $a$. Thus, the effective $c0$
lattice and electronic lattice are identical. Furthermore, it is
found in Ref. \cite{IIIb} that the sites occupied by $c0$
pseudoparticles (and $c0$ pseudoparticle holes) are the same as
the sites singly occupied by rotated electrons (and doubly
occupied and unoccupied by rotated electrons). While the $c0$
pseudoparticles {\it do not feel} the statistical-interactions of
the remaining pseudoparticles, the $\nu >0$ composite $\alpha\nu$
pseudoparticles {\it feel} the statistical interactions of the
$c0$ pseudoparticles, as confirmed by Eq. (\ref{gan}). The form of
the entries given in Eq. (\ref{gan}) confirms that the latter
composite pseudoparticles are neither fermions nor bosons
\cite{Haldane91}. The statistical interactions of the composite
pseudoparticles result in part from the property that the width of
the bare-momentum domain $2q_{\alpha\nu} =
2\pi[1/a_{\alpha\nu}-1/L]$ is different for different values of
the pseudoparticle numbers.

The $\alpha\nu$ pseudoparticle bare-momentum distribution
functions $N_{\alpha\nu}(q)$ play an important role in the
pseudoparticle description \cite{I,II}. These functions are for
all energy eigenstates the eigenvalues of the following
pseudoparticle bare-momentum distribution operators,

\begin{equation}
\hat{N}_{\alpha\nu}(q)= b^{\dag
}_{q,\,\alpha\nu}\,b_{q,\,\alpha\nu} \, . \label{Nanop}
\end{equation}
The bare-momentum distribution functions $N_{\alpha\nu}(q)$ read
$N_{\alpha\nu} (q_j)= 1$ for occupied discrete bare-momentum
values $q_j$ and $N_{\alpha\nu} (q_j)= 0$ for unoccupied discrete
bare-momentum values $q_j$. Each LWS is uniquely specified by the
values of the set of distribution functions $\{N_{\alpha\nu}(q)\}$
such that $\nu =0,1,2,...$ for $\alpha =c$ and $\nu =1,2,...$ for
$\alpha =s$. Physical quantities such as the energy, depend on the
values of these distribution functions and numbers through the
rapidity momentum functional $k(q)$ and rapidity functionals
$\Lambda_{c\nu}(q)$ and $\Lambda_{s\nu}(q)$. The value of these
functionals is uniquely provided by solution of the following
functional integral equations \cite{I,II},

\begin{eqnarray}
k(q) & = & q - {1\over \pi}\sum_{\nu =1}^{\infty}
\int_{-q_{s\nu}}^{q_{s\nu}}\,dq'\, N_{s\nu}(q')\arctan\Bigl({\sin
k(q)-\Lambda_{s\nu}(q') \over \nu U/4t}\Bigr)
\nonumber \\
& - & {1\over \pi}\sum_{\nu =1}^{\infty}
\int_{-q_{c\nu}}^{q_{c\nu}}\,dq'\, N_{c\nu}(q') \arctan\Bigl({\sin
k(q)-\Lambda_{c\nu}(q') \over \nu U/4t}\Bigr) \, , \label{Tapco1}
\end{eqnarray}

\begin{eqnarray}
k_{c\nu}(q) & = & q + {1\over \pi}
\int_{q_{c0}^{-}}^{q_{c0}^{+}}\,dq'\,
N_{c0}(q')\arctan\Bigl({\Lambda_{c\nu}(q)-\sin k(q')\over \nu
U/4t}\Bigr)
\nonumber \\
& + & {1\over 2\pi}\sum_{\nu' =1}^{\infty} \int_{-q_{c,\,\nu
'}}^{q_{c\nu'}}\,dq'\, N_{c\nu'}(q')\,
\Theta_{\nu,\,\nu'}\Bigl({\Lambda_{c\nu}(q)-\Lambda_{c\nu'}(q')
\over U/4t}\Bigr) \, ; \hspace{0.5cm} \nu >0 \, , \label{Tapco2}
\end{eqnarray}
and

\begin{eqnarray}
0 & = & q - {1\over \pi} \int_{q_{c0}^{-}}^{q_{c0}^{+}}\,dq'\,
N_{c0}(q')\arctan\Bigl({\Lambda_{s\nu}(q)-\sin k(q')
\over \nu U/4t}\Bigr)\nonumber \\
& + & {1\over 2\pi}\sum_{\nu' =1}^{\infty} \int_{-q_{s\nu
'}}^{q_{s\nu'}}\,dq'\, N_{s\nu'}(q')\Theta_{\nu, \,\nu'}
\Bigl({\Lambda_{s\nu}(q)-\Lambda_{s\nu'}(q')\over U/4t}\Bigr) \, .
\label{Tapco3}
\end{eqnarray}
Here

\begin{equation}
k_{c\nu}(q) = 2\,{\rm Re}\,\{\arcsin (\Lambda_{c\nu} (q) + i \nu
U/4t)\} \, ; \hspace{0.5cm} \nu >0 \, , \label{kcn}
\end{equation}
is the $c\nu$ rapidity-momentum functional and the limiting
bare-momentum values $q_{c0}^{\pm}$ and $q_{\alpha\nu}$ where
$\alpha =c,\,s$ and $\nu=1,2,...$ are given in Eqs. (\ref{qcev})
and (\ref{qcodd}) and in Eq. (\ref{qag}), respectively, of
Appendix A. The function $\Theta_{\nu,\,\nu'} (x)$ appearing in
Eqs. (\ref{Tapco2}) and (\ref{Tapco3}) is given in Eq.
(\ref{Theta}) of Appendix B. The equations
(\ref{Tapco1})-(\ref{Tapco3}) correspond to a functional
representation of the thermodynamic Bethe-ansatz equations
introduced by Takahashi \cite{Takahashi}. The rapidity-momentum
functional is real and the rapidity functionals are the real part
of Takahashi's ideal strings \cite{Takahashi,I}. It is useful to
introduce the following $c0$ rapidity functional,

\begin{equation}
\Lambda_{c0}(q) \equiv \sin k (q) \, , \label{Lc}
\end{equation}
where $k(q)$ is the rapidity-momentum functional.

%%%%%%%%%%%%%%%%%%%%%%%%%%%%%%%%%%%%%%%%%%%%%%%%%%%%%%%%%%%%%%%%%%%%%%%%%%
\subsection{HOLONS, SPINONS, $c0$ PSEUDOPARTICLES AND THE YANG HOLON AND
HL SPINON ELEMENTARY OPERATORS}

In this subsection we consider a holon, spinon, and $c0$
pseudoparticle description for the whole Hilbert space of the 1D
Hubbard model whose validity is shown elsewhere \cite{I}. Such
elementary quantum objects correspond to specific rotated-electron
site occupations. This description of all energy eigenstates in
terms of occupancy configurations of three elementary quantum
objects only, is useful for the studies of this paper.

Let us distinguish the total $\eta$ spin (and spin) value, which
we denote by $S_c$ (and $S_s$) and the corresponding $\eta$-spin
(and spin) projection, which we denote by $S_c^z$ (and $S_s^z$),
from the $\eta$ spin (and spin) carried by the elementary quantum
objects. We call $s_c$ (and $s_s$) the $\eta$ spin (and spin)
carried by the holons, spinons, and other elementary objects and
$\sigma_c$ (and $\sigma_s$) their $\eta$-spin (and spin)
projection. The operators ${\hat{M}}_{c,\,\sigma_{c}}$ and
${\hat{M}}_{s,\,\sigma_{s}}$ which count the number of the
$\sigma_{c}=\pm 1/2$ holons and $\sigma_{s}=\pm 1/2$ spinons have
the following form \cite{I},

\begin{eqnarray}
{\hat{M}}_{c,\,-1/2} & = & {\hat{V}}^{\dag}(U/t)\,\sum_{j}
c_{j\uparrow}^{\dag}\,c_{j\uparrow}\,
c_{j\downarrow}^{\dag}\,c_{j\downarrow}\,{\hat{V}}(U/t) \, ; \nonumber \\
{\hat{M}}_{c,\,+1/2} & = & {\hat{V}}^{\dag}(U/t)\,\sum_{j}
c_{j\uparrow}\,c_{j\uparrow}^{\dag}\,
c_{j\downarrow}\,c_{j\downarrow}^{\dag}\,{\hat{V}}(U/t) \, ,
\label{Mc-+}
\end{eqnarray}
and

\begin{eqnarray}
{\hat{M}}_{s,\,-1/2} & = & {\hat{V}}^{\dag}(U/t)\,\sum_{j}
c_{j\downarrow}^{\dag}\,c_{j\uparrow}\,
c_{j\uparrow}^{\dag}\,c_{j\downarrow}\,{\hat{V}}(U/t) \, ; \nonumber \\
{\hat{M}}_{s,\,+1/2} & = & {\hat{V}}^{\dag}(U/t)\,\sum_{j}
c_{j\uparrow}^{\dag}\,c_{j\downarrow}\,
c_{j\downarrow}^{\dag}\,c_{j\uparrow}\,{\hat{V}}(U/t) \, ,
\label{Ms-+}
\end{eqnarray}
respectively. Here, the operator $\sum_{j}
c_{j\uparrow}^{\dag}\,c_{j\uparrow}\,
c_{j\downarrow}^{\dag}\,c_{j\downarrow}$ counts the number of
electron doubly-occupied sites, $\sum_{j}
c_{j\uparrow}\,c_{j\uparrow}^{\dag}\,
c_{j\downarrow}\,c_{j\downarrow}^{\dag}$ counts the number of
electron empty sites, $\sum_{j}
c_{j\downarrow}^{\dag}\,c_{j\uparrow}\,
c_{j\uparrow}^{\dag}\,c_{j\downarrow}$ counts the number of
spin-down electron singly-occupied sites, and $\sum_{j}
c_{j\uparrow}^{\dag}\,c_{j\downarrow}\,
c_{j\downarrow}^{\dag}\,c_{j\uparrow}$ counts the number of
spin-up electron singly-occupied sites. The operator
${\hat{V}}(U/t)$ on the right-hand side of Eqs. (\ref{Mc-+}) and
(\ref{Ms-+}) is uniquely defined for all values of $U/t$ by Eqs.
(\ref{SV})-(\ref{HHDtil}). The new physics brought about by the
relations of Eqs. (\ref{Mc-+}) and (\ref{Ms-+}) is that the
electron - rotated-electron unitary transformation generates the
holons and spinons whose occupancy configurations describe the
exact energy eigenstates. Throughout this paper we denote such
holons and spinons according to their value of $\sigma_c =\pm 1/2$
and $\sigma_s =\pm 1/2$, respectively. The unitary rotation is
such that for all values of $U/t$ the number of emerging $-1/2$
holons, $+1/2$ holons, $-1/2$ spinons, and $+1/2$ spinons equals
precisely the number of rotated-electron doubly occupied sites,
empty sites, spin-down singly occupied sites, and spin-up singly
occupied sites, respectively.

The holons have $s_c=1/2$, $s_s=0$, and $\sigma_s =0$. Thus, there
are two types of holons, which have $\sigma_c=-1/2$ and
$\sigma_c=+1/2$ and carry charge $-2e$ and $+2e$, respectively.
These objects are on-site spin-singlet rotated-electron pairs and
rotated-electron unoccupied sites, respectively. An important
property is that rotated-electron double occupation $D_r$ equals
the value $M_{c,\,-1/2}$ for the number of $-1/2$ holons. The
rotated-electron double occupation $D_r=M_{c,\,-1/2}$ plays an
important role in the description of the few-electron spectral
properties of the quantum problem. For few-electron spectral
functions the weight distribution resulting from transitions to
the Hilbert subspace spanned by excited states of rotated-electron
double occupation $D_r$ corresponds to the $D_r^{th}$ upper
Hubbard band, where $D_r=1,2,...$ \cite{V}. Moreover, the spin-up
and spin-down rotated-electron singly occupied sites correspond to
the $\sigma_s=+1/2$ and $\sigma_s =-1/2$ spinons, respectively.
The $s_s =1/2$ spinons have no charge degrees of freedom and thus
describe the spin degrees of freedom of rotated-electron singly
occupied sites only. The charge degrees of freedom of the rotated
electrons (and rotated-electronic holes) of these singly occupied
sites are described by the chargeons of charge $-e$ (and
antichargeons of charge $+e$) \cite{I}. These quantum objects are
part of the charge degrees of freedom, but do not contribute to
the $\eta$-spin and spin $SU(2)$ algebras. In the case of the
description of the transport of charge in terms of electrons (and
electronic holes) the elementary carriers of charge are the
chargeons of charge $-e$ and $-1/2$ holons of charge $-2e$ (and
the antichargeons of charge $+e$ and $+1/2$ holons of charge
$+2e$). The $c0$ pseudoparticle has no $\eta$-spin and no spin
degrees of freedom and is a composite quantum object which
contains a chargeon and an antichargeon. However, in how transport
of charge is concerned, the chargeon and antichargeon correspond
to alternative descriptions of the $c0$ pseudoparticle. When the
transport of charge is described in terms of electrons (and
electronic holes) the $c0$ pseudoparticle couples to charge probes
through the chargeon (and antichargeon) and carries elementary
charge $-e$ (and $+e$).

All energy eigenstates of the 1D Hubbard model can be described in
terms of occupancy configurations of holons, spinons, and $c0$
pseudoparticles. On the other hand, in Ref. \cite{Carmelo97} it
was found that all energy eigenstates associated with the 1D
Hubbard model Bethe-ansatz solution \cite{Lieb,Takahashi}, can be
described in terms of occupancy configurations of $\alpha\nu$
pseudoparticles. By merging these two representations, one finds
that the $c\nu$ pseudoparticles such that $\nu\neq 0$ and the
$s\nu$ pseudoparticles are $2\nu$-holon and $2\nu$-spinon
composite objects, respectively \cite{I}. The composite $c\nu$
pseudoparticles (and $s\nu$ pseudoparticles) are $s_c =\sigma_c=0$
(and $s_s =\sigma_s =0$) objects without spin (and charge) degrees
of freedom. These composite quantum objects contain an equal
number $\nu$ of $-1/2$ holons and $+1/2$ holons (and $-1/2$
spinons and $+1/2$ spinons). When the transport of charge is
described in terms of electrons (and electronic holes), the
$2\nu$-holon composite $c\nu$ pseudoparticles couple to charge
probes through their $\nu$ $-1/2$ holons of charge $-2e$ (and
their $\nu$ $+1/2$ holons of charge $+2e$). Thus these composite
quantum objects carry elementary charge $-2\nu\,e$ (and
$+2\nu\,e$).

The $\pm 1/2$ holons (and $\pm 1/2$ spinons) which are not part of
$2\nu$-holon composite $c\nu$ pseudoparticles (and $2\nu$-spinon
composite $s\nu$ pseudoparticles) are called $\pm 1/2$ Yang holons
(and $\pm 1/2$ HL spinons). In the designations {\it HL spinon}
and {\it Yang holon}, HL stands for Heilmann and Lieb and Yang
refers to C. N. Yang, respectively, who are the authors of Refs.
\cite{HL,Yang89}. Note that the holons (and spinons) which are
part of $s_c=0$ (and $s_s=0$) composite $c\nu$ pseudoparticles
(and $s\nu$ pseudoparticles) remain invariant under application of
the off-diagonal generators of the $\eta$-spin (and spin) $SU(2)$
algebras given in Eq. (\ref{Sc}) (and (\ref{Ss})). Application of
these generators produces $\eta$-spin (and spin) flips in $\pm
1/2$ Yang holons (and $\pm 1/2$ HL spinons). Thus, the $-1/2$ Yang
holons and $-1/2$ HL spinons are created by application onto the
LWSs of suitable off-diagonal generators of the $\eta$-spin and
spin $SU(2)$ algebras, respectively, given in Eqs. (\ref{Sc}) and
(\ref{Ss}). The generators (\ref{Sc}) and (\ref{Ss}) can be
written in terms of rotated-electron operators as,

\begin{equation}
d^{\dag}_{q_c,\,-1/2} = \sum_{j=1}^{N_a}(-1)^j
{\tilde{c}}_{j,\,\downarrow}^{\dagger}\,{\tilde{c}}_{j,\,\uparrow}^{\dagger}
\, ; \hspace{1cm} d_{q_c,\,-1/2} =  \sum_{j=1}^{N_a}(-1)^j
{\tilde{c}}_{j,\,\uparrow}\,{\tilde{c}}_{j,\,\downarrow} \, .
\label{dc-e-re}
\end{equation}
and

\begin{equation}
d^{\dag}_{q_{s},\,-1/2} =
\sum_{j=1}^{N_a}{\tilde{c}}_{j,\,\downarrow}^{\dagger}\,
{\tilde{c}}_{j,\,\uparrow} \, ; \hspace{1cm} d_{q_{s},\,-1/2} =
\sum_{j=1}^{N_a}{\tilde{c}}_{j,\,\uparrow}^{\dagger}\,
{\tilde{c}}_{j,\,\downarrow} \, , \label{ds-e-re}
\end{equation}
respectively. In these equations $q_c=\pi/a$ and $q_s=0$ is the
momentum. The operators (\ref{dc-e-re}) and (\ref{ds-e-re}) are
invariant under the electron - rotated-electron unitary
transformation \cite{I}. Therefore, they have precisely the same
expression in terms of electron and rotated-electron creation and
annihilation operators.

Note that within the operational representation of Eqs.
(\ref{dc-e-re}) and (\ref{ds-e-re}) the Yang $+1/2$ holons and HL
$+1/2$ spinons are not explicitly considered. Indeed, the numbers
$L_{c,\,+1/2}$ of $+1/2$ Yang holons and $L_{s,\,+1/2}$ of $+1/2$
HL spinons are fully determined by the numbers $L_{c,\,-1/2}$ of
Yang $-1/2$ holons, $L_{s,\,-1/2}$ of HL $-1/2$ spinons, and the
set $\{N_{\alpha\nu}\}$ of the different $\alpha\nu$ pseudofermion
branches. This justifies why here we consider the operators of
Eqs. (\ref{dc-e-re}) and (\ref{ds-e-re}) as creation and
annihilation operators for $-1/2$ Yang holons and $-1/2$ HL
spinons, respectively. When applied onto LWSs, the operators
$d^{\dag}_{q_{\alpha},\,-1/2}$ produce energy eigenstates with
finite values for the numbers $L_{c,\,-1/2}$ and/or $L_{s,\,-1/2}$
of the following form,

\begin{equation}
\vert \{L_{c,\,-1/2},\,L_{s,\,-1/2}\}\rangle = \prod_{\alpha
=c,\,s}\frac{(d^{\dag}_{q_{\alpha},\,-1/2})^{L_{\alpha,\,-1/2}}}{
\sqrt{L_{\alpha}}}\vert LWS\rangle \, . \label{IRREG}
\end{equation}
Here $\vert LWS\rangle$ is the LWS that corresponds to the state
$\vert \{L_{c,\,-1/2},\,L_{s,\,-1/2}\}\rangle$ and
$L_{\alpha}=L_{\alpha,\,+1/2}+L_{\alpha,\,-1/2}=2S_{\alpha}$ where
$S_{\alpha}$ is the state $\eta$-spin value ($\alpha =c$) and spin
value ($\alpha =s$). The energy eigenstates (\ref{IRREG}) are not
described by the Bethe-ansatz solution \cite{I,Essler}.

The pseudoparticle bare momentum $q$ is such that $q\in
(-q_{\alpha\nu},\,+q_{\alpha\nu})$ where the limiting
bare-momentum values $\pm q_{\alpha\nu}$ associated with the
limits of the $\alpha\nu$ pseudoparticle {\it Brillouin zone} are
given in Eq. (\ref{qag}) of Appendix A. According to the results
of Ref. \cite{II}, the $\pm 1/2$ Yang holons, $\pm 1/2$ HL
spinons, and $\nu > 0$ composite $\alpha\nu$ pseudoparticles of
bare-momentum values $q=\pm q_{\alpha\nu}$ are non-interacting and
localized quantum objects. Such behavior results from the
invariance of these quantum objects under the electron -
rotated-electron unitary transformation. This means that the $\pm
1/2$ Yang holons, $\pm 1/2$ HL spinons, and $\nu > 0$ composite
$\alpha\nu$ pseudoparticles of bare-momentum values $q=\pm
q_{\alpha\nu}$ are the same quantum objects as the corresponding
rotated objects. Thus, these objects are localized and do not
contribute to the transport of charge or spin. However, in general
a $\nu > 0$ composite $\alpha\nu$ pseudoparticle is different from
a $\nu > 0$ rotated composite $\alpha\nu$ pseudoparticle. The only
exception is precisely for bare momentum values $q$ such that
$q\rightarrow \pm q_{\alpha\nu}$. As the bare momentum approaches
its limiting values, $q\rightarrow \pm q_{\alpha\nu}$, the $\nu >
0$ composite $\alpha\nu$ pseudoparticle and the $\nu > 0$ rotated
composite $\alpha\nu$ pseudoparticle become the same quantum
object. Importantly, the electrons and rotated electrons involved
in processes associated with creation of $\pm 1/2$ Yang holons,
$\pm 1/2$ HL spinons, and $\nu > 0$ composite $\alpha\nu$
pseudoparticles of bare-momentum values $q=\pm q_{\alpha\nu}$ are
also the same quantum object, {\it i.e.} remain also invariant
under the electron - rotated-electron unitary transformation. This
always refers to localized electrons. It follows that the
transport of charge (and spin) is associated with the $c0$
pseudoparticle and $q\neq \pm q_{c\nu}$ composite $c\nu$
pseudoparticle quantum charge fluids (and $q\neq \pm q_{s\nu}$
composite $s\nu$ pseudoparticle quantum spin fluids) where
$\nu=1,2,...$.

Within the pseudoparticle, Yang holon, and HL spinon operational
description, the $\pm 1/2$ holon ($\alpha =c$) and $\pm 1/2$
spinon ($\alpha =s$) number operators ${\hat{M}}_{\alpha,\,\pm
1/2}$ given in Eqs. (\ref{Mc-+}) and (\ref{Ms-+}) are written in
terms of pseudoparticle operators and $\pm 1/2$ Yang holon
($\alpha =c$) or $\pm 1/2$ HL spinon ($\alpha =s$) number
operators ${\hat{L}}_{\alpha,\,\pm 1/2}$ as follows,

\begin{equation}
{\hat{M}}_{\alpha,\,\pm 1/2} = {\hat{L}}_{\alpha,\,\pm 1/2} +
\sum_{\nu=1}^{\infty}\,\sum_{q=-q_{\alpha,\,\nu
}}^{+q_{\alpha,\,\nu }}\nu\,{\hat{N}}_{\alpha\nu} (q) \, ,
\label{Mop}
\end{equation}
where the pseudoparticle bare-momentum distribution operators
$\hat{N}_{\alpha\nu}(q)$ are provided in Eq. (\ref{Nanop}). The
operator ${\hat{L}}_{\alpha,\,\pm 1/2}$ can be written as,

\begin{equation}
{\hat{L}}_{\alpha ,\,\pm 1/2} =
\sqrt{{\vec{\hat{S}}}_{\alpha}.{\vec{\hat{S}}}_{\alpha} + 1/4} -
1/2 \mp {\hat{S}}_{\alpha}^z \, ; \hspace{1cm} \alpha = c,s \, .
\label{NYhHLsop}
\end{equation}
Here ${\hat{S}}_{\alpha}^z$ is the diagonal generator of the
$\eta$-spin ($\alpha =c$) and spin ($\alpha =s$) algebras whose
expression in terms of electronic operators is provided in Eq.
(\ref{Sz}). The spin $\sigma$ electron number operator commutes
with the electron - rotated-electron unitary operator. Thus, the
operator ${\hat{S}}_{\alpha}^z$ has the same expression in terms
of electron and rotated-electron creation and annihilation
operators. The same occurs for the $\eta$-spin ($\alpha =c$) and
spin ($\alpha =s$) operator
${\vec{\hat{S}}}_{\alpha}.{\vec{\hat{S}}}_{\alpha}$. This operator
can be expressed in terms of rotated-electron creation and
annihilation operators by use of Eqs. (\ref{dc-e-re}) and
(\ref{ds-e-re}). Therefore, Eq. (\ref{NYhHLsop}) provides the
expression of the operator ${\hat{L}}_{\alpha,\,\pm 1/2}$ both in
terms of electron and rotated-electron creation and annihilation
operators.

An electronic ensemble space is spanned by all energy eigenstates
with the same values for the numbers $N_{\uparrow}$ and
$N_{\downarrow}$. An important concept is that of {\it CPHS
ensemble space} \cite{II}. This is a subspace spanned by all
energy eigenstates with the same values for the numbers of $\pm
1/2$ holons and $\pm 1/2$ spinons $\{M_{\alpha,\,\pm 1/2}\}$,
where $\alpha =c,s$. In general, an electronic ensemble space
contains several CPHS ensemble spaces. Moreover, usually a CPHS
ensemble space includes different {\it CPHS ensemble subspaces}. A
CPHS ensemble subspace is spanned by all energy eigenstates with
the same values for the sets of numbers $N_{c0}$,
$\{N_{\alpha\nu}\}$, and $\{L_{\alpha,\,-1/2}\}$ such that $\alpha
=c,s$ and $\nu =1,2,...$. (According to the notation of Ref.
\cite{II}, CPHS stands for $c$ pseudoparticle, holon, and spinon.)

%%%%%%%%%%%%%%%%%%%%%%%%%%%%%%%%%%%%%%%%%%%%%%%%%%%%%%%%%%%%%%%%%%%%%%%%%%
\subsection{THE GENERAL ENERGY SPECTRUM AND THE MOMENTUM OPERATOR}

The generators of the $\eta$-spin and spin $SU(2)$ algebras given
in Eqs. (\ref{dc-e-re}) and (\ref{ds-e-re}) commute with the $c0$
pseudoparticle and composite $\alpha\nu$ pseudoparticle creation
and annihilation operators of Eq. (\ref{elop}). We emphasize that
this property has the following important effect: All
$2S_{\alpha}$ energy eigenstates obtained from a given regular
energy eigenstate have the same pseudoparticle momentum
distribution functions $N_{c0} (q)$ and $\{N_{\alpha\nu}(q)\}$ for
all branches $\alpha =c,s$ and $\nu =1,2,...$. Thus these states
are described by similar pseudoparticle occupancy configurations
and only differ in the relative numbers of $+1/2$ Yang holons and
$-1/2$ Yang holons ($\alpha =c$), or/and in the relative numbers
of $+1/2$ HL spinons and $-1/2$ HL spinons ($\alpha =s$). This
reveals that the coupled functional equations
(\ref{Tapco1})-(\ref{Tapco3}), which involve the pseudoparticle
momentum distribution functions and do not depend on the
$L_{\alpha ,\,\sigma_{\alpha}}$ numbers, describe both LWSs and
non-LWSs. The eigenvalues $E$ of the energy eigenstates of the
Hamiltonian (\ref{H}) can be written in the following form,

\begin{equation}
E=E_{SO(4)} + \sum_{\alpha =c,\,s}\mu_{\alpha }\,S^{\alpha}_z \, ,
\label{E}
\end{equation}
where

\begin{equation}
E_{SO(4)} = E_H + {U\over 2}\,\Bigl[\,M_c - 2M_{c,\,-1/2}
-{N_a\over 2}\Bigr] \, , \label{ESO4}
\end{equation}
and

\begin{eqnarray}
E_H & = & -2t {L\over 2\pi} \int_{q_{c0}^{-}}^{q_{c0}^{+}} dq\,
N_{c0} (q)\, \cos k(q)
\nonumber \\
& + & 4t {L\over 2\pi}\sum_{\nu=1}^{\infty} \int_{-q_{c,\,\nu
}}^{+q_{c\nu}} dq\, N_{c\nu} (q)\,{\rm Re}\,\Bigl\{\sqrt{1 -
(\Lambda_{c\nu} (q) + i \nu\,U/4t)^2}\Bigr\} +U\,L_{c,\,-1/2}\, .
\label{EH}
\end{eqnarray}
On the right-hand side of Eq. (\ref{E}) the numbers
$S_c^z=-{1\over 2}[N_a-N]$ and $S_s^z= -{1\over
2}[N_{\uparrow}-N_{\downarrow}]$ are the eigenvalues of the
diagonal generators given in Eq. (\ref{Sz}) and $\mu_c =2\mu$ and
$\mu_s =2\mu_0\,H$ are the same quantities as on the right-hand
side of Eq. (\ref{H}).

The values of the rapidity-momentum functional $k(q)$ and rapidity
functionals $\Lambda_{\alpha\nu}(q)$ are the same for all
$2S_{\alpha}+1$ states in the same tower. Such functionals are
eigenvalues of operators which commute with the off-diagonal
generators of $\eta$-spin and spin algebras. This is consistent
with the $\eta$-spin and spin $SU(2)$ symmetries, which imply that
the Hamiltonian (\ref{HH}) commutes with these generators and thus
the energy (\ref{ESO4}) is the same for the set of $2S_{\alpha}+1$
states belonging to the same $\eta$-spin ($\alpha =c$) or spin
($\alpha =s$) tower. The above operators obey equations similar to
Eqs. (\ref{Tapco1})-(\ref{Tapco3}), with the $\alpha\nu$
bare-momentum distribution functions $N_{\alpha\nu}(q)$ replaced
by the corresponding operators $\hat{N}_{\alpha\nu}(q)$ given in
Eq. (\ref{Nanop}). The pseudoparticle Hamiltonian expression is
also obtained by replacing in the energy expressions
(\ref{E})-(\ref{EH}) the distribution functions $N_{\alpha\nu}(q)$
by the operators $\hat{N}_{\alpha\nu}(q)$, and the
rapidity-momentum and rapidity functionals by the corresponding
operators.

The momentum operator given in Eq. (\ref{Popel}) can be expressed
in terms of pseudoparticle and $-1/2$ holon operators as follows
\cite{I},

\begin{eqnarray}
\hat{P} & = & \sum_{q=q_{c0}^{-}}^{q_{c0}^{+}} \hat{N}_{c0} (q)\,
q + \sum_{\nu =1}^{\infty}\sum_{q=-q_{s\nu }}^{+q_{s\nu}}
\hat{N}_{s\nu} (q)\, q  + \sum_{\nu
=1}^{\infty}\sum_{q=-q_{c,\,\nu }}^{+q_{c,\,\nu }} \,
\hat{N}_{c\nu} (q)\, [{\pi\over a} -q]
+ {\pi\over a}\,\hat{M}_{c,\,-1/2} \nonumber \\
& = & \sum_{q=q_{c0}^{-}}^{q_{c0}^{+}}\, \hat{N}_{c0} (q)\, q +
\sum_{\nu =1}^{\infty}\sum_{q=-q_{s\nu }}^{+q_{s\nu}}
\hat{N}_{s\nu} (q)\, q  + \sum_{\nu
=1}^{\infty}\sum_{q=-q_{c,\,\nu }}^{+q_{c,\,\nu }} \hat{N}_{c\nu}
(q)\, [(1+\nu){\pi\over a} -q] + {\pi\over a}\,\hat{L}_{c,\,-1/2}
\, . \label{Pop}
\end{eqnarray}
The momentum operator (\ref{Pop}) commutes with the electron -
rotated-electron unitary operator $V(U/t)$ \cite{I}. The momentum
eigenvalues can be straightforwardly written by replacing on the
right-hand side of Eq. (\ref{Pop}) the pseudoparticle and holon
number operators by the corresponding eigenvalues.

Note that the number $L_{c,\,-1/2}$ and corresponding operator
${\hat{L}}_{c,\,-1/2}$ appearing on the right-hand side of Eqs.
(\ref{EH}) and (\ref{Pop}), respectively, are beyond the
Bethe-ansatz solution. Indeed, the states which span the Hilbert
subspace associated with that solution have no $-1/2$ Yang holons
and thus $L_{c,\,-1/2}=0$ in Eq. (\ref{EH}) for these states.
However, the energy spectrum defined by Eqs. (\ref{E})-(\ref{EH})
and the momentum-operator expression (\ref{Pop}) refer to the
whole Hilbert space of the 1D Hubbard model. On the other hand, as
a result of the $SO (4)$ symmetry of the Hamiltonian (\ref{HH}),
the corresponding energy spectrum (\ref{ESO4}) does not depend on
the value of the number $L_{c,\,-1/2}$. This property is confirmed
by analysis of expressions (\ref{ESO4}) and (\ref{EH}) and by
noting that
$M_{c,\,-1/2}=L_{c,\,-1/2}+\sum_{\nu=1}^{\infty}\,\nu\,N_{c\nu}$.

%%%%%%%%%%%%%%%%%%%%%%%%%%%%%%%%%%%%%%%%%%%%%%%%%%%%%%%%%%%%%%%%
\section{THE GROUND-STATE NORMAL-ORDERED PSEUDOPARTICLE OPERATIONAL DESCRIPTION}

In this section we consider the ground-state normal-ordered
pseudoparticle operational description which is needed for the
introduction of the pseudofermion description. We are mostly
interested in excited states generated from the ground state by
processes involving changes in the occupancy configurations of a
finite number of $\alpha\nu$ pseudoparticles, $-1/2$ Yang holons,
and $-1/2$ HL spinons. The pseudofermion description introduced in
the ensuing section corresponds to the Hilbert subspace spanned by
such excited states. In the thermodynamic limit the few-electron
excitations are contained in that subspace.

%%%%%%%%%%%%%%%%%%%%%%%%%%%%%%%%%%%%%%%%%%%%%%%%%%%%%%%%%%%%%%%%
\subsection{NORMAL-ORDERED NUMBER OPERATORS AND USEFUL GROUND-STATE QUANTITIES}

Throughout this paper the symbol $:\hat{O}:$ refers to the
ground-state normal-ordered expression of a general operator
$\hat{O}$. Such a ground-state normal-ordered expression is given
by that operator minus its ground-state expectation value. We
introduce the ground-state normal-ordered $\alpha\nu$
pseudoparticle bare-momentum distribution operator,

\begin{equation}
:\hat{N}_{\alpha\nu}(q): = \hat{N}_{\alpha\nu}(q) -
N^{0}_{\alpha\nu}(q) \, , \label{noNanop}
\end{equation}
and the $-1/2$ Yang holon ($\alpha =c$) and $-1/2$ HL spinon
($\alpha =s$) ground-state normal-ordered number operator,

\begin{equation}
:\hat{L}_{\alpha,\,-1/2}:=\hat{L}_{\alpha,\,-1/2}-L^0_{\alpha,\,-1/2}
=\hat{L}_{\alpha,\,-1/2} \, . \label{noNYhHLsop}
\end{equation}
Here the operators $\hat{N}_{\alpha\nu}(q)$ and
$\hat{L}_{\alpha,\,-1/2}$ are defined in Eqs. (\ref{Nanop}) and
(\ref{NYhHLsop}), respectively, $N^{0}_{\alpha\nu} (q)$ is the
ground-state $\alpha\nu$ pseudoparticle distribution function
whose expressions are given in Eqs. (\ref{Nc0})-(\ref{Ncnsn0}) of
Appendix A, and $L^0_{\alpha ,\,-1/2}=0$ is the $-1/2$ Yang holon
($\alpha =c$) and $-1/2$ HL spinon ($\alpha =s$) ground-state
number. It follows from the expressions given in Eq.
(\ref{N*csnu}) of Appendix A for the ground-state values of the
number $N^*_{\alpha\nu}$, whose general expressions are given in
Eqs. (\ref{N*}), (\ref{Nhag}), and (\ref{Nhcsn}) of the same
Appendix, that in the case of the ground state the effective
$\alpha\nu$ lattice constants (\ref{aan}) are given by
\cite{IIIb},

\begin{equation}
a_{c0}^0 = a \, ; \hspace{1cm} a_{c\nu}^0 = {1\over \delta} \, ;
\hspace{1cm} a_{s1}^0 = {1\over n_{\uparrow}} \, ; \hspace{1cm}
a_{s\nu}^0 = {1\over m} \, , \label{acanGS}
\end{equation}
where $\delta =(1/a-n)$ is the doping concentration. We note that
the meaning of the divergences in the value of the constants
$a_{\alpha\nu}^0$ defined in Eq. (\ref{acanGS}) is that the
corresponding effective $\alpha\nu$ lattice has no sites, {\it
i.e.} $N^{0,*}_{\alpha\nu}=0$ and, therefore, does not exist for
the ground state. This is the case of the effective $c\nu$
lattices for half filling when $\nu>0$ and of the effective $s\nu$
lattices for zero spin density when $\nu>1$. It follows that such
singularities just indicate the collapse of the corresponding
effective $\alpha\nu$ lattice. This is one of the reasons why some
of our expressions refer to electronic densities $0<n<1/a$ and
spin densities $0<m<n$, such that all ground-state effective
$\alpha\nu$ lattice constants (\ref{acanGS}) have finite values.

From use of expressions (\ref{acanGS}) one can write the
ground-state number $N^{0,*}_{\alpha\nu}$ given in Eq.
(\ref{N*csnu}) of Appendix A as,

\begin{equation}
N^{0,*}_{\alpha\nu}={L\over a_{\alpha\nu}^0} \, . \label{N*anuGS}
\end{equation}
Except for terms of order $1/L$, the limiting bare-momentum values
defined in Eqs. (\ref{rangeqjc})-(\ref{qag}) of Appendix A
simplify and are given by,

\begin{equation}
q^0_{\alpha\nu} = {\pi\over a_{\alpha\nu}^0} \, ,
\label{qcanGSefa}
\end{equation}
where the ground-state effective-lattice constants
$a_{\alpha\nu}^0$ are these given in Eq. (\ref{acanGS}).
Comparison of Eqs. (\ref{acanGS}) and (\ref{qcanGSefa}) leads to
the following ground-state expressions for the $\alpha\nu$
pseudoparticle limiting bare-momentum values,

\begin{eqnarray}
q^0_{c0} & = & \pi/a \, ; \hspace{1cm} q^0_{s1} =
k_{F\uparrow} \, ;  \nonumber \\
q^0_{c\nu} & = & [\pi/a -2k_F] \, , \hspace{0.3cm} \nu >0 \, ;
\hspace{1cm} q^0_{s\nu} = [k_{F\uparrow}-k_{F\downarrow}] \, ,
\hspace{0.3cm} \nu >1 \, . \label{qcanGS}
\end{eqnarray}
In most situations, one can disregard the $1/L$ corrections and
use the bare-momentum limiting values given in Eq. (\ref{qcanGS}).

For each specific energy eigenstate the rapidity-momentum
functional $k (q)$ and the rapidity functionals
$\Lambda_{\alpha\nu} (q)$ become mere functions of $q$ and are
called rapidity-momentum and rapidity functions, respectively. We
denote the ground-state rapidity-momentum function by $k^0 (q)$
and we call $\Lambda_{c\nu}^0 (q)$ and $\Lambda_{s\nu}^0 (q)$ the
ground-state rapidity functions. These ground-state functions are
computed by solution of the integral equations obtained by
introducing in Eqs. (\ref{Tapco1})-(\ref{Tapco3}) the ground-state
distribution functions (\ref{Nc0})-(\ref{Ncnsn0}) given in
Appendix A. The solution of these equations can be written in
terms of the inverse functions of $k^0 (q)$ and
$\Lambda_{\alpha\nu}^0 (q)$ which we call ${\bar{q}}_{c} (k)$ and
${\bar{q}}_{\alpha\nu} (\Lambda)$, respectively. Here $k$ and
$\Lambda$ are the rapidity-momentum coordinate and rapidity
coordinate, respectively. For the pseudoparticle description these
functions refer to the ground state only. However, in the case of
the pseudofermion description introduced in the ensuing section
these functions apply both to the ground state and excited states.
It follows that these functions play an important role in the
latter description. These are odd functions such that,

\begin{equation}
{\bar{q}}_{c} (k) = - {\bar{q}}_{c} (-k) \, ; \hspace{0.5cm}
{\bar{q}}_{\alpha\nu} (\Lambda) = {-\bar{q}}_{\alpha\nu}
(-\Lambda) \, ; \hspace{0.5cm} \alpha =c,\,s \, ; \hspace{0.3cm}
\nu =1,2, ... \, . \label{qcqan}
\end{equation}
The domains of the rapidity-momentum coordinate $k$ and rapidity
coordinate $\Lambda$ are such that $-\pi/a\leq k\leq +\pi/a$ and
$-\infty \leq \Lambda\leq \infty$, respectively. If follows that,

\begin{equation}
k^0 \Bigl(\pm{\pi\over a}\Bigr)=\pm{\pi\over a} \, ;
\hspace{0.5cm} \Lambda^0_{c0}\Bigl(\pm{\pi\over a}\Bigr) = 0 \, ;
\hspace{0.5cm} \Lambda^0_{\alpha\nu}\Bigr(\pm
q^0_{\alpha\nu}\Bigl)=\pm\infty \, ; \hspace{0.5cm} \alpha =c,\,s
\, ; \hspace{0.3cm} \nu =1,2, ...  \, , \label{Rqan}
\end{equation}
where we introduced the ground-state $c0$ rapidity function,

\begin{equation}
\Lambda^0_{c0}(q) = \sin k^0 (q) \, . \label{Lc0}
\end{equation}
The $c0$ rapidity function (\ref{Lc0}) is the ground-state value
of the rapidity functional defined in Eq. (\ref{Lc}). The
relations (\ref{Rqan}) are equivalent to,

\begin{equation}
{\bar{q}}_{c} \Bigl(\pm{\pi\over a}\Bigr) = \pm{\pi\over a} \, ;
\hspace{0.5cm} {\bar{q}}_{c\nu} \Bigl(\pm\infty\Bigr) = \pm
[{\pi\over a} - 2k_F] \, ; \hspace{0.5cm} {\bar{q}}_{s\nu}
\Bigl(\pm\infty\Bigr) = \pm [k_{F\uparrow} -k_{F\downarrow}] \, ;
\hspace{0.5cm} \nu =1,2, ... \, . \label{qcqanLIM}
\end{equation}

The {\it Fermi points} $\pm 2k_F$ and $\pm k_{F\downarrow}$ of the
$c0$ pseudoparticles and $s1$ pseudoparticles, respectively,
correspond to the following {\it rapidity Fermi points},

\begin{equation}
Q \equiv k^{0}(2k_F) \, , \hspace{1cm} B \equiv
\Lambda^{0}_{s1}(k_{F\downarrow}) \, . \label{QB}
\end{equation}
Then,

\begin{equation}
{\bar{q}}_{c} (\pm 2k_F) = \pm Q \, ; \hspace{0.5cm}
{\bar{q}}_{s1} (\pm k_{F\downarrow}) = \pm B \, . \label{qcqsF}
\end{equation}
Interestingly, one can write the above functions ${\bar{q}}_{c}
(k)$, ${\bar{q}}_{c\nu} (\Lambda)$, and ${\bar{q}}_{s\nu}
(\Lambda)$ in terms of the two-pseudofermion phase shifts
$\tilde{\Phi}_{\alpha\nu,\,\alpha'\nu'}$ expressed in terms of the
rapidity-momentum coordinate $k$ and rapidity coordinate
$\Lambda$, whose physical meaning is discussed in the ensuing
section. These functions read,

\begin{equation}
{\bar{q}}_c (k) = k + \int_{-Q}^{+Q}dk'\,\widetilde{\Phi
}_{c0,\,c0} \left(k',k\right) \, , \label{kcGS}
\end{equation}

\begin{equation}
{\bar{q}}_{c\nu} (\Lambda ) = 2\,{\rm Re}\,\{\arcsin \Bigl(\Lambda
+ i \nu U/4t\Bigr)\} -\int_{-Q}^{+Q}dk'\,\widetilde{\Phi
}_{c0,\,c\nu} \left(k',\Lambda\right)  \, ; \hspace{0.5cm} \nu
=1,2, ... \, , \label{GcnGS}
\end{equation}

\begin{equation}
{\bar{q}}_{s\nu} (\Lambda ) = \int_{-Q}^{+Q}dk'\, \widetilde{\Phi
}_{c0,\,s\nu} \left(k',\Lambda\right)  \, ; \hspace{0.5cm} \nu
=1,2, ... \, , \label{GsnGS}
\end{equation}
respectively. The two-pseudofermion phase shifts play an important
role in the few-electron spectral properties and appear in the
expression of the momentum carried by the pseudofermions. The
two-pseudofermion phase shifts
$\tilde{\Phi}_{\alpha\nu,\,\alpha'\nu'}$ expressed in terms of the
rapidity-momentum coordinate $k$ and rapidity coordinate $\Lambda$
are mathematically defined by the following equations,

\begin{equation}
\tilde{\Phi}_{c0,\,c0}(k,k') = \bar{\Phi
}_{c0,\,c0}\left({4t\,\sin k\over U}, {4t\,\sin k'\over U}\right)
\, ; \hspace{1cm} \tilde{\Phi}_{c0,\,\alpha\nu}(k,\Lambda') =
\bar{\Phi }_{c0,\,\alpha\nu}\left({4t\,\sin k\over U},
{4t\,\Lambda'\over U}\right) \, , \label{tilPcc}
\end{equation}

\begin{equation}
\tilde{\Phi}_{\alpha\nu,\,c0}(\Lambda,k') = \bar{\Phi
}_{\alpha\nu,\,c0}\left({4t\,\Lambda\over U}, {4t\,\sin k'\over
U}\right) \, ; \hspace{1cm}
\tilde{\Phi}_{\alpha\nu,\,\alpha\nu'}\left(\Lambda,\Lambda'\right)
= \bar{\Phi }_{\alpha\nu,\,\alpha\nu'}\left({4t\,\Lambda\over U}
,{4t\,\Lambda'\over U}\right) \, , \label{tilPanan}
\end{equation}
where the two-pseudofermion phase shifts $\bar{\Phi
}_{\alpha\nu,\,\alpha'\nu'}(r,r')$ are defined by the integral
equations (\ref{Phis1c})-(\ref{Phisncn}) of Appendix B. These are
expressed in terms of the variable $r$ defined in the arguments of
the functions on the right-hand side of Eqs.
(\ref{tilPcc})-(\ref{tilPanan}). The values of the parameters $Q$
and $B$ introduced in Eq. (\ref{QB}) are controlled by the
two-pseudofermion phase shifts and are computed by self-consistent
solution of the following equations,

\begin{equation}
2k_F = Q + \int_{-Q}^{+Q}dk\,\widetilde{\Phi }_{c0,\,c0}
\left(k,Q\right) \, ; \hspace{1cm} k_{F\downarrow} =
\int_{-Q}^{+Q}dk\, \widetilde{\Phi }_{c0,\,s1} \left(k,B\right) \,
, \label{Beq}
\end{equation}
respectively. At zero spin density the parameter $Q$ changes from
$Q=k_F$ in the limit $U/t\rightarrow 0$ to $Q=2k_F$ as
$U\rightarrow\infty$. At half filling and zero spin density, the
$U/t$ dependence of the parameter $Q$ is singular at $U/t=0$. It
reads $Q=\pi/2a$ at $U/t=0$ and is given by $Q=\pi/a$ for all
finite values of $U/t$. The $U/t>0$ value, $Q=2k_F=\pi/a$, is
associated with a full $c0$ pseudoparticle band both in
bare-momentum $q$ and in rapidity-momentum $k$ spaces and thus
implies insulator behavior. In contrast, the $U/t=0$ value
$Q=\pi/2a$ corresponds to a metallic band which is half filled in
the rapidity-momentum space. Such a singular behavior is
associated with the Mott-Hubbard transition \cite{Lieb,spectral}.
At zero spin density the parameter $B$ is given by $B=\infty$ and
vanishes in the limit of spin density $m\rightarrow n$, as the
fully polarized ferromagnetic state is approached.

Unfortunately, it is in general difficult to obtain closed form
expressions for the ground-state functions $k^0 (q)$ and
$\Lambda_{\alpha\nu}^0 (q)$ by inverting the functions defined in
Eqs. (\ref{kcGS})-(\ref{GsnGS}). This can be achieved for specific
limits of the parameter space only. For instance, from the use of
Eqs. (\ref{kcGS})-(\ref{GsnGS}) we can obtain for zero spin
density $m=0$, values of the electronic density $0\leq n\leq 1/a$,
and limiting on-site repulsion values $U/t\rightarrow 0$ and
$U/t>>1$ the following closed form expressions for the
ground-state functions $k^0 (q)$, $\Lambda_{c0}^0 (q)$,
$\Lambda_{c\nu}^0 (q)$, and $\Lambda_{s1}^0 (q)$,

\begin{eqnarray}
k^0 (q) & = & {q\over 2} \, ; \hspace{0.5cm} \vert q\vert
\leq 2k_F \, , \hspace{0.5cm} U/t\rightarrow 0 \, ;  \nonumber \\
& = & {\rm sgn} (q)\,[\vert q\vert - k_F] \, ; \hspace{1cm} 2k_F
\leq
\vert q\vert < \pi/a \, , \hspace{0.5cm} U/t\rightarrow 0 \, ; \nonumber \\
& = & {\rm sgn} (q)\,\pi/a \, ; \hspace{1cm} \vert q\vert = \pi/a
\, , \hspace{0.5cm} U/t\rightarrow 0\, ; \nonumber \\ & = & q -
{4tn\over U}\,\ln (2)\,{\sin (q\,a)\over a} \, ; \hspace{0.5cm}
\vert q\vert \leq \pi/a \, , \hspace{0.2cm} U/t
>> 1 \, , \label{k0lim}
\end{eqnarray}

\begin{eqnarray}
\Lambda^0_{c0} (q) & = & \sin\Bigl({q\,a\over 2}\Bigr) \, ;
\hspace{0.5cm} \vert q\vert
\leq 2k_F \, , \hspace{0.5cm} U/t\rightarrow 0 \, ;  \nonumber \\
& = & {\rm sgn} (q)\,\sin\Bigl((\vert q\vert - k_F)\,a\Bigr) \, ;
\hspace{1cm} 2k_F \leq
\vert q\vert < \pi/a \, , \hspace{0.5cm} U/t\rightarrow 0 \, ; \nonumber \\
& = & 0 \, ; \hspace{1cm} \vert q\vert = \pi/a \, , \hspace{0.5cm}
U/t\rightarrow 0\, ; \nonumber \\ & = & \sin (q\,a) - {2tn\over
U}\,\ln (2)\,\sin (2q\,a) \, ; \hspace{0.5cm} \vert q\vert \leq
\pi/a \, , \hspace{0.2cm} U/t
>> 1 \, , \label{Gc0lim}
\end{eqnarray}

\begin{eqnarray}
\Lambda^0_{c\nu} (q) & = & {\rm sgn} (q)\,\sin\Bigl({(\vert q\vert
+ \pi n)\,a\over 2}\Bigr) \, ; \hspace{0.5cm}  0<\vert q\vert<
(\pi/a-2k_F) \, , \hspace{0.5cm} U/t\rightarrow 0  \nonumber \\
& = & 0 \, ; \hspace{0.5cm} q = 0 \, , \hspace{0.5cm}
U/t\rightarrow 0
\nonumber \\
& = & \pm\infty \, ; \hspace{0.5cm} q = \pm (\pi/a-2k_F) \, ,
\hspace{0.5cm} U/t\rightarrow 0
\nonumber \\
& = & {4\nu t\over U}\,\tan \bigl({q\,a\over 2\delta}\Bigr) \, ;
\hspace{0.5cm} 0\leq \vert q\vert \leq (\pi/a-2k_F) \, ,
\hspace{0.2cm} U/t >> 1 \, , \label{Gcnlim}
\end{eqnarray}
for $\nu >0$, and

\begin{eqnarray}
\Lambda^0_{s1} (q) & = & \sin (q\,a) \, ; \hspace{0.5cm} \vert
q\vert < k_F \, , \hspace{0.5cm} U/t\rightarrow 0  \nonumber \\
& = & \pm\infty \, ; \hspace{0.5cm} q =\pm k_F \, , \hspace{0.5cm}
U/t\rightarrow 0 \nonumber \\ & = & {8t\over \pi U}\, {\rm
arcsinh} \Bigl(\tan ({q\,a\over n})\Bigr) \, ; \hspace{0.5cm}
\vert q\vert \leq k_F \, , \hspace{0.2cm} U/t >> 1 \, ,
\label{Gs1lim}
\end{eqnarray}
respectively. We note that for zero-spin density the $\nu
>1$ ground-state rapidity function $\Lambda^0_{s\nu} (q)$
vanishes for all values of $n$ and $U/t$. For $m=0$, the
bare-momentum domain width of the $s\nu$ pseudoparticles belonging
to branches such that $\nu >1$ vanishes and the corresponding
effective $s\nu$ lattice has no sites and thus collapses. This
collapsing is associated with the divergence of the effective
$s\nu$ lattice constant given in Eq. (\ref{acanGS}) as
$m\rightarrow 0$ when $\nu >1$. The same occurs for the $c\nu$
pseudoparticles belonging to branches such that $\nu >0$ when half
filling is approached.

For any energy eigenstate $\vert\psi\rangle$ the normal-ordered
bare-momentum distribution and number operators defined in Eqs.
(\ref{noNanop}) and (\ref{noNYhHLsop}), respectively, obey the
following eigenvalue equations,

\begin{equation}
:\hat{N}_{\alpha\nu}(q):\vert\psi\rangle = \Delta
N_{\alpha\nu}(q)\vert\psi\rangle \, ; \hspace{1cm}
:\hat{L}_{\alpha,\,-1/2}:\vert\psi\rangle = \Delta
L_{\alpha,\,-1/2}\vert\psi\rangle \, . \label{noNanEQ}
\end{equation}
Here $\Delta N_{\alpha\nu} (q)$ is the $\alpha\nu$ pseudoparticle
bare-momentum distribution function deviation and $\Delta
L_{\alpha ,\,-1/2}$ is the deviation in the number of $-1/2$ Yang
holons ($\alpha =c$) or of $-1/2$ HL spinons ($\alpha =s$). These
deviations are given by

\begin{equation}
\Delta N_{\alpha\nu} (q) \equiv N_{\alpha\nu} (q) -
N^{0}_{\alpha\nu} (q) \, ; \hspace{1cm} \Delta L_{\alpha ,\,-1/2}
\equiv L_{\alpha ,\,-1/2} - L^{0}_{\alpha ,\,-1/2} \, .
\label{DNq}
\end{equation}
These values describe deviations of occupancy configurations of
excited states relative to the ground-state occupancy
configurations described by the bare-momentum distribution
functions and numbers given in Eqs. (\ref{Nc0})-(\ref{Ncnsn0}) of
Appendix A. For these excited states, the $\alpha\nu$
pseudoparticle bare-momentum distribution function and the $-1/2$
Yang holon and $-1/2$ HL spinon numbers read,

\begin{equation}
N_{\alpha\nu} (q) = N^{0}_{\alpha\nu} (q) + \Delta N_{\alpha\nu}
(q) \, ; \hspace{1cm} L_{\alpha ,\,-1/2} = L^{0}_{\alpha ,\,-1/2}
+ \Delta L_{\alpha ,\,-1/2} \, . \label{N0DNq}
\end{equation}

From use of the ground-state distribution and number values given
in Eqs. (\ref{Nc0})-(\ref{Ncnsn0}) of Appendix A, we find the
following operational relations,

\begin{equation}
:\hat{N}_{c\nu}(q): = \hat{N}_{c\nu}(q) \, ; \hspace{0.5cm} \nu
>0 \, ; \hspace{1cm} :\hat{N}_{s\nu}(q): = \hat{N}_{s\nu}(q)
\, ; \hspace{0.5cm} \nu >1 \, ; \hspace{1cm}
:\hat{L}_{\alpha,\,-1/2}:=\hat{L}_{\alpha,\,-1/2} \, .
\label{nul0}
\end{equation}
Relations (\ref{nul0}) are justified by the absence of the
corresponding quantum objects in the initial ground state.

%%%%%%%%%%%%%%%%%%%%%%%%%%%%%%%%%%%%%%%%%%%%%%%%%%%%%%%%%%%%%%%%
\subsection{THE NORMAL-ORDERED PSEUDOPARTICLE MOMENTUM AND ENERGY FUNCTIONALS}

Let us introduce the momentum and energy functionals associated
with the Hilbert subspace spanned by excited states described by
small deviations (\ref{DNq}). The bare-momentum distribution
deviations of these states involve a small density of
pseudoparticles, $-1/2$ Yang holons, and $-1/2$ HL spinons. In the
case of momentum, such a functional is obtained directly from use
of the operator expression (\ref{Pop}). Introduction of the
corresponding normal-ordered expression and replacement of the
bare-momentum number operators and holon number operators by their
eigenvalues leads to the following momentum functional,

\begin{equation}
\Delta P = \sum_{q=q_{c0}^{-}}^{q_{c0}^{+}}\, \Delta N _{c0} (q)\,
q + \sum_{\nu =1}^{\infty}\sum_{q=-q_{s\nu }}^{+q_{s\nu}}\, \Delta
N_{s\nu} (q)\, q  + \sum_{\nu =1}^{\infty}\sum_{q=-q_{c,\,\nu
}}^{+q_{c,\,\nu }} \, \Delta N_{c\nu} (q)\, [(1+\nu){\pi\over a}
-q] + {\pi\over a}\,\Delta L_{c,\,-1/2} \, . \label{noP}
\end{equation}
The momentum functional (\ref{noP}) is linear in the bare-momentum
distribution function deviations for all excited states
independently of whether the values of these deviations are small
or large. Thus, this expression is valid even for large values of
these deviations.

As for the Fermi-liquid quasiparticles \cite{Pines,Baym}, while
the momentum functional (\ref{noP}) is linear in the bare-momentum
distribution function deviations, the corresponding energy
functional includes linear, quadratic, and higher-order terms in
these deviations. Such an energy functional is derived by solution
of the integral equations (\ref{Tapco1})-(\ref{Tapco3}) for
bare-momentum distribution functions of the general form
(\ref{N0DNq}). This leads to deviation expansion expressions for
the rapidity-momentum and rapidity functionals in terms of the set
of bare-momentum distribution function deviations $\Delta
N_{c0}(q)$, $\{\Delta N_{c\nu}(q)\}$, and $\{\Delta N_{s\nu}(q)\}$
where $\nu =1,2,...$. According to Eqs. (\ref{E})-(\ref{EH}), the
energy spectrum of the excited states can be expressed in terms of
these functionals. Use in the general energy expressions
(\ref{E})-(\ref{EH}) of the systematic expansion in the
pseudoparticle bare-momentum distribution deviations and holon and
spinon number deviations of the above functionals leads to a
general finite-energy Landau-liquid energy functional of the
following form,

\begin{equation}
\Delta E = \sum_{i=1}^{\infty}\Delta E_i \, . \label{EF}
\end{equation}
The terms of order $i$ larger than one describe the residual
interactions of the pseudoparticles. In contrast, the energy
functional (\ref{EF}) is linear in the $-1/2$ Yang holon and
$-1/2$ HL spinon number deviations for all excited states
independently of whether the values of these deviations are small
or large. Such behavior follows from the non-interacting character
of these quantum objects \cite{II}. In the case of excited states
whose bare-momentum distribution function deviations involve a
small but finite density of pseudoparticles, the physics is
described by the energy terms on the right-hand side of Eq.
(\ref{EF}) of all scattering orders $i=1,2,...$. However, in the
limit when the density of these pseudoparticles becomes vanishing,
only the first two terms become relevant. According to the results
obtained in Refs. \cite{II,Carmelo97}, the first and second-order
terms on the right-hand side of Eq. (\ref{EF}) are of the
following general form,

\begin{equation}
\Delta E_1 = \omega_0 + \sum_{q=-q_{c0}^0}^{+q_{c0}^0}
dq\,\epsilon_{c0} (q)\,\Delta N_{c0} (q) +
\sum_{q=-q_{s1}^0}^{+q_{s1}^0} dq\,\epsilon_{s1}(q)\,\Delta
N_{s1}(q) + \sum_{\alpha =c,\,s}\,\sum_{\nu =1+\delta_{\alpha
,\,s}}^{\infty} \sum_{q=-q_{\alpha\nu}^0}^{+q_{\alpha\nu}^0}
dq\,\epsilon^0_{\alpha\nu}(q)\,\Delta N_{\alpha\nu}(q) \, ,
\label{E1}
\end{equation}
and

\begin{equation}
\Delta E_2 = {1\over L}\,\sum_{\alpha =c,s}\,\sum_{\nu
=1-\delta_{\alpha,\,c}}^{\infty}\,
\sum_{-q^0_{\alpha\nu}}^{+q^0_{\alpha\nu}}\,\sum_{\alpha'
=c,s}\sum_{\nu' =1-\delta_{\alpha',\,c}}^{\infty}\,
\sum_{-q^0_{\alpha',\,\nu'}}^{+q^0_{\alpha',\,\nu'}}\,{1\over
2}\,f_{\alpha\nu,\, \alpha'\nu'}(q,q')\,\Delta
N_{\alpha\nu}(q)\,\Delta N_{\alpha',\,\nu'}(q') \, , \label{E2}
\end{equation}
where $\Delta N_c (q)$, $\Delta N_{c\nu} (q)$, and $\Delta
N_{s\nu} (q)$ are the pseudoparticle bare-momentum distribution
function deviations given in Eq. (\ref{DNq}) and $\omega_0$ is an
energy functional linear in the $-1/2$ holon, $-1/2$ spinon, and
$s1$ pseudoparticle number deviations $\Delta M_{c ,\,-1/2}$
$\Delta M_{s ,\,-1/2}$, and $\Delta N_{s1}$, respectively. (The
latter functional is given below in Eq. (\ref{om0}) of Sec. V,
with the $s1$ pseudofermion number deviation equaling the
corresponding $s1$ pseudoparticle number deviation.) The
coefficients of the $i=1$ linear terms are the $\alpha\nu$
pseudoparticle energy bands $\epsilon_{c0}(q)$,
$\epsilon_{s1}(q)$, and $\epsilon_{\alpha\nu}^0(q)$ studied in
Refs. \cite{I,II,Carmelo97}. The coefficients of the $i=2$
quadratic terms are the pseudoparticle $f$ functions
$f_{\alpha\nu,\,\alpha'\nu'}(q,q')$ studied in Ref.
\cite{Carmelo97}.

In contrast to what the energy terms (\ref{E1})-(\ref{E2}) may
suggest, we emphasize that the energy (\ref{EF}) is not an
expansion in $1/L$. The small parameters in such an expansion are
the pseudoparticle bare-momentum distribution function deviations,
as in the case of the quasiparticle Fermi-liquid energy functional
\cite{Pines,Baym}. Indeed, when these function deviations involve
a small but finite density of pseudoparticles, the energy terms of
order $i$ on the right-hand side of Eq. (\ref{EF}) are not of
order $[1/L]^i$. Instead, in that case all terms of the energy
expansion (\ref{EF}) are of the same order $[1/L]^{-1}=L$. As
discussed in the ensuing section, few-electron excitations are
associated with excited states generated from the ground state by
a finite number of pseudoparticle, holon, and spinon processes.
The deviations of these states involve a vanishing density of
pseudoparticles, holons, and spinons. However, in order to achieve
the correct microscopic description of the few-electron spectral
properties, one must consider the limit where a vanishing density
of pseudoparticles is approached rather than considering that such
a density is zero from the very beginning. As the density of
pseudoparticles involved in the bare-momentum distribution
deviations decreases and approaches zero, the energy terms on the
right-hand side of Eq. (\ref{EF}) of large scattering order $i$,
also vanish and do not not contribute as they become of order
$[1/L]^i$.

The $f$ functions on the right-hand side of Eq. (\ref{E2}), are
associated with the two-pseudoparticle residual interactions and
have the same role as those of Fermi-liquid theory
\cite{Pines,Baym}. Indeed, for small values of the energy $\omega$
and electronic densities $n$ and spin densities $m$ such that
$0<n<1/a$ and $0<m<n$, respectively, the few-electron spectral
properties are controlled by the residual two-pseudoparticle
interactions described by the $i=2$ terms (\ref{E2}) of the energy
functional (\ref{EF}). In the corresponding low-energy Hilbert
subspace, only the $c0$ and $s1$ pseudoparticle branches have
finite occupancies. In this case, as the limit of vanishing
density of pseudoparticles is approached, the general energy
functional (\ref{EF}) acquires the form of the energy spectrum of
a two-component $c0$ and $s1$ conformal field theory \cite{Frahm}.
For the relation of the $c\equiv c,0$ and $s\equiv s,1$
pseudoparticle description to two-component conformal-field theory
see Refs. \cite{Carmelo91,Carmelo92}. The weight distribution of
the few-electron spectral functions can in this case be studied by
conformal field theory \cite{Carmelo97pp}.

We thus conclude that the pseudoparticles have residual
interactions associated with the terms of the general energy
functional (\ref{EF}) of order $i>1$. The pseudoparticle residual
interactions control the low-energy few-electron spectral
properties of the quantum liquid, as confirmed by the studies of
Refs. \cite{Carmelo91,Carmelo92}. In this paper, we introduce an
alternative representation in terms of non-interactiong
pseudofermions. For such pseudofermions the bare momentum is
replaced by a momentum functional whose coefficients are
two-pseudofermion phase shifts. In this case, the relevant energy
spectrum is non-interacting and thus has only linear terms in the
momentum distribution function deviations. For the pseudofermion
description the few-electron spectral properties are controlled by
such two-pseudofermion phase shifts \cite{V}. Such shifts are
associated with first-order momentum distribution function
deviations only, whereas the pseudoparticle residual interactions
involve scattering orders $i$ associated with terms of the energy
functional (\ref{EF}) such that $i>1$.

%%%%%%%%%%%%%%%%%%%%%%%%%%%%%%%%%%%%%%%%%%%%%%%%%%%%%%%%%%%%%%%%
\subsection{THE CONCEPT OF J-CPHS ENSEMBLE SUBSPACE}

Let us introduce the quantum number $\iota ={\rm sgn} (q) 1=\pm 1$
which refers to the number of right ($\iota =+1$) and left ($\iota
=-1$) pseudoparticle movers. The number $N_{\alpha\nu,\,\iota}$ of
$\alpha\nu$ pseudoparticles of $\iota $ character is a good
quantum number. We thus introduce the $\alpha\nu$ pseudoparticle
current number $J_{\alpha\nu}$ such that,

\begin{equation}
J_{\alpha\nu} = {1\over 2}\sum_{\iota =\pm
1}(\iota)\,N_{\alpha\nu,\,\iota} \, ; \hspace{1cm}
N_{\alpha\nu,\,\iota} = {N_{\alpha\nu}\over 2} + \iota\,
J_{\alpha\nu} \, . \label{Nani}
\end{equation}

Each CPHS ensemble subspace contains in general several subspaces
with different values for the sets of current numbers $J_{c0}$ and
$\{J_{\alpha\nu}\}$ such that $\alpha =c,s$ and $\nu =1,2,...$. We
call these subspaces {\it J-CPHS ensemble subspaces}. For a given
J-CPHS ensemble subspace one can introduce the J-CPHS ground
state. This energy eigenstate has compact bare-momentum
occupancies for the $\alpha\nu$ pseudoparticle bands. Moreover, a
partial J-CPHS ground state is a state with such a type of
occupancy configuration for the $\alpha\nu$ pseudoparticle
branches such that $\alpha\nu\neq c0$ and $\alpha\nu\neq s1$ only.
These concepts are associated with the right ($\iota =+1$) and
left ($\iota =-1$) {\it Fermi points}, which separate the
$\alpha\nu$ pseudoparticle occupied from the unoccupied regions of
bare-momentum space. Ignoring terms of order $1/L$, these {\it
Fermi bare-momentum} values read,

\begin{equation}
q_{F\alpha\nu,\,\iota} = \iota\,{2\pi\over L}
N_{\alpha\nu,\,\iota} \, . \label{qiFan}
\end{equation}

The ground state is a particular case of J-CPHS ground state such
that the {\it Fermi points} of Eq. (\ref{qiFan}) read,

\begin{equation}
q^0_{F\alpha\nu,\,\iota} = \iota\,q^0_{F\alpha\nu} \, ,
\label{qiFcs}
\end{equation}
where

\begin{equation}
q^0_{Fc0} = 2k_F \, ; \hspace{0.5cm} q^0_{Fs1} = k_{F\downarrow}
\, ; \hspace{0.5cm} q^0_{Fc\nu} = 0 ; \hspace{0.5cm} q^0_{Fs\nu} =
0 \, , \hspace{0.5cm} \nu > \delta_{\alpha ,\,s} \, .
\label{q0Fcs}
\end{equation}
In most situations one can disregard the $1/L$ corrections and use
the bare-momentum {\it Fermi values} provided in Eqs.
(\ref{qiFan})-(\ref{q0Fcs}).

The pseudofermion description introduced in the ensuing section,
corresponds to J-CPHS ensemble subspaces spanned by energy
eigenstates differing from the ground-state by the occupancy
configuration of a finite number of $\alpha\nu$ pseudoparticles,
$-1/2$ Yang holons, and $-1/2$ HL spinons. For such subspaces the
pseudoparticle {\it Fermi band momenta} given in Eq. (\ref{qiFan})
can be written as follows,

\begin{equation}
q_{F\alpha\nu,\,\iota} = \iota\,q^0_{F\alpha\nu} + \Delta
q_{F\alpha\nu,\,\iota} \label{HSqiFan}
\end{equation}
where $q^0_{F\alpha\nu}$ is the $\alpha\nu$ ground-state {\it
Fermi bare momentum} whose values are given in Eq. (\ref{q0Fcs})
and thus,

\begin{eqnarray}
q_{Fc0,\,\iota} & = & \iota\,2k_F + \Delta q_{Fc0,\,\iota} \, ;
\hspace{1cm} q_{Fs1,\,\iota} = \iota\,k_{F\downarrow} + \Delta
q_{Fs1,\,\iota} \, ; \nonumber \\ q_{Fc\nu,\,\iota} & = & \Delta
q_{Fc\nu,\,\iota} \, , \hspace{0.5cm} \nu
> 0 \, ; \hspace{1cm} q_{Fs\nu,\,\iota} = \Delta q_{Fs\nu,\,\iota} \, , \hspace{0.5cm}
\nu
> 1 \, . \label{HSqiFcs}
\end{eqnarray}
Here

\begin{equation}
\Delta q_{F\alpha\nu,\,\iota} = \iota\,{2\pi\over L} \Delta
N_{\alpha\nu,\,\iota} = \iota\,{2\pi\over L}\Bigl[{\Delta
N_{\alpha\nu}\over 2} + \iota\,\Delta J_{\alpha\nu}\Bigr] \, .
\label{DqiFan}
\end{equation}

%%%%%%%%%%%%%%%%%%%%%%%%%%%%%%%%%%%%%%%%%%%%%%%%%%%%%%%%%%%%%%%%
\section{THE PSEUDOFERMION DESCRIPTION}

In this section we introduce the pseudofermion operational
description. This includes introduction of the pseudoparticle -
pseudofermion unitary transformation and of the Hilbert subspace
associated with that transformation. In such a subspace, the
rapidity functionals have a particular form. Moreover, we obtain
the anticommutation relations of the pseudofermions.

%%%%%%%%%%%%%%%%%%%%%%%%%%%%%%%%%%%%%%%%%%%%%%%%%%%%%%%%%%%%%%%%
\subsection{THE FUNCTIONAL CHARACTER OF THE PSEUDOFERMION MOMENTUM}

The $\alpha\nu$ pseudofermion carries momentum ${\bar{q}}_j$ given
by,

\begin{equation}
{\bar{q}}_j = {\bar{q}} (q_j) = q_j + {Q_{\alpha\nu} (q_j)\over L}
= {2\pi\over L}I^{\alpha\nu}_j + {Q_{\alpha\nu} (q_j)\over L} \, ;
\hspace{0.5cm} j=1,2,...,N_{\alpha\nu}^* \, . \label{barqan}
\end{equation}
Here $Q_{\alpha\nu}(q_j)/L$ is the momentum functional,

\begin{eqnarray}
{Q_{\alpha\nu} (q_j)\over L} = {2\pi\over L}
\sum_{\alpha'=c,s}\,\sum_{\nu'=\delta_{\alpha',\,s}}^{\infty}\,
\sum_{j'=1}^{N^*_{\alpha'\nu'}}\,\Phi_{\alpha\nu,\,\alpha'\nu'}(q_j,q_{j'})\,
\Delta N_{\alpha'\nu'}(q_{j'})  \, . \label{Qcan1j}
\end{eqnarray}
The functions $\Phi_{\alpha\nu,\,\alpha'\nu'}(q,q')$ on the
right-hand side of this equation are the {\it two-pseudofermion
phase shifts} expressed in terms of the bare-momentum $q$. These
phase shifts are given by,

\begin{equation}
\Phi_{\alpha\nu,\,\alpha'\nu'}(q,q') = \bar{\Phi
}_{\alpha\nu,\,\alpha'\nu'}
\left({4t\,\Lambda^{0}_{\alpha\nu}(q)\over U},
{4t\,\Lambda^{0}_{\alpha\nu}(q')\over U}\right) \, ,
\label{Phi-barPhi}
\end{equation}
where $\bar{\Phi }_{\alpha\nu,\,\alpha'\nu'} (r ,\,r')$ are the
two-pseudofermion phase shifts expressed in terms of the variable
$r$. The latter phase shifts are the unique solutions of the
integral equations (\ref{Phis1c})-(\ref{Phisncn}) of Appendix B.
The value of the momentum functional given in Eq. (\ref{Qcan1j})
is controlled by shake-up two-pseudofermion phase-shift processes
resulting from the changes in the quantum-object occupancy
configurations described by the deviations $\Delta
N_{\alpha'\nu'}(q_{j'})$ on the right-hand side of that equation.

The $\alpha\nu$ pseudofermion is related to the corresponding
$\alpha\nu$ pseudoparticle by a mere unitary transformation
involving the discrete bare-momentum values $q_j$ and such that,

\begin{equation}
q_j \rightarrow {\bar{q}}_j \, , \label{pfermions}
\end{equation}
where ${\bar{q}}_j$ is the discrete momentum defined in Eq.
(\ref{barqan}). The number of $\alpha\nu$ pseudoparticles,
$N_{\alpha\nu}$, equals that of $\alpha\nu$ pseudofermions, which
we denote by ${\cal{N}}_{\alpha\nu}$. Moreover, we introduce the
$\alpha\nu$ pseudofermion momentum distribution function
${\cal{N}}_{\alpha\nu} ({\bar{q}}_j)$ such that
${\cal{N}}_{\alpha\nu} ({\bar{q}}_j) = N_{\alpha\nu}
(q_j({\bar{q}}_j))$, where $q_j=q_j({\bar{q}}_j)$ is the inverse
function of (\ref{barqan}).

Note that the momentum functional (\ref{Qcan1j}), is such that
$Q_{\alpha,\,\nu}(q)/L=0$ for the initial ground state. Indeed,
the pseudofermion description refers to the ground-state normal
ordered 1D Hubbard model. Thus, there is a specific $\alpha\nu$
pseudofermion description for each initial ground state. For the
latter state the bare momentum $q_j = [2\pi / L]I^{\alpha\nu}_j$
equals the pseudofermion momentum ${\bar{q}}_j = q_j +
Q_{\alpha\nu} (q_j)/L$. This justifies the designation {\it bare
momentum} for $q_j$. Thus, in the case of the ground state the
$\alpha\nu$ pseudoparticles are identical to the $\alpha\nu$
pseudofermions. It follows that the ground state is invariant
under the pseudoparticle - pseudofermion unitary transformation
and plays the role of the vacuum of the pseudofermion theory.

As for the case of the pseudoparticle representation, the
pseudofermion description corresponds to large values of the
Hubbard chain length $L$ and is thus compatible with Takahashi's
string hypothesis \cite{Takahashi,I}. However, while the
pseudoparticle representation corresponds to the whole Hilbert
space, the pseudofermion description refers to a Hilbert subspace
spanned by the initial ground state and the following types of
states:\vspace{0.5cm}

(A) - Excited states generated from the initial ground state by a
finite number of pseudofermion processes involving creation or
annihilation of $c0$ and $s1$ pseudofermions, creation of
pseudofermions belonging to other $\alpha\nu\neq c0$ and
$\alpha\nu\neq s1$ branches, and creation of $-1/2$ Yang holons
and $-1/2$ HL spinons. In the thermodynamic limit, the momentum
and bare-momentum distribution function deviations associated with
such excited states obey the following relations,

\begin{eqnarray}
\sum_{j=1}^{N^*_{\alpha\nu}}\Delta {\cal{N}}_{\alpha\nu}
({\bar{q}}_j)/L & = & \sum_{j=1}^{N^*_{\alpha\nu}}\Delta
N_{\alpha\nu} (q_j)/L  = \Delta {\cal{N}}_{\alpha\nu}/L = \Delta
N_{\alpha\nu}/L\rightarrow 0 \, ; \hspace{1.0cm}
\nu\geq\delta_{\alpha ,\,s} \, , \hspace{0.5cm}
\alpha = c,s \, ; \nonumber \\
\Delta L_{\alpha ,\,-1/2}/L & = & L_{\alpha ,\,-1/2}/L \rightarrow
0 \, ; \hspace{1.0cm} \alpha = c,s \, . \label{DNqzero}
\end{eqnarray}
\vspace{0.5cm}

(B) - Excited states generated from the initial ground state by a
finite number of $c0$ and $s1$ pseudofermion particle-hole
processes.\vspace{0.5cm}

Throughout this paper, we call such states, excited states A and
excited states B, respectively. We emphasize that in general the
excited states involve both. However, for few-electron excitations
which do not change the electronic numbers the low-energy states
are of type B only. Moreover, for simplicity we are considering
densities in the ranges $0<n<1/a$ and $0<m<n$. Our analysis also
holds for other values of the densities, yet for the half-filling
$n=1/a$ or zero-magnetization $m=0$ phases the excitation subspace
is more reduced.

The unitary transformation associated with Eq. (\ref{pfermions}),
is well defined for the Hilbert subspace spanned by the initial
ground state and the excited states A and B. Fortunately, in the
thermodynamic limit application onto the ground state of a
few-electron operator generates excited states A and B only. By
{\it few-electron} operators, we mean here operators which can be
written as a product of ${\cal N}$ electron creation and/or
annihilation operators and ${\cal N}/N_a\rightarrow 0$ as
$N_a\rightarrow\infty$. The pseudofermion functional theory
introduced in this paper, can be applied to the study of the
few-electron spectral weight distributions for all values of
energy \cite{V,spectral,spectral0}. We find below that the
pseudofermions have a non-interacting character. This allows the
evaluation of few-electron spectral functions for all values of
energy \cite{V,spectral}. The $\alpha\nu$ pseudoparticle residual
interactions are cancelled by the momentum transfer term
$Q_{\alpha,\,\nu}(q)/L$ of Eq. (\ref{Qcan1j}). The information
recorded in the pseudoparticle interactions is transferred over to
the momentum two-pseudofermion phase shifts of the momentum term
$Q_{\alpha\nu}(q)/L$. The pseudofermion theory is of first order
in $1/L$ both for the energy and momentum spectra. In contrast,
the methods that use the pseudoparticle residual interactions lead
to expressions for these spectral functions which are valid for
small values of the energy $\omega$ only
\cite{Carmelo91,Carmelo92,Carmelo97}. Therefore, the pseudofermion
description introduced below corresponds to a breakthrough in what
the study of the finite-energy spectral properties of the 1D
Hubbard model is concerned.

That only excited states A and B have finite overlap with
few-electron excitations, is confirmed by direct evaluation of
matrix elements between few-electron excitations and excited
states \cite{V}. For instance, if $Q_{\alpha\nu} (q)/L=0$ for the
excited states B, the contribution to the one-electron spectral
function would come from one particle-hole pseudofermion processes
only, and the spectral functions would have the familiar
$\delta$-function structure. This is not true for our case where
$Q_{\alpha\nu} (q)/L\neq 0$ for these states. In this case, there
are contributions from many particle-hole processes in the $c0$
and $s1$ pseudofermion bands as well. The largest weight comes
from the one particle-hole pseudofermion processes, and increasing
the number of particle-hole processes, the additional weight
decreases rapidly. Most of the weight associated with the excited
states B is generated by one, two, and three particle-hole
pseudofemion processes \cite{V}. In the general case of
few-electron spectral functions, the contribution from excited
states B involving an infinite number of particle-hole processes
in the $c0$ and $s1$ pseudofermion bands vanishes in the
thermodynamic limit. The same holds for the excited states A. The
excited states generated from the ground state by an infinite
number of processes of the types reported in (A) and (B) have
vanishing overlap with few-electron excitations as
$L\rightarrow\infty$.

The cancellation of the $\alpha\nu$ pseudoparticle residual
interactions by the momentum transfer term $Q_{\alpha,\,\nu}(q)/L$
of Eq. (\ref{Qcan1j}) is related to the form of the rapidity
functionals $\Lambda_{\alpha\nu}(q)$ and rapidity-momentum
functional $k(q)$ in the Hilbert subspace spanned by the excited
states A and B. Introduction of the pseudoparticle bare-momentum
distribution functions of general form given in Eq. (\ref{N0DNq})
in the rapidity functional integral equations
(\ref{Tapco1})-(\ref{Tapco3}) and their expansion in the small
deviations (\ref{DNq}), permits explicit solution of these
equations. This procedure leads to expressions for the rapidity
functionals $\Lambda_{\alpha\nu}(q)$ and rapidity-momentum
functional $k(q)$ in terms of the bare-momentum distribution
function deviations introduced in Eq. (\ref{DNq}). Solution of the
integral equations (\ref{Tapco1})-(\ref{Tapco3}) for distributions
of the general form (\ref{N0DNq}) leads to first-order in the
deviations to expressions for the rapidity-momentum functional and
rapidity functionals of the following form,

\begin{equation}
k (q) = k^0\Bigl({\bar{q}} (q)\Bigr) \, ; \hspace{1cm}
\Lambda_{\alpha\nu}(q) = \Lambda_{\alpha\nu}^0\Bigl({\bar{q}}
(q)\Bigr) \, ; \hspace{1cm} \alpha =c \, ,  \hspace{0.3cm} \nu
=0,1,2,... \, ; \hspace{0.5cm} \alpha =s \, , \hspace{0.3cm} \nu =
1,2,... \, . \label{FL}
\end{equation}
Here $\Lambda_{\alpha\nu}^{0}(q)$ and $k^0 (q)$ are the
corresponding ground state functions. On the right-hand side of
Eq. (\ref{FL}), ${\bar{q}} (q)$ is the $\alpha\nu$ {\it momentum
functional} given in Eq. (\ref{barqan}) with $q_j$ replaced by the
continuum momentum $q$.

It is remarkable that in the Hilbert subspace spanned by the
excited states A and B the functionals $\Lambda_{\alpha\nu}(q)$
and $k(q)$ equal the corresponding ground-state functions
$\Lambda_{\alpha\nu}^0(q)$ and $k^0(q)$, respectively, with the
bare momentum $q$ replaced by the momentum functional
(\ref{barqan}). This property is behind the non-interacting
character of the pseudofermions. The two-pseudofermion phase
shifts $\bar{\Phi }_{\alpha\nu,\,\alpha'\nu'} (r ,\,r')$ defined
by Eqs. (\ref{Phis1c})-(\ref{Phisncn}) of Appendix B which appear
in the expression of the momentum functional (\ref{Qcan1j})
through the relation (\ref{Phi-barPhi}) play a central role in the
pseudofermion description of the few-electron spectral properties.
Within that description the expression of many physical quantities
involve such two-pseudofermion phase shifts. For instance, the
functions defined by Eqs. (\ref{kcGS})-(\ref{GsnGS}) and the
expressions of the pseudofermion energy bands given below involve
the two-pseudofermion phase shifts. Furthermore, these phase
shifts control the matrix elements between few-electron
excitations and the excited states and thus also the few-electron
spectral weight distributions \cite{V}.

According to Eq. (\ref{EH}) the energy spectrum depends on the
quantum object occupancy configurations through the rapidity and
rapidity-momentum functionals. It is this functional character
that is behind the pseudoparticle residual interactions and leads
to the energy expansion given in Eqs. (\ref{EF})-(\ref{E2}).
However, by re-expression of these functionals in terms of the
pseudoferminon momentum ${\bar{q}}$, the general energy spectrum
(\ref{E})-(\ref{EH}) can be written in terms of pseudofermion
momentum distribution functions ${\cal{N}}_{c0} ({\bar{q}}_j)$ and
${\cal{N}}_{c\nu} ({\bar{q}}_j)$ as,

\begin{eqnarray}
E & = & -2t \sum_{j=1}^{N_a}\, {\cal{N}}_{c0} ({\bar{q}}_j)\, \cos
k^0({\bar{q}}_j) + 4t \sum_{\nu=1}^{\infty}
\sum_{j=1}^{N^*_{c\nu}}\, {\cal{N}}_{c\nu} ({\bar{q}}_j)\, {\rm
Re}\,\Bigl\{\sqrt{1 -
(\Lambda_{c\nu}^0 ({\bar{q}}_j) + i \nu U/4t)^2}\Bigr\} \nonumber \\
& + & {U\over 2}\Bigl[\,M_c -
\sum_{\nu=1}^{\infty}2\nu\,{\cal{N}}_{c\nu} -{N_a\over 2}\Bigr] +
\sum_{\alpha}\mu_{\alpha }S^{\alpha}_z \, . \label{ESS}
\end{eqnarray}
This expression is valid for the Hilbert subspace spanned by the
initial ground state and the excited states A and B. The term
$\sum_{\alpha}\mu_{\alpha }S^{\alpha}_z$ is the same as on the
right-hand side of Eq. (\ref{E}) and ${\cal{N}}_{c\nu}=N_{c\nu}$
is the number of $c\nu$ pseudofermions. The form of the general
energy spectrum (\ref{ESS}) justifies why the shake-up effects
associated with the phase shifts of the functional (\ref{Qcan1j})
occur in the case of the pseudofermions in the momentum instead of
in the energy. The dependence of the general energy spectrum
(\ref{ESS}) on the momentum occupancy configuration values occurs
through the arguments of the ground-state rapidity and
rapidity-momentum functions. Thus these functions play the role of
non-interacting spectra, since they have the same form both for
the initial ground state and excited states A and B. The shake-up
effects associated with the two-pseudofermion phase shifts are
thus {\it felt} by the non-interactiong pseudofermions as mere
changes in the momentum occupancies through the twisted boundary
conditions imposed by the elementary processes which generate the
excited state from the ground state. This property is behind the
non-interacting character of the pseudofermions. It is also behind
the fact that within the pseudofermion description the functions
${\bar{q}}_{c} (k)$ and ${\bar{q}}_{\alpha\nu} (\Lambda)$ defined
in Eqs. (\ref{kcGS})-(\ref{GsnGS}) refer to both the ground state
and the excited states. In contrast, within the pseudoparticle
representation these functions refer to the ground state only.

The pseudoparticle bare-momentum $q_j$ description is naturally
provided by the Bethe-ansatz equations within Takahashi's string
hypothesis \cite{Takahashi,Carmelo97,I}. We recall that the
pseudoparticle discrete bare-momentum values $q_j$ are of form
given in Eq. (\ref{qj}) of Appendix A and according to Eq.
(\ref{differ0}) are such that $q_{j+1}-q_j = 2\pi/L$. The single
discrete bare-momentum values $q_j$ are integer multiples of
$2\pi/L$ or of $\pi/L$ \cite{I} and bare-momentum contributions of
order $[1/L]^j$ such that $j>1$ are outside the validity of the
pseudoparticle description. These bare-momentum contributions have
no physical meaning and must be considered as zero. Importantly,
the same is required for the pseudofermion momentum discrete
values ${\bar{q}}_j$ given in Eq. (\ref{barqan}). These discrete
values are also at least of order of $1/L$ and contributions of
order $[1/L]^j$ such that $j>1$ must be considered as equaling
zero. For instance, as confirmed in Appendix C, the discrete
momentum level separation,

\begin{equation}
{\bar{q}}_{j+1}-{\bar{q}}_j= {2\pi\over L} + {Q_{\alpha\nu}
(q_{j+1})- Q_{\alpha\nu} (q_j)\over L} \approx {2\pi\over L} \, ,
\label{differ}
\end{equation}
is such that the second term on the right-hand side of Eq.
(\ref{differ}) is of order $[1/L]^2$, where $Q_{\alpha\nu}(q)/L$
is the momentum functional given in Eq. (\ref{Qcan1j}). Thus, to
first order in $1/L$ one finds that ${\bar{q}}_{j+1}-{\bar{q}}_j=
2\pi/L$, as for the corresponding discrete bare-momentum level
separation given in Eq. (\ref{differ0}) of Appendix A. However,
this does not imply that to first order in $1/L$ the pseudofermion
momentum equals the bare-momentum. Indeed, we emphasize that the
momenta $Q_{\alpha\nu} (q_j)/L$ on the right-hand side of Eq.
(\ref{pfermions}) are of order $1/L$ and play a central role in
the control of the few-electron spectral weight distribution by
the non-perturbative many-electron shake-up effects \cite{V}. We
note that the level separation ${\bar{q}}_{j+1}-{\bar{q}}_j=
2\pi/L$ is valid locally in the discrete momentum space. By that
we mean the following: If in the present thermodynamic limit two
momentum values ${\bar{q}}_j$ and ${\bar{q}}_{j'}$ differ by a
small yet finite momentum difference $\Delta
{\bar{q}}={\bar{q}}_j-{\bar{q}}_{j'}$, then in general $\Delta
{\bar{q}}\neq {2\pi\over L}[j-j']$. In contrast, for the
corresponding bare-momentum values it holds that $\Delta q=
{2\pi\over L}[j-j']$. Therefore, for small but non-vanishing
momentum separation the difference $[Q_{\alpha\nu} (q_{j})-
Q_{\alpha\nu} (q_{j'})]/L$ is not anymore of order $[1/L]^2$ and
has physical significance.

%%%%%%%%%%%%%%%%%%%%%%%%%%%%%%%%%%%%%%%%%%%%%%%%%%%%%%%%%%%%%%%%
\subsection{PSEUDOFERMION OPERATOR ALGEBRA}

The elementary creation and annihilation operators of the
$\alpha\nu$ pseudofermions can be expressed in terms of the
corresponding operators of the $\alpha\nu$ pseudoparticles as
follows,

\begin{equation}
f^{\dag }_{{\bar{q}}_j,\,\alpha\nu} =
{\Hat{V}}^{\dag}_{\alpha\nu}\,b^{\dag
}_{q_j,\,\alpha\nu}\,{\Hat{V}}_{\alpha\nu} \, ; \hspace{1cm}
f_{{\bar{q}}_j,\,\alpha\nu} =
{\Hat{V}}^{\dag}_{\alpha\nu}\,b_{q_j,\,\alpha\nu}\,{\Hat{V}}_{\alpha\nu}
\, . \label{f}
\end{equation}
Here ${\Hat{V}}_{\alpha\nu}$ is a unitary operator that we call
the {\it $\alpha\nu$ pseudoparticle - $\alpha\nu$ pseudofermion
unitary operator}. This operator acts onto the Hilbert subspace
spanned by the initial ground state and excited states A and B. It
shifts the bare-momentum value $q_j$ by $Q_{\alpha\nu}(q_j)/L$ and
thus is given by,

\begin{equation}
{\hat{V}}^{\dag}_{\alpha\nu} =
\exp\bigl\{\sum_{j=1}^{N^*_{\alpha\nu}}\,b^{\dag
}_{q_j+Q_{\alpha\nu}(q_j)/L,\,\alpha\nu}\,b_{q_j,\,\alpha\nu}\Bigr\}
\, ; \hspace{1cm} {\hat{V}}_{\alpha\nu} =
\exp\bigl\{\sum_{j=1}^{N^*_{\alpha\nu}}\,b^{\dag
}_{q_j-Q_{\alpha\nu}(q_j)/L,\,\alpha\nu}\,b_{q_j,\,\alpha\nu}\Bigr\}
\label{Van} \, .
\end{equation}

The momentum distribution function ${\cal{N}}_{\alpha\nu}
({\bar{q}}_j)$ is the eigenvalue of the operator,

\begin{equation}
{\hat{\cal{N}}}_{\alpha\nu} ({\bar{q}}_j) = f^{\dag
}_{{\bar{q}}_j,\,\alpha\nu}\,f_{{\bar{q}}_j,\,\alpha\nu} \, ;
\hspace{1cm} {\cal{N}}_{\alpha\nu} ({\bar{q}}_j) = N_{\alpha\nu}
(q_j({\bar{q}}_j)) \, . \label{Nqbarop}
\end{equation}
Here $N_{\alpha\nu} (q_j)$ stands for the $\alpha\nu$
pseudoparticle bare-momentum distribution function and
$q_j({\bar{q}}_j)$ is the inverse of the function
(\ref{pfermions}). The function $N_{\alpha\nu} (q_j)$ is the
eigenvalue of the corresponding bare-momentum number operator
(\ref{Nanop}). Keeping only the physical contributions that
correspond to terms up to first order in $1/L$, the function
$q_j({\bar{q}}_j)$ is given by,

\begin{equation}
q_j= q_j ({\bar{q}}_j) = {\bar{q}}_j - {Q_{\alpha\nu}
({\bar{q}}_j)\over L} = {\bar{q}}_j - {2\pi\over L}
\sum_{\alpha'=c,s}\,\sum_{\nu'=\delta_{\alpha',\,s}}^{\infty}\,
\sum_{j'=1}^{N^*_{\alpha'\nu'}}\,\Phi_{\alpha\nu,\,\alpha'\nu'}^f
({\bar{q}}_j,{\bar{q}}_{j'})\, \Delta {\cal{N}}_{\alpha'\nu'}
({\bar{q}}_{j'}) \, . \label{INVqj}
\end{equation}
Here, $\Delta {\cal{N}}_{\alpha'\nu'} ({\bar{q}}_{j'})$ is the
deviation of the function (\ref{Nqbarop}) relative to its
ground-state value. Since the functional (\ref{Qcan1j}) vanishes
for the ground state, note that $q_j = {\bar{q}}_j$ for that
state. Moreover, on the right-hand side of Eq. (\ref{INVqj})
$\Phi_{\alpha\nu,\,\alpha'\nu'}^f ({\bar{q}},{\bar{q}}')$ is the
two-pseudfermion phase shift. It is defined as,

\begin{equation}
\Phi_{\alpha\nu,\,\alpha'\nu'}^f ({\bar{q}},{\bar{q}}') =
\Phi_{\alpha\nu,\,\alpha'\nu'}
\Bigl(q({\bar{q}}),q({\bar{q}}')\Bigr) = \bar{\Phi
}_{\alpha\nu,\,\alpha'\nu'}
\left({4t\,\Lambda^{0}_{\alpha\nu}(q({\bar{q}}))\over U},
{4t\,\Lambda^{0}_{\alpha\nu}(q({\bar{q}}')))\over U}\right) \, ,
\label{Phi-barPhipf}
\end{equation}
where the function $q=q({\bar{q}})$ is the continuum version of
the function (\ref{INVqj}), $\Phi_{\alpha\nu,\,\alpha'\nu'}(q,q')$
is given in Eq. (\ref{Phi-barPhi}), and $\bar{\Phi
}_{\alpha\nu,\,\alpha'\nu'} (r ,\,r')$ is the two-pseudofermion
phase shift expressed in the variable $r$ defined by the integral
equations (\ref{Phis1c})-(\ref{Phisncn}) of Appendix B.

Often in the expressions given in previous sections we replaced
the pseudoparticle bare-momentum summations by integrals and the
corresponding discrete bare-momentum values $q_j$ by a continuum
bare-momentum variable $q$. Since according to Eq. (\ref{differ0})
of Appendix A, the difference $q_{j+1}-q_j = 2\pi/L$ is constant
for all values of $j$, the use of such a continuum representation
involves the replacement,

\begin{equation}
\sum_{j=1}^{N^*_{\alpha\nu}}\equiv\sum_{q=-q_{\alpha\nu
}}^{+q_{\alpha\nu}}\rightarrow {L\over 2\pi}\int_{-q_{\alpha\nu
}}^{+q_{\alpha\nu}}\,dq \, , \label{sumint0}
\end{equation}

In the Hilbert subspace spanned by the excited states A and B, the
rapidity functional $\Lambda_{\alpha\nu}(q)$ and rapidity-momentum
functional $k(q)$ equal the corresponding ground-state rapidity
function $\Lambda_{\alpha\nu}^0(q)$ and rapidity-momentum function
$k^0(q)$, respectively, with the bare-momentum $q$ replaced by the
momentum ${\bar{q}}$. It follows that in such a subspace the
limiting values of the continuum momentum ${\bar{q}}$ are given by
the ground-state limiting values $\pm q_{\alpha\nu }^0$ given in
Eqs. (\ref{qcanGSefa}) and (\ref{qcanGS}). (This is confirmed in
Sec. VII.) Thus, to replace the discrete momentum values by a
continuum momentum variable ${\bar{q}}$, one must replace the
summations by the following integrals,

\begin{equation}
\sum_{j=1}^{N^*_{\alpha\nu}}\equiv\sum_{\bar{q}=-q_{\alpha\nu
}^0}^{+q_{\alpha\nu}^0}\rightarrow {L\over
2\pi}\int_{-q_{\alpha\nu }^0}^{+q_{\alpha\nu}^0}\,d{\bar{q}}
\,{dq({\bar{q}})\over d{\bar{q}}}\, . \label{sumint}
\end{equation}
We then introduce the momentum distribution function,

\begin{equation}
{\bar{\cal{N}}}_{\alpha\nu} ({\bar{q}}) = {dq({\bar{q}})\over
d{\bar{q}}}\,{\cal{N}}_{\alpha\nu} ({\bar{q}}) \, . \label{barNq}
\end{equation}
The function $q = q({\bar{q}})$ on the right-hand side of Eq.
(\ref{sumint}) is the inverse of the function given in Eq.
(\ref{barqan}) and is given in Eq. (\ref{INVqj}) with
${\bar{q}}_j$ replaced by ${\bar{q}}$. Its derivative is given by,

\begin{equation}
{dq({\bar{q}})\over d{\bar{q}}} = 1 -
\sum_{\alpha'=c,s}\,\sum_{\nu'=\delta_{\alpha',\,s}}^{\infty}\,
\int_{-q_{\alpha'\nu' }^0}^{+q_{\alpha'\nu'}^0}\,d{\bar{q}'}\,
{d\,\Phi_{\alpha\nu,\,\alpha'\nu'}^f({\bar{q}},\,{\bar{q}'})\over
d{\bar{q}}}\,\Delta {\bar{\cal{N}}}_{\alpha'\nu'} ({\bar{q}'}) \,
. \label{INVderq}
\end{equation}

The second term on the right-hand side of this equation is of
first order in the momentum distribution function deviations. For
the non-interacting pseudofermion description only momentum and
energy contributions up to first order in these deviations are
physical. As a result, in the case of momentum distribution
function deviations $\Delta {\bar{\cal{N}}}_{\alpha\nu}
({\bar{q}})$ one can consider that,

\begin{equation}
\Delta {\bar{\cal{N}}}_{\alpha\nu} ({\bar{q}}) = \Delta
{\cal{N}}_{\alpha\nu} ({\bar{q}}) \, , \label{barDNq}
\end{equation}
where in contrast to the case of Eq. (\ref{barNq}) we used
$dq({\bar{q}})/d{\bar{q}}=1$.

The $\alpha\nu$ pseudoparticle number operator,

\begin{equation}
{\hat{N}}_{\alpha\nu} = \sum_{j=1}^{N^*_{\alpha\nu}}\,b^{\dag
}_{q_j,\,\alpha\nu}\,b_{q_j,\,\alpha\nu} =
\sum_{q=-q_{\alpha\nu}}^{+q_{\alpha\nu}}\,b^{\dag
}_{q,\,\alpha\nu}\,b_{q,\,\alpha\nu} = {L\over
2\pi}\int_{-q_{\alpha\nu}}^{+q_{\alpha\nu}}\,dq \,b^{\dag
}_{q,\,\alpha\nu}\,b_{q,\,\alpha\nu} \, , \label{Nop}
\end{equation}
is invariant under the pseudoparticle - pseudofermion
transformation. It equals the number of $\alpha\nu$ pseudofermions
operator,

\begin{equation}
{\hat{\cal{N}}}_{\alpha\nu} =
\sum_{j=1}^{N^*_{\alpha\nu}}\,f^{\dag
}_{{\bar{q}}_j,\,\alpha\nu}\,f_{{\bar{q}}_j,\,\alpha\nu} =
\sum_{{\bar{q}}=-q_{\alpha\nu }^0}^{+q_{\alpha\nu}^0}\,f^{\dag
}_{{\bar{q}},\,\alpha\nu}\,f_{{\bar{q}},\,\alpha\nu} = {L\over
2\pi}\int_{-q_{\alpha\nu }^0}^{+q_{\alpha\nu}^0}\,d{\bar{q}}
\,{dq({\bar{q}})\over d{\bar{q}}}\,f^{\dag
}_{{\bar{q}},\,\alpha\nu}\,f_{{\bar{q}},\,\alpha\nu} \, .
\label{Nbarop}
\end{equation}
The $c\nu$ pseudoparticle charge and $s_c$ and $\sigma_c$ values
considered in Sec. III are also invariant under the same
transformation. The same occurs for the $s\nu$ pseudoparticle
$s_s$ and $\sigma_s$ values. The pseudoparticle - pseudofermion
transformation also leaves invariant the $\pm 1/2$ Yang holons and
$\pm 1/2$ HL spinons. The $\pm 1/2$ holon (and $\pm 1/2$ spinon)
composite character of the $c\nu$ pseudoparticles (and $s\nu$
pseudoparticles) also remains invariant under that transformation.
It follows that the $c\nu$ pseudofermions (and $s\nu$
pseudofermions) are $s_c=0$ (and $s_s=0$) composite objects of an
equal number $\nu=1,2,...$ of $-1/2$ holons and $+1/2$ holons (and
$-1/2$ spinons and $+1/2$ spinons). Consistently, the $\pm 1/2$
holon ($\alpha =c$) and $\pm 1/2$ spinon ($\alpha =s$) number
operators ${\hat{M}}_{\alpha,\,\pm 1/2}$ given in Eq. (\ref{Mop})
can be rewritten in terms of pseudofermion operators as follows,

\begin{equation}
{\hat{M}}_{\alpha,\,\pm 1/2} = {\hat{L}}_{\alpha,\,\pm 1/2} +
\sum_{\nu=1}^{\infty}\,\sum_{q=-q_{\alpha\nu }^0}^{+q_{\alpha\nu
}^0}\nu\,{\hat{\cal{N}}}_{\alpha\nu} (\bar{q}) \, , \label{Mopf}
\end{equation}
where the pseudfermion momentum distribution operators
${\hat{\cal{N}}}_{\alpha\nu} (\bar{q})$ are given in Eq.
(\ref{Nqbarop}). Like in Eq. (\ref{Mop}), the operator
${\hat{L}}_{\alpha,\,\pm 1/2}$ on the right-hand side of Eq.
(\ref{Mopf}) is the $\pm 1/2$ Yang holon ($\alpha =c$) and $\pm
1/2$ HL spinon ($\alpha =s$) number operator given in Eq.
(\ref{NYhHLsop}). All results reported in Sec. III concerning
pseudoparticle charge and spin transport are also valid for the
corresponding pseudofermions. For instance, for finite values of
$U/t$ the transport of charge (and spin) is associated with the
$c0$ pseudofermion and $c\nu$ pseudofermion quantum charge fluids
(and $s\nu$ pseudofermion quantum spin fluids).

It is found in Ref. \cite{IIIb} and discussed in Appendix A that
the bare-momentum $q$ is the conjugate of the spatial coordinate
$x_j =a_{\alpha\nu}\,j$ associated with the effective $\alpha\nu$
lattice, where $j=1,2,...,N^*_{\alpha\nu}$. As in the case of the
charge (or spin) carried by the pseudoparticles and of their
composite character in terms of chargeaons and antichargeons, $\pm
1/2$ holons, or $\pm 1/2$ spinons, also the effective $\alpha\nu$
lattice remains invariant under the $\alpha\nu$ pseudoparticle -
$\alpha\nu$ pseudofermion unitary transformation. Indeed, the
momentum functional $Q_{\alpha\nu}(q)/L$ which controls the
pseudoparticle - pseudofermion transformation (\ref{pfermions})
does not affect the underlying effective $\alpha\nu$ lattice. That
momentum functional just imposes a twisted boundary condition such
that each $\alpha\nu$ pseudofermion hopping from site
$N^*_{\alpha\nu}-1$ to site $0$ of such an effective lattice will
acquire a phase $e^{iQ_{\alpha\nu}(q)}$. From combination of that
analysis with the expression for the general momentum functional
$Q_{\alpha\nu}(q)/L$ given in Eq. (\ref{Qcan1j}), we find that the
phase shift $\Phi_{\alpha\nu,\,\alpha'\nu'}^f
({\bar{q}},{\bar{q}}')$ is such that creation into the system of a
$\alpha'\nu'$ pseudofermion of momentum ${\bar{q}}'$ imposes a
twisted boundary condition on the wave function of the $\alpha\nu$
pseudofermion branch. This condition is such that when the
$\alpha\nu$ pseudofermion of momentum ${\bar{q}}$ hops from site
$N_{\alpha\nu}^*-1$ to site $0$ of the effective $\alpha\nu$
lattice, it will acquire a phase
$e^{i\,2\pi\,\Phi_{\alpha\nu,\,\alpha'\nu'}^f
({\bar{q}},{\bar{q}}')}$. The studies of Ref. \cite{V} reveal that
within the pseudofermion description the few-electron spectral
properties are fully controlled by such two-pseudofermion twisted
boundary conditions associated with the phase shifts
(\ref{Phi-barPhipf}).

As for the case of the pseudoparticles, it is useful to introduce
the local $\alpha\nu$ pseudofermion creation operator $f^{\dag
}_{x_j,\,\alpha\nu}$ and annihilation operator
$f_{x_j,\,\alpha\nu}$. These operators are related to the
operators $f^{\dag }_{\bar{q},\,\alpha\nu}$ and
$f_{\bar{q},\,\alpha\nu}$, respectively, obtained from the
corresponding pseudoparticle operators through the relations given
in Eq. (\ref{f}), as follows,

\begin{equation}
f^{\dag }_{\bar{q},\,\alpha\nu} =
{1\over\sqrt{L}}\sum_{j=1}^{N^*_{\alpha\nu}}e^{i\bar{q}\,x_j}\,
f^{\dag }_{x_j,\,\alpha\nu} \, ; \hspace{1cm}
f_{\bar{q},\,\alpha\nu} =
{1\over\sqrt{L}}\sum_{j=1}^{N^*_{\alpha\nu}}e^{-i\bar{q}\,x_j}\,
f_{x_j,\,\alpha\nu} \, ,  \label{elopf}
\end{equation}
where the summations refer to the sites of the effective
$\alpha\nu$ lattice. The local $\alpha\nu$ pseudofermion creation
(and annihilation) operator $f^{\dag }_{x_j,\,\alpha\nu}$ (and
$f_{x_j,\,\alpha\nu}$) creates (and annihilates) a $\alpha\nu$
pseudofermion at the effective $\alpha\nu$ lattice site of spatial
coordinate $x_j =a_{\alpha\nu}^0\,j$, where
$j=1,2,...,N^*_{\alpha\nu}$ and $a_{\alpha\nu}^0$ is the effective
$\alpha\nu$ lattice constant given in Eq. (\ref{acanGS}). (Since
the pseudofermion representation refers to the Hilbert subspace
spanned by the initial ground state and the excited states A and
B, except for $1/L$ corrections we can consider that the
corresponding effective $\alpha\nu$ lattice constants are the
ground-state constants $a_{\alpha\nu}^0$.) Thus, the conjugate
variable of the momentum ${\bar{q}}_j$ of the $\alpha\nu$
pseudofermion branch is the space coordinate $x_j$ of the
corresponding effective $\alpha\nu$ lattice. The local $\alpha\nu$
pseudoparticles and corresponding local $\alpha\nu$ pseudofermions
have the same effective $\alpha\nu$ lattice. It follows that the
local pseudoparticle and local pseudofermion site distribution
configurations which describe the ground state and the excited
states A and B are the same. These configurations are expressed in
terms of rotated-electron site distribution configurations in Ref.
\cite{IIIb}.

While the local $\alpha\nu$ pseudoparticles and corresponding
local $\alpha\nu$ pseudofermions {\it live} in the same effective
$\alpha\nu$ lattice, the values of the set of discrete
bare-momentum values $\{q_j\}$ and momentum values
$\{{\bar{q}}_j\}$ such that $j=1,2,...,N^*_{\alpha\nu}$ are
different and related by Eq. (\ref{INVqj}). There is an one-to-one
relation between these two sets of discrete values, which keep the
same order because there is no level crossing. This property
follows from the discrete bare-momentum and momentum separation
given in Eq. (\ref{differ0}) of Appendix A and Eq. (\ref{differ}),
respectively.

Finally, we consider the anticommutation relations of the
pseudofermion operators. It is confirmed in Ref. \cite{V} that
such relations play a major role in the evaluation of matrix
elements between energy eigenstates. Let us consider the general
situation when the momenta ${\bar{q}}$ and ${\bar{q}}'$ of the
operators $f^{\dag }_{{\bar{q}},\,\alpha\nu}$ and
$f_{{\bar{q}}',\,\alpha\nu}$, respectively, correspond to
different J-CPHS ensemble subspaces. The anticommutator $\{f^{\dag
}_{{\bar{q}},\,\alpha\nu},\,f_{{\bar{q}}',\,\alpha'\nu'}\}$ can be
expressed in terms of the local-pseudofermion anticommutators
$\{f^{\dag }_{x_j,\,\alpha\nu},\,f_{x_{j'},\,\alpha'\nu'}\}$
associated with spatial coordinates $x_j$ and $x_{j'}$ of the
effective $\alpha\nu$ and $\alpha'\nu'$ lattices, respectively, as
follows,

\begin{equation}
\{f^{\dag
}_{{\bar{q}},\,\alpha\nu},\,f_{{\bar{q}}',\,\alpha'\nu'}\} =
{1\over
L}\,\sum_{j=1}^{N^*_{\alpha\nu}}\,\sum_{j'=1}^{N^*_{\alpha'\nu'}}\,
e^{i({\bar{q}}\,x_j-{\bar{q}}'\,x_{j'})}\, \{f^{\dag
}_{x_j,\,\alpha\nu},\,f_{x_{j'},\,\alpha'\nu'}\} \, .
\label{pfacrjj}
\end{equation}
After performing the $j$ and $j'$ summations we find that the
$\alpha\nu$ pseudofermion operators obey the following algebra,

\begin{equation}
\{f^{\dag
}_{{\bar{q}},\,\alpha\nu},\,f_{{\bar{q}}',\,\alpha'\nu'}\} =
\delta_{\alpha,\,\alpha'}\,\delta_{\nu ,\,\nu'}\,{1\over
L}\,e^{-i({\bar{q}}-{\bar{q}}')\,a/
2}\,e^{i(Q_{\alpha\nu}(q)-{Q'}_{\alpha\nu}
(q'))/2}\,{\sin\Bigl([Q_{\alpha\nu} (q)-{Q'}_{\alpha\nu}(q')]/
2\Bigr)\over\sin ([{\bar{q}}-{\bar{q}}']\,a/2)} \, , \label{pfacr}
\end{equation}
and the anticommutators between two $\alpha\nu$ pseudofermion
creation or annihilation operators vanish. On the right-hand side
of Eq. (\ref{pfacr}) the functionals $Q_{\alpha\nu}(q)$ and
${Q'}_{\alpha\nu}(q')$ whose general expression is given in Eq.
(\ref{Qcan1j}) refer to the J-CPHS ensemble subspaces where the
momenta ${\bar{q}}$ and ${\bar{q}}'$ refer to, respectively. Note
that when $\sin ([Q_{\alpha\nu} (q)-{Q'}_{\alpha\nu} (q)]/2)$
vanishes the anticommutation relation (\ref{pfacr}) is the usual
one, $\{f^{\dag
}_{{\bar{q}},\,\alpha\nu},\,f_{{\bar{q}}',\,\alpha'\nu'}\} =
\delta_{\alpha,\,\alpha'}\,\delta_{\nu
,\,\nu'}\,\delta_{{\bar{q}},\,{\bar{q}}'}$. In contrast, for
finite values of that quantity there is an overall
two-pseudofermion phase shift which arises from a non-perturbative
shake-up effect associated with the functional character of the
momentum $Q_{\alpha\nu}(q)/L$ given in Eq. (\ref{Qcan1j}).

A case of particular importance is when one of the two J-CPHS
ensemble subspaces associated with the momenta ${\bar{q}}$ and
${\bar{q}}'$ is that of the initial ground state. In such a case
${Q'}_{\alpha\nu}(q')=0$ for the ground-state J-CPHS ensemble
subspace and thus the anticommutation relation (\ref{pfacr})
simplifies to,

\begin{equation}
\{f^{\dag
}_{{\bar{q}},\,\alpha\nu},\,f_{{\bar{q}}',\,\alpha'\nu'}\} =
\delta_{\alpha,\,\alpha'}\,\delta_{\nu ,\,\nu'}\,{1\over
L}\,e^{-i({\bar{q}}-{\bar{q}}')\,a/
2}\,e^{iQ_{\alpha\nu}(q)/2}\,{\sin\Bigl(Q_{\alpha\nu} (q)/
2\Bigr)\over\sin ([{\bar{q}}-{\bar{q}}']\,a/2)} \, .
\label{pfacrGS}
\end{equation}

%%%%%%%%%%%%%%%%%%%%%%%%%%%%%%%%%%%%%%%%%%%%%%%%%%%%%%%%%%%%%%%%
\section{THE PSEUDOFERMION ENERGY AND MOMENTUM SPECTRA AND THE
WAVE-FUNCTION FACTORIZATION OF THE NORMAL-ORDERED 1D HUBBARD
MODEL}

In this section, we find that the description of the quantum
problem in terms of the non-interacting pseudofermions leads in
the thermodynamic limit to a wave-function factorization for the
excited states A and B contained in the excitations generated by
application onto the ground state of few-electron operators. Such
a factorization refers to the ground-state normal-ordered 1D
Hubbard model.

%%%%%%%%%%%%%%%%%%%%%%%%%%%%%%%%%%%%%%%%%%%%%%%%%%%%%%%%%%%%%%%%
\subsection{THE PSEUDOFERMION ENERGY AND MOMENTUM SPECTRA}

Provided that all scattering orders are considered, the
pseudoparticle energy functional (\ref{EF}) describes excitations
which involve a small, but finite density of pseudoparticles.
However, in the thermodynamic limit, the few-electron excitations
are contained in a smaller Hilbert subspace. This subspace is
spanned by the excited states A and B. These states are generated
from the ground state by processes which involve a finite number
of pseudofermions (and pseudoparticles). In the thermodynamic
limit, this corresponds to a vanishing density of these objects
rather than to a small but finite density. In the Hilbert subspace
spanned by the ground state and the excited states A and B the
energy spectrum is of the form given in Eq. (\ref{ESS}). This
spectrum can be expressed in terms of continuum momentum integrals
with the result,

\begin{eqnarray}
E & = & -2t {L\over 2\pi} \int_{-q_{c0}^0}^{+q_{c0}^0}
d{\bar{q}}\, {\bar{\cal{N}}}_{c0} ({\bar{q}})\, \cos k^0(\bar{q})
+ 4t {L\over 2\pi}\sum_{\nu=1}^{\infty} \int_{-q_{c\nu
}^0}^{+q_{c\nu}^0} d{\bar{q}}\, {\bar{\cal{N}}}_{c\nu}
({\bar{q}})\,{\rm Re}\,\Bigl\{\sqrt{1 -
(\Lambda_{c\nu}^0 (\bar{q}) + i \nu U/4t)^2}\Bigr\} \nonumber \\
& + & {U\over 2}\Bigl[\,M_c -
\sum_{\nu=1}^{\infty}2\nu\,{\cal{N}}_{c\nu} -{N_a\over 2}\Bigr] +
\sum_{\alpha}\mu_{\alpha }S^{\alpha}_z \, , \label{ESS2}
\end{eqnarray}
where ${\bar{\cal{N}}}_{c0} ({\bar{q}})$ and
${\bar{\cal{N}}}_{c\nu} ({\bar{q}})$ are the pseudofermion
momentum distribution functions of Eq. (\ref{barNq}).

In contrast to the pseudoparticle energy functional
(\ref{EF})-(\ref{E2}) in terms of the pseudoparticle bare-momentum
distribution function deviations, the energy spectrum derived from
expressions (\ref{ESS}) and (\ref{ESS2}) only includes first-order
terms in the pseudofermion momentum distribution function
deviations. In the Hilbert subspace that the energy spectrum
(\ref{ESS})-(\ref{ESS2}) refers to, only such first-order
pseudofermion momentum distribution function deviation
contributions have physical meaning. The information recorded in
the pseudoparticle interactions is transferred over to the
momentum two-pseudofermion phase shifts of the momentum term
$Q_{\alpha\nu}(q)/L$. The pseudofermion discrete momentum values
are of $1/L$ order and store this information.

According to Eq. (\ref{Qcan1j}), the processes which generate the
excited states A and B from the initial ground state lead to a
collective momentum shift of all the $c0$ pseudofermions and $s1$
pseudofermions of the initial-state {\it Fermi sea}. That momentum
shift reads,

\begin{equation}
[\pi_{\alpha\nu}^0 + Q_{\alpha\nu} (q_j)]/L \, ; \hspace{0.5cm}
\pi_{\alpha\nu}^0 = 0,\,\pm\pi \, ; \hspace{0.5cm} \alpha =
c0,\,s1 \, , \label{Q+pi}
\end{equation}
where the bare-momentum shift $\pi_{\alpha\nu}^0/L$ is given by,

\begin{eqnarray}
\pi_{c0}^0 & = & 0 \, ; \hspace{0.5cm} \sum_{\alpha
=c,\,s}\sum_{\nu=1}^{\infty} \Delta N_{\alpha\nu} \hspace{0.25cm}
{\rm even} \, ;  \hspace{1.0cm} \pi_{c0}^0=\pm\pi \, ;
\hspace{0.5cm} \sum_{\alpha =c,\,s}\sum_{\nu=1}^{\infty} \Delta
N_{\alpha\nu}
\hspace{0.25cm} {\rm odd} \, ; \nonumber \\
\pi_{s1}^0 & = & 0 \, ; \hspace{0.5cm} \Delta N_{c0}+\Delta N_{s1}
\hspace{0.25cm} {\rm even} \, ; \hspace{1.0cm} \pi_{s1}^0=\pm\pi
\, ; \hspace{0.5cm} \Delta N_{c0}+\Delta N_{s1} \hspace{0.25cm}
{\rm odd} \, . \label{pic0s1}
\end{eqnarray}
The excited states A and B are generated from the ground state by
two virtual excitations: (1) a collective momentum shift of the
{\it Fermi sea} $c0$ pseudofermions and $s1$ pseudofermions; (2)
for the states A, a finite number of $c0$ pseudofermion and $s1$
pseudofermion creation or annihilation processes and of
$\alpha\nu\neq c0$ and $\alpha\nu\neq s1$ pseudofermion, $-1/2$
Yang holon, and $-1/2$ HL spinon creation processes; for the
states B, a finite number of $c0$ pseudofermion and $s1$
pseudofermion particle-hole processes. Consideration of both the
energy associated with the virtual excitations (1) and (2), gives
the energy spectrum of few-electron excitations associated with
the energy (\ref{ESS})-(\ref{ESS2}). Such an energy spectrum
corresponds to the ground-state normal-ordered 1D Hubbard model
and is additive in the $-1/2$ holon, $-1/2$ spinons, and
$\alpha\nu$ pseudofermion energies. It reads,

\begin{eqnarray}
\Delta E & = & \omega_0 + \sum_{j=1}^{N_a}\,\Delta {\cal{N}}_{c0}
({\bar{q}}_j) \,\epsilon_{c0} ({\bar{q}}_j) +
\sum_{j=1}^{N^*_{s1}}\,\Delta {\cal{N}}_{s1} ({\bar{q}}_j)
\,\epsilon_{s ,\,1}({\bar{q}}_j) + \sum_{\alpha =c,\,s}\,\sum_{\nu
=1+\delta_{\alpha ,\,s}}^{\infty}\,
\sum_{j=1}^{N^*_{\alpha\nu}}\,\Delta {\cal{N}}_{\alpha\nu}
({\bar{q}}_j)\,\epsilon^0_{\alpha\nu}({\bar{q}}) \nonumber
\\
& = & \omega_0 + {L\over 2\pi} \int_{-q_{c0}^0}^{+q_{c0}^0}
d{\bar{q}}\, \Delta {\cal{N}}_{c0} ({\bar{q}})\,\epsilon_{c0}
({\bar{q}}) + {L\over 2\pi}\sum_{\nu=1}^{\infty} \int_{-q_{s1
}^0}^{+q_{s1}^0} d{\bar{q}}\,\Delta {\cal{N}}_{s1}
({\bar{q}})\,\epsilon_{s ,\,1}({\bar{q}}) \nonumber \\ & + &
{L\over 2\pi}\sum_{\alpha =c,\,s}\,\sum_{\nu =1+\delta_{\alpha
,\,s}}^{\infty}\,\int_{-q_{\alpha\nu }^0}^{+q_{\alpha\nu}^0}
d{\bar{q}}\,\Delta {\cal{N}}_{\alpha\nu} ({\bar{q}})
\,\epsilon^0_{\alpha\nu}({\bar{q}}) \, , \label{E1pf}
\end{eqnarray}
where the energy parameter $\omega_0$ is given by,

\begin{equation}
\omega_0 = 2\mu\, \Delta M_{c,\,-1/2} + 2\mu_0\,H\, [\Delta
M_{s,\,-1/2}-\Delta {\cal{N}}_{s1}] \, . \label{om0}
\end{equation}
Here $\Delta M_{\alpha ,\,-1/2}$ are the deviations in the numbers
of $-1/2$ holons ($\alpha =c$) and of $-1/2$ spinons ($\alpha =s$)
and $\Delta {\cal{N}}_{s1}=\Delta N_{s1}$ is the deviation in the
number of $s1$ pseudofermions. The limiting-momentum values
$q_{c0}^0$, $q_{s1}^0$, and $q_{\alpha\nu}^0$ appearing in the
integrations of Eqs. (\ref{ESS2}) and (\ref{E1pf}) are given in
Eqs. (\ref{qcanGSefa}) and (\ref{qcanGS}). Moreover, on the
right-hand side of Eq. (\ref{E1pf}) the functions $\epsilon_{c0}
({\bar{q}})$, $\epsilon_{s ,\,1}({\bar{q}})$, and
$\epsilon^0_{\alpha\nu}({\bar{q}})$ are the pseudofermion energy
bands. These energy bands equal the corresponding pseudoparticle
energy bands appearing on the right-hand side of Eq. (\ref{E1}),
provided that the bare momentum $q$ is replaced by the momentum
$\bar{q}$. The latter bands are plotted in Figs. 6 to 9 of Ref.
\cite{II} as a function of the bare momentum for zero spin density
and several values of electronic density and on-site Coulombian
repulsion.

The shape of the pseudofermion energy bands is controlled by the
two-pseudofermion phase shifts and reads,

\begin{equation}
\epsilon_{c0} (\bar{q}) = -{U\over 2} -2t\cos k^{0}(\bar{q}) +
2t\int_{-Q}^{+Q}dk\,\widetilde{\Phi }_{c0,\,c0}
\left(k,k^{0}(\bar{q})\right)\,\sin k + \mu - \mu_0 H \, ,
\label{epc}
\end{equation}

\begin{equation}
\epsilon^0_{c\nu} (\bar{q}) = -\nu\, U + 4t\,{\rm Re}\,\Bigl\{
\sqrt{1 - [\Lambda^{0}_{c\nu}(\bar{q}) - i\nu {U\over
4t}]^2}\Bigr\} + 2t\int_{-Q}^{+Q}dk\,\widetilde{\Phi }_{c0,\,c\nu}
\left(k,\Lambda^{0}_{c\nu}(\bar{q})\right)\,\sin k \, ,
\label{epcn}
\end{equation}

\begin{equation}
\epsilon^0_{s\nu} (\bar{q}) = 2t\int_{-Q}^{+Q}dk\, \widetilde{\Phi
}_{c0,\,s\nu} \left(k,\Lambda^{0}_{s\nu}(\bar{q})\right)\,\sin k
\, , \label{epsn}
\end{equation}
and

\begin{equation}
\epsilon_{s1} (\bar{q}) = \epsilon_{s1}^0 (\bar{q}) + 2\mu_0 H \,
. \label{eps1}
\end{equation}
The two-pseudofermion phase shifts on the right-hand side of these
equations are defined in Eqs. (\ref{tilPcc})-(\ref{tilPanan}). The
parameter $Q$ is given in Eq. (\ref{QB}) and the functions
$k^{0}(\bar{q})$ and $\Lambda^{0}_{c\nu}(\bar{q})$ are the inverse
of the functions defined in Eqs. (\ref{kcGS}) and (\ref{GcnGS}),
respectively. The zero-energy levels of the energy dispersions
(\ref{epc})-(\ref{eps1}) are such that,

\begin{equation}
\epsilon_{c0} (\pm 2k_F) = \epsilon_{s1} (\pm k_{F\downarrow})=
\epsilon_{c\nu}^0 (\pm [\pi/a -2k_F])=\epsilon_{s\nu}^0 (\pm
[k_{F\uparrow}-k_{F\downarrow}])=0 \, . \label{eplev0}
\end{equation}
The dependence on the electronic density $n$, spin density $m$,
and $U/t$ of the chemical potential $\mu$ and magnetic field $H$
appearing in Eqs. (\ref{epc}) and (\ref{eps1}) can be expressed in
terms of the two-pseudofermion phase shifts as,

\begin{equation}
\mu = {U\over 2} +2t\cos Q - 2t\int_{-Q}^{+Q}dk\,\widetilde{\Phi
}_{c0,\,c0} \left(k,Q\right)\,\sin k - t\int_{-Q}^{+Q}dk\,
\widetilde{\Phi }_{c0,\,s1} \left(k,B\right)\,\sin k \, ,
\label{mu}
\end{equation}
and

\begin{equation}
H = -{t\over \mu_0}\int_{-Q}^{+Q}dk\, \widetilde{\Phi }_{c0,\,s1}
\left(k,B\right)\,\sin k \, , \label{Hm0}
\end{equation}
respectively.

We emphasize that the virtual-excitation (1) energy spectrum
vanishes and thus the general deviation-linear energy spectrum
(\ref{E1pf}) amounts to the contributions from the excitation (2).
By construction of the pseudofermion description subspace, the
virtual excitation (2) of the states A and B, involves changes in
the occupancy configurations of a finite number of quantum
objects. However, the excitation (1) is a collective momentum
shift of all the ground-state $c0$ pseudofermions and $s1$
pseudofermions. In the thermodynamic limit, the value of the
corresponding ground-state numbers ${\cal{N}}_{c0}=N$ and
${\cal{N}}_{s1}=N_{\downarrow}$ approaches infinity. Thus, the
self-consistency of the non-interacting pseudofermion theory
implies that the energy of the excitation (1) vanishes in that
limit, so that the total energy is additive in the corresponding
pseudofermion energies. Let us confirm that as
$L\rightarrow\infty$ this holds true. The $c0$ and $c1$
pseudofermion momentum distribution function deviations of Eq.
(\ref{E1pf}) can be written as,

\begin{equation}
\Delta {\cal{N}}_{\alpha\nu} ({\bar{q}}_j) = \Delta
{\cal{N}}^{(1)}_{\alpha\nu} ({\bar{q}}_j) + \Delta
{\cal{N}}^{(2)}_{\alpha\nu} ({\bar{q}}_j) \, ; \hspace{1cm}
\alpha\nu = c0,\,s1 \, , \label{Nqcs1-12}
\end{equation}
where $\Delta {\cal{N}}^{(1)}_{\alpha\nu} ({\bar{q}}_j)$ and
$\Delta {\cal{N}}^{(2)}_{\alpha\nu} ({\bar{q}}_j)$ are associated
with the collective pseudofermion {\it Fermi sea} excitation (1)
and excitation (2), respectively. Within the first order $1/L$
pseudofermion description the former deviation reads,

\begin{eqnarray}
\Delta {\cal{N}}^{(1)}_{\alpha\nu} ({\bar{q}}) & = &
{\cal{N}}^0_{\alpha\nu} (q+ {[\pi_{\alpha\nu}^0 +
Q_{\alpha\nu}(q)]\over
L})-{\cal{N}}^0_{\alpha\nu} (q) \nonumber \\
& = & {[Q_{\alpha\nu}(q)+\pi_{\alpha\nu}^0]\over L}{\partial
{\cal{N}}^0_{\alpha\nu} ({\bar{q}})\over \partial {\bar{q}}} =
-{\rm sgn} (\bar{q})\,{[\pi_{\alpha\nu}^0 + Q_{\alpha\nu}({\rm
sgn} (\bar{q})q^0_{F\alpha\nu})]\over N_a}\,\delta
(q^0_{F\alpha\nu}-\vert\,\bar{q}\vert)  \, ; \hspace{0.5cm} \alpha
= c0,\,s1 \, . \label{DevEt}
\end{eqnarray}
Here, the momentum {\it Fermi value} $q^0_{F\alpha\nu}$ of the
$c0$ and $s1$ branches is given in Eq. (\ref{q0Fcs}) and
$\pi_{\alpha\nu}^0/L$ is the corresponding bare-momentum shift of
Eq. (\ref{pic0s1}). To first order in $1/L$, use of Eq.
(\ref{DevEt}) in the energy spectrum (\ref{E1pf}) leads to the
following excitation (1) energy spectrum,

\begin{equation}
\Delta E^{(1)}_{c0,\,s1} =\sum_{\alpha\nu
=c0,s1}\sum_{{{\bar{q}}_j}= -q^0_{\alpha\nu }}^{+q^0_{\alpha\nu }}
\,\Delta {\cal{N}}^{(1)}_{\alpha\nu} (\bar{q})
\epsilon_{\alpha\nu} (\bar{q})= -\sum_{\iota =\pm
1}\sum_{\alpha\nu =c0,s1}\iota \,{[\pi_{\alpha\nu}^0 +
Q_{\alpha\nu}(\iota\,q^0_{F\alpha\nu})]\over
N_a}\,\epsilon_{\alpha\nu} (q^0_{F\alpha\nu})=0 \, .
\label{VanishDEt}
\end{equation}
In order to obtain the result (\ref{VanishDEt}) we have used the
symmetry $\epsilon_{\alpha\nu} (q)=\epsilon_{\alpha\nu} (-q)$ and
Eq. (\ref{eplev0}) such that $\epsilon_{\alpha\nu}
(q^0_{F\alpha\nu})=0$ for $\alpha\nu=c0,s1$. We recall that the
occupancies of other $\alpha\nu\neq c0,\,s1$ pseudofermion
branches vanish for the ground state. Since the excitation (1)
involves all ground-state $c0$ and $s1$ pseudofermions, the
evaluation of the $1/L$ higher-order contributions to the
corresponding energy spectrum involves non-linear deviation terms.
We find that such $1/L$ higher-order energy contributions also
vanish as $L\rightarrow\infty$. Thus, the self-consistency
condition that the excitation (1) energy spectrum vanishes as the
system length $L$ approaches infinity is fulfilled.

The energy (\ref{om0}) on the right-hand side of Eq. (\ref{E1pf})
controls the finite-energy physics. The remaining energy terms
correspond to gapless contributions provided that the involved
pseudofermions correspond to the momentum values of the
energy-band arguments of Eq. (\ref{eplev0}). For most excited
states, the latter terms also lead to finite-energy contributions.
The typical value of such energy contributions is of the order of
the pseudofermion energy dispersion band-width per pseudofermion
involved in the excited states. We note that the energy spectrum
(\ref{E1pf}) of the excited states A and B, can have any finite
value associated with the regions of the ($k,\,\omega$) plane
where the few-electron spectral functions have finite spectral
weight \cite{V,spectral}.

Also, the excitation momentum spectrum can be written in terms of
the pseudofermion momentum distribution function deviations. It is
given by,

\begin{eqnarray}
\Delta P & = & {\pi\over a}\,\Delta M_{c,\,-1/2} +
\sum_{j=1}^{N_a}\,\Delta {\cal{N}}_{c0} ({\bar{q}}_j)
\,{\bar{q}}_j + \sum_{\nu
=1}^{\infty}\,\sum_{j=1}^{N^*_{s\nu}}\,\Delta {\cal{N}}_{s\nu}
({\bar{q}}_j) \,{\bar{q}}_j + \sum_{\nu =1}^{\infty}\,
\sum_{j=1}^{N^*_{c\nu}}\,\Delta {\cal{N}}_{c\nu}
({\bar{q}}_j)\,[{\pi\over a} -{\bar{q}}_j]
\nonumber \\
& = & {\pi\over a}\,\Delta M_{c,\,-1/2} + {L\over 2\pi}
\int_{q_{c0}^{-}}^{q_{c0}^{+}} d{\bar{q}}\,\Delta {\cal{N}}_{c0}
({\bar{q}})\,{\bar{q}} \nonumber \\
& + & {L\over 2\pi}\,\sum_{\nu =1}^{\infty}\,\int_{-q_{s\nu
}^0}^{+q_{s\nu}^0} d{\bar{q}}\,\Delta {\cal{N}}_{s\nu}
({\bar{q}})\,{\bar{q}} + {L\over 2\pi}\,\sum_{\nu =1}^{\infty}\,
\int_{-q_{c\nu }^0}^{+q_{c\nu}^0} d{\bar{q}}\,\Delta
{\cal{N}}_{c\nu} ({\bar{q}})\,[{\pi\over a} -{\bar{q}}_j] \, .
\label{noPpf}
\end{eqnarray}

The important point is that the large-$L$ pseudofermion
operational description introduced in this paper, associated with
the momentum distribution function deviation first-order energy
spectrum (\ref{E1pf}) and momentum spectrum (\ref{noPpf}),
contains full information about the few-electron spectral
properties. Besides the expression in terms of pseudofermion
operators of the diagonal (in the basis of the energy eigenstates)
operators associated with these spectra, such a description is
suitable for the evaluation of few-electron spectral functions for
finite values of energy \cite{V}.

When acting in the Hilbert subspace spanned by the initial ground
state and excited states A and B, the ground-state normal-ordered
1D Hubbard model and momentum operator can be written in terms of
pseudofermion, $-1/2$ Yang holon, and $-1/2$ HL spinon operators
as follows,

\begin{equation}
:\hat{H}: = \sum_{\alpha =c,\,s}\,\sum_{\nu
=\delta_{\alpha,\,s}}^{\infty}\, \sum_{j=1}^{N^*_{\alpha\nu}}
\,\epsilon_{\alpha\nu}({\bar{q}}_j)\,:f^{\dag
}_{{\bar{q}}_j,\,\alpha\nu}:\,:f_{{\bar{q}}_j,\,\alpha\nu}: +
\sum_{\alpha =c,s}\,\epsilon_{L\alpha,\,-1/2}\,{\hat{L}}_{\alpha
,\,-1/2}
 \, , \label{Hno1pf}
\end{equation}
and

\begin{eqnarray}
:\hat{P}:\, & = & \sum_{j=1}^{N_a}\,{\bar{q}}_j\,
:f^{\dag}_{{\bar{q}}_j,\,c0}:\,:f_{{\bar{q}}_j,\,c0}: + \sum_{\nu
=1}^{\infty}\,\sum_{j=1}^{N^*_{s\nu}}\,{\bar{q}}_j :f^{\dag
}_{{\bar{q}}_j,\,s\nu}:\,:f_{{\bar{q}}_j,\,s\nu}: \nonumber \\
& + & \sum_{\nu
=1}^{\infty}\,\sum_{j=1}^{N^*_{c\nu}}\,[(1+\nu){\pi\over a}
-{\bar{q}}_j]\,:f^{\dag
}_{{\bar{q}}_j,\,c\nu}:\,:f_{{\bar{q}}_j,\,c\nu}: + {\pi\over
a}\,{\hat{L}}_{c ,\,-1/2} \, , \label{noPoppf}
\end{eqnarray}
respectively, where $N^*_{c0}=N_a$ and the operator
${\hat{L}}_{\alpha ,\,-1/2}$ is given in Eq. (\ref{NYhHLsop}). On
the right-hand side of Eq. (\ref{Hno1pf}), the $c0$ and $s1$
pseudofermion energy bands are given in Eqs. (\ref{epc}) and
(\ref{eps1}), respectively, and the $\alpha\nu\neq c0$ and
$\alpha\nu\neq s1$ pseudofermion energy bands and $-1/2$ Yang
holon and $-1/2$ HL spinon energies read \cite{II},

\begin{equation}
\epsilon_{c\nu} (\bar{q}) = \epsilon_{c\nu}^0 (\bar{q}) + 2\nu\mu
\, ; \hspace{1cm} \epsilon_{s\nu} (\bar{q}) = \epsilon_{s\nu}^0
(\bar{q}) + 2\nu\mu_0 H \, ; \hspace{0.5cm}
\nu\geq\delta_{\alpha,\,s} \, , \label{epcnsn}
\end{equation}
and

\begin{equation}
\epsilon_{Lc,\,-1/2} = 2\mu \, ; \hspace{0.5cm}
\epsilon_{Ls,\,-1/2} = 2\mu_0\,H \, , \label{epHS}
\end{equation}
respectively. The energy dispersions $\epsilon_{c\nu}^0 (\bar{q})$
and $\epsilon_{s\nu}^0 (\bar{q})$ of Eq. (\ref{epcnsn}) are
defined in Eqs. (\ref{epcn}) and (\ref{epsn}), respectively. The
ground-state normal-ordered Hamiltonian (\ref{Hno1pf}) and
momentum operator (\ref{noPoppf}) correspond to the energy and
momentum spectra given in Eqs. (\ref{E1pf}) and (\ref{noPpf}),
respectively.

%%%%%%%%%%%%%%%%%%%%%%%%%%%%%%%%%%%%%%%%%%%%%%%%%%%%%%%%%%%%%%%%
\subsection{WAVE-FUNCTION FACTORIZATION OF THE NORMAL-ORDERED 1D HUBBARD
MODEL}

It is well known that both the ground state wave function and the
wave function of the excited states of the 1D Hubbard model can in
the $U/t\rightarrow\infty$ limit be constructed as a product of a
spin-less fermion wave function and a squeezed spin wave function
\cite{Ogata,Penc95,Penc96}. In our pseudofermion, Yang holon, and
HL spinon language this factorization means that in such a limit
the expression of the momentum and energy of these states is
linear in the $\alpha\nu$ pseudofermion momentum distribution
functions and in the $-1/2$ Yang holon and $-1/2$ HL spinon
numbers.

Let us show that for finite values of $U/t$ the energy (\ref{E})
is not linear in the $\alpha\nu$ pseudofermion momentum
distribution functions ${\bar{\cal{N}}}_{\alpha\nu} (\bar{q})$.
Our analysis refers to the few-electron excitation subspace. This
confirms that the above type of factorization does not occur for
finite values of the on-site repulsion. For the pseudofermion
Hilbert subspace spanned by the ground state and excited states A
and B the form of the general energy spectrum (\ref{E}) simplifies
to,

\begin{equation}
E = E^0 + \Delta E \, ; \hspace{1cm} E^0 = E_H^0 + {U\over
2}\Bigl[\,{N_a\over 2}-{\cal{N}}_{c0}\Bigr]+
\sum_{\alpha}\mu_{\alpha }S^{\alpha}_z \, , \label{Epfss}
\end{equation}
where $\Delta E$ is the general energy excitation spectrum given
in Eq. (\ref{E1pf}), $E_0$ stands for the ground-state energy,
${\cal{N}}_{c0}$ is the number of $c0$ pseudofermions, and the
term $\sum_{\alpha}\mu_{\alpha }S^{\alpha}_z$ is the same as on
the right-hand side of Eq. (\ref{E}). From combination of the
ground-state occupancy configurations provided in Appendix A with
the general energy spectrum given in Eq. (\ref{EH}) we find that
the energy $E_H^0$ on the right-hand side of Eq. (\ref{Epfss})
reads,

\begin{equation}
E_H^0 = -2t {L\over 2\pi} \int_{-{\pi\over a}}^{+{\pi\over a}}
d\bar{q}\, {\bar{\cal{N}}}_{c0}^0 (\bar{q})\,\cos k^0 (\bar{q}) \,
, \label{EGS}
\end{equation}
where $ {\bar{\cal{N}}}_{c0}^0 (\bar{q})$ is the ground-state $c0$
pseudofermion momentum distribution function. (We recall that
$q=\bar {q}$ for the ground state.) From use of Eqs. (\ref{kcGS}),
(\ref{tilPcc}), and (\ref{Phi-barPhipf}) we arrive to the
following expressions,

\begin{equation}
k^0 ({\bar{q}}) = \bar{q}-\int_{-{\pi\over a}}^{+{\pi\over a}}
d\bar{q}\, {\partial k^0 ({\bar{q}}')\over\partial {\bar{q}}'}\,
{\bar{\cal{N}}}_{c0}^0 ({\bar{q}}')\,\Phi_{c0,\,c0}^f
\left({\bar{q}}',\bar{q}\right) \, ; \hspace{0.5cm} {\partial k^0
({\bar{q}})\over\partial {\bar{q}}} = 1-\int_{-{\pi\over
a}}^{+{\pi\over a}} d\bar{q}\, {\partial k^0
({\bar{q}}')\over\partial {\bar{q}}'}\, {\bar{\cal{N}}}_{c0}^0
({\bar{q}}')\,{\partial\,\Phi_{c0,\,c0}^f
({\bar{q}}',{\bar{q}})\over\partial {\bar{q}}} \, .
\label{kcGSINV}
\end{equation}
By iterative solution of the equations given in (\ref{kcGSINV})
one can derive a functional representation for the
rapidity-momentum function $k^0 (\bar{q})$ in terms of the
ground-state distribution function ${\bar{\cal{N}}}_{c0}^0
(\bar{q})$. Use of such a functional on the right-hand side of Eq.
(\ref{EGS}) leads to the following non-linear energy functional in
the ground-state momentum distribution function
${\bar{\cal{N}}}_{c0}^0 (\bar{q})$,

\begin{eqnarray}
E_H^0 & = & -2t {L\over 2\pi} \int_{-{\pi\over a}}^{+{\pi\over a}}
d\bar{q}\, {\bar{\cal{N}}}_{c0}^0
(\bar{q})\,\sum_{j=0}^{\infty}\,{(-1)^j\over
(2j)!}\,\Bigl[\,\bar{q} - \int_{-{\pi\over a}}^{+{\pi\over a}}
d{\bar{q}}_1\, {\bar{\cal{N}}}_{c0}^0
({\bar{q}}_1)\,\Phi_{c0,\,c0}^f
({\bar{q}}_1,{\bar{q}}) \times \nonumber \\
& & \Bigl(1 + \sum_{l=1}^{\infty}\,\prod_{i=1}^l\,\int_{-{\pi\over
a}}^{+{\pi\over a}}\, d{\bar{q}}_{i+1}\, {\bar{\cal{N}}}_{c0}^0
({\bar{q}}_{i+1})\, {\partial\,\Phi_{c0,\,c0}^f
({\bar{q}}_{i+1},{\bar{q}}_i)\over\partial
{\bar{q}}_i}\Bigr)\Bigr]^{2j} \, . \label{EH0}
\end{eqnarray}
Here the indices $i=1,2,...$ of the momentum ${\bar{q}}_i$ label
independent integration continuum variables rather than discrete
momentum values. The form of expression (\ref{EH0}) confirms that
for finite values of $U/t$ the energy (\ref{EGS}) is highly
non-linear in the ground-state pseudofermion momentum distribution
function ${\bar{\cal{N}}}_{c0}^0 (\bar{q})$. Therefore, the above
type of factorization does not occur in general for the 1D Hubbard
model. However, one can confirm that in the limit
$U/t\rightarrow\infty$ the energy functional (\ref{EH0}) becomes
linear in ${\bar{\cal{N}}}_{c0}^0 (\bar{q})$. Indeed, one finds
from analysis of the integral equations provided in Appendix B
that for spin density $m=0$ the two-pseudofermion phase shift on
the right-hand side of Eq. (\ref{EH0}) is such that
$\Phi_{c0,\,c0}^f ({\bar{q}},{\bar{q}}')\rightarrow 0$ as
$U/t\rightarrow\infty$. Thus, in that limit the ground-state
expression (\ref{EH0}) simplifies to,

\begin{equation}
E_H^0 = -2t {L\over 2\pi} \int_{-{\pi\over a}}^{+{\pi\over a}}
d\bar{q}\, {\bar{\cal{N}}}_{c0}^0 (\bar{q})\,\cos (\bar{q}) \, ;
\hspace{0.5cm} U/t \rightarrow \infty \, . \label{EH0INF}
\end{equation}
This property also holds for the excited states and is behind the
full factorization of the wave functions used in the studies of
Refs. \cite{Ogata,Penc95,Penc96}.

Fortunately, the evaluation of few-electron spectral functions can
be achieved without the full factorization of the wave functions.
Such a problem can be solved by use of the ground-state
normal-ordered 1D Hubbard model. When expressed in terms of
pseudofermion operators, that normal-ordered Hamiltonian and
associated momentum operator are given in Eqs. (\ref{Hno1pf}) and
(\ref{noPoppf}), respectively. These quantum problems correspond
to the energy and momentum spectra given in Eqs. (\ref{E1pf}) and
(\ref{noPpf}), respectively. The ground-state normal-ordered
description is equivalent to specifying the energy eigenstates by
means of the momentum distribution function deviations $\Delta
{\bar{\cal{N}}}_{\alpha\nu}(\bar{q})=\Delta
{\cal{N}}_{\alpha\nu}(\bar{q})$ instead of in terms of the
corresponding full momentum distribution functions
${\bar{\cal{N}}}_{\alpha\nu}(\bar{q})$.

In the thermodynamic limit, there is a wave-function factorization
for the excited states belonging to the Hilbert subspace of the
pseudofermion normal-ordered Hamiltonian (\ref{Hno1pf}). This
factorization follows from the form of the expressions
(\ref{E1pf}) and (\ref{noPpf}) for the energy and momentum,
respectively. The energy spectrum defined in Eqs. (\ref{E1pf}) and
(\ref{om0}) can be written as,

\begin{equation}
\Delta E = \sum_{\alpha =c,\,s}\,\sum_{\nu =\delta_{\alpha
,\,s}}^{\infty}\,\Delta E_{\alpha\nu} + 2\mu\, \Delta L_{c,\,-1/2}
+ 2\mu_0\,H\, \Delta L_{s,\,-1/2} \, , \label{DE1pf}
\end{equation}
where the contributions from each branch are linear in the
$\alpha\nu$ pseudofermion momentum distribution function
deviations and read,

\begin{equation}
\Delta E_{\alpha\nu} = \sum_{j=1}^{N^*_{\alpha\nu}}\,\Delta
{\cal{N}}_{\alpha\nu}
({\bar{q}}_j)\,\epsilon_{\alpha\nu}({\bar{q}}) = {L\over
2\pi}\,\int_{-q_{\alpha\nu }^0}^{+q_{\alpha\nu}^0}
d{\bar{q}}\,\Delta {\cal{N}}_{\alpha\nu} ({\bar{q}})
\,\epsilon_{\alpha\nu}({\bar{q}}) \, . \label{DEan}
\end{equation}
Here the pseudofermion energy bands are given in Eqs. (\ref{epc}),
(\ref{eps1}), and (\ref{epcnsn}). The $-1/2$ Yang holons, $-1/2$
HL spinons, and each pseudofermion branch leads to a different
energy term linear in the corresponding number deviation or
momentum distribution function deviation. We emphasize that the
virtual-excitation (1) energy $\Delta E^{(1)}_{c0,\,s1}$ of Eq.
(\ref{VanishDEt}) associated with the momentum shift of Eq.
(\ref{Q+pi}) which involves all $c0$ pseudofermions and $s1$
pseudofermions of the ground-state {\it Fermi sea}, is also
additive in the energy contributions from each of these quantum
objects. Indeed, according to Eqs. (\ref{DevEt}) and
(\ref{VanishDEt}), the energy contribution from each $c0$
pseudofermion and $s1$ pseudofermion involved in such a collective
excitation vanishes. As the total energy $\Delta
E^{(1)}_{c0,\,s1}$ of the same excitation also vanishes, it is
indeed additive in the energies of all involved quantum objects.
Therefore, the wave function of the energy eigenstates of the
normal-ordered Hamiltonian can be expressed in the pseudofermion
subspace as a product of wave functions. Each wave function
corresponds to a different pseudofermion branch. In excited states
with finite $-1/2$ Yang holon and $-1/2$ HL spinon occupancy there
is also a wave function for these objects. This factorization is
associated with the deep physical meaning of the pseudoparticle -
pseudofermion transformation (\ref{sumint0}): it transfers the
information recorded in the pseudoparticle interactions over to
the pseudofermion momentum, providing a non-interacting character
to the latter objects.

In contrast, for the pseudoparticle representation the energy
functional (\ref{EF})-(\ref{E2}) includes bare-momentum
distribution function deviation non-linear terms associated with
the pseudoparticle residual interactions. For instance, the
quadratic energy term (\ref{E2}) contains bare-momentum summations
over products of deviations of the form $f_{\alpha\nu;
\alpha'\nu'}(q,q')\,\Delta N_{\alpha\nu}(q)\,\Delta
N_{\alpha'\nu'}(q')$. Such $f$ function terms are associated with
the residual two-pseudoparticle interactions. The occurrence of
these energy terms mixes contributions from different branches. It
follows that the pseudoparticle energy spectrum is not additive in
the $\alpha\nu$ pseudoparticle branch contributions, in contrast
to the pseudofermion energy spectrum given in Eqs. (\ref{E1pf})
and (\ref{DE1pf}). Thus, in this case the wave function of the
excited states does not factorize in the form of a product of
pseudoparticle wave functions.

The number of wave functions contributing to the factorized wave
function of a given energy eigenstate depends on the occupancy
configurations of that state. Only the $\alpha\nu$ pseudofermion
branches with finite occupancy in the state contribute the wave
function. This contribution is in the form of a $\alpha\nu$ wave
function factor. The same applies to the occupancy of $-1/2$ Yang
holons and $-1/2$ HL spinons.

%%%%%%%%%%%%%%%%%%%%%%%%%%%%%%%%%%%%%%%%%%%%%%%%%%%%%%%%%%%%%%%%
\section{TRANSFORMATION LAWS UNDER THE PSEUDOPARTICLE -
PSEUDOFERMION SUBSPACE UNITARY ROTATION}

In this section, we find the transformation laws under the
pseudoparticle - pseudofermion Hilbert subspace unitary rotation
of several quantum objects and quantities and discuss the physical
meaning of these laws. For instance, the $\alpha\nu$
pseudoparticle number operator (\ref{Nop}), is invariant under
such a transformation and thus equals the corresponding
$\alpha\nu$ pseudoparticle number operator (\ref{Nbarop}). Also
the Yang holons and HL spinons, the charge and spin carried by the
pseudoparticles, and the effective pseudoparticle lattices remain
invariant under such a transformation.

The pseudoparticle - pseudofermion unitary transformation is
generated by the operator (\ref{Van}), which only shifts the
pseudoparticle bare-momentum values. Thus, given a pseudoparticle
of bare-momentum $q$, the transformation law $q\rightarrow\bar{q}$
of its bare momentum $q$ also defines the transformation law of
such a quantum object under the pseudoparticle - pseudofermion
unitary transformation. For instance, if a specific value of the
bare momentum $q$ remains invariant under that transformation,
then the pseudoparticle carrying bare momentum $q$ is the same
quantum object as the corresponding pseudofermion.

%%%%%%%%%%%%%%%%%%%%%%%%%%%%%%%%%%%%%%%%%%%%%%%%%%%%%%%%%%%%%%%%
\subsection{TRANSFORMATION LAWS OF PSEUDOPARTICLES AT THE BARE-MOMENTUM
LIMITING VALUES}

The transformation laws of the limiting pseudoparticle
bare-momentum values $\pm q_{\alpha,\nu}$ under the pseudoparticle
- pseudofermion transformation defined in Eq. (\ref{sumint0})
provide interesting information about the physics described by the
$\alpha\nu$ pseudofermions. For the excited states A and B, the
limiting $\alpha\nu$ pseudoparticle bare-momentum values $\iota
\,q_{\alpha\nu}$, where $\iota =\pm 1$, can be written as,

\begin{equation}
\iota\,q_{\alpha\nu} = \iota\,q^0_{\alpha\nu} + \iota\,\Delta
q_{\alpha\nu} \, ; \hspace{0.5cm} \alpha =c,\,s \, ;
\hspace{0.3cm} \nu=1,2,... \, ; \hspace{0.3cm} \iota = \pm 1 \, .
\label{q0Dqag}
\end{equation}
Here $q^0_{\alpha\nu}$ is the ground-state limiting bare-momentum
value given in Eqs. (\ref{qcanGSefa}) and (\ref{qcanGS}).

We start by considering the case of $\nu
>0$ pseudoparticle branches. In that case the deviations
$\Delta q_{\alpha\nu}$ have the following form,

\begin{equation}
\iota\,\Delta q_{\alpha\nu}  = \iota\,{\pi\over L}\Bigr[\Delta
N_{\alpha\nu} + \Delta M_{\alpha} - \sum_{\nu'=1}^{\infty}
\Bigl(\nu + \nu' - \vert\nu - \nu'\vert\Bigl) \Delta
N_{\alpha\nu'}\Bigl] \, ; \hspace{0.5cm} \alpha =c,\,s \, ;
\hspace{0.3cm} \nu=1,2,... \, ; \hspace{0.3cm} \iota = \pm 1 \, ,
\label{Dq0Dqag}
\end{equation}
where $\Delta N_{\alpha\nu}$ (and $\Delta N_{\alpha\nu'}$) is the
deviation in the number of $\alpha\nu$ pseudoparticles (and
$\alpha\nu'$ pseudoparticles) and $\Delta M_{\alpha}$ stands for
the deviation in the holon number $M_c=[N_a -N_{c0}]$ or spinon
number $M_s=N_{c0}$. These expressions are easily obtained by
combination of Eqs. (\ref{N*}), (\ref{Nhcsn}), (\ref{aan}), and
(\ref{qag}) of Appendix A.

In Appendix C we find that at $q=\pm q_{\alpha\nu}$ and for $\nu
>0$ the functional defined in Eq. (\ref{Qcan1j}) equals the
following value for all excited states A and B,

\begin{eqnarray}
{Q_{\alpha\nu} (\iota\,q_{\alpha\nu})\over L} & = &
\sum_{\alpha'=c,s}\,\sum_{\nu'=1-\delta_{\alpha',\,c}}^{\infty}\,
\int_{-q_{\alpha'\nu'}}^{q_{\alpha'\nu'}} dq'\,
\Phi_{\alpha\nu,\,\alpha '\,\nu'}(\iota\,q_{\alpha\nu},q ')
\, \Delta N_{\alpha'\nu'}(q') \nonumber \\
& = & -\iota\,\Delta q_{\alpha\nu} \, ; \hspace{0.5cm} \alpha
=c,\,s \, ; \hspace{0.3cm} \nu=1,2,... \, ; \hspace{0.3cm} \iota =
\pm 1 \, , \label{Qcan1ql}
\end{eqnarray}
where the quantity $\Delta q_{\alpha\nu}$ is given in Eq.
(\ref{Dq0Dqag}). Here it is assumed that the bare momentum $q'$
associated with the pseudoparticle bare-momentum distribution
function deviations $\Delta N_{\alpha'\nu'}(q')$ on the right-hand
side of Eq. (\ref{Qcan1ql}) belongs to the domain $q'\in
(-q_{\alpha'\nu'},\,+q_{\alpha'\nu'})$ and can be such that
$q'\rightarrow \pm q_{\alpha'\nu'}$ but $q'\neq q_{\alpha'\nu'}$.
Interestingly, the value of the momentum functional $Q_{\alpha\nu}
(q)/L$ at $q=\iota\,q_{\alpha\nu}$ is such that it precisely
cancels the term $\iota\,\Delta q_{\alpha\nu}$ appearing in the
limiting bare-momentum expression on the right-hand side of Eq.
(\ref{q0Dqag}). Thus at the limiting bare-momentum values $q=\pm
q_{\alpha\nu}$ given in Eq. (\ref{q0Dqag}) the value of the
momentum functional ${\bar{q}}(q)$ defined in Eq. (\ref{barqan})
is independent of the value of the deviation $\Delta
q_{\alpha\nu}$ on the right-hand side of Eq. (\ref{q0Dqag}) which
defines the final excited state. By use of this result in Eq.
(\ref{barqan}) we find that,

\begin{equation}
\iota\,q_{\alpha\nu} = \iota\,q^0_{\alpha\nu} + \iota\,\Delta
q_{\alpha\nu} \, \rightarrow \, \iota\, {\bar{q}}_{\alpha\nu} =
\iota\,q^0_{\alpha\nu} \, ; \hspace{0.5cm} \alpha =c,\,s \, ;
\hspace{0.3cm} \nu=1,2,... \, ; \hspace{0.3cm} \iota = \pm 1 \, .
\label{pfconqan}
\end{equation}

Thus, we conclude that for $\nu>0$ pseudoparticle branches the
limiting momentum values of the corresponding $\alpha\nu$
pseudofermion bands equal  for all excited states A and B the
ground-state limiting bare-momentum values given in Eqs.
(\ref{qcanGSefa}) and (\ref{qcanGS}), {\it i.e.} $\pm
{\bar{q}}_{\alpha\nu} = \pm q^0_{\alpha\nu}$. While the width of
the bare-momentum {\it Brillouin zone} of the $\alpha\nu$
pseudoparticles, $2q_{\alpha\nu}$, has according to Eqs.
(\ref{q0Dqag}) and (\ref{Dq0Dqag}) an exotic dependence on the
state occupancy configurations, the value of the corresponding
momentum width of the $\alpha\nu$ pseudofermion {\it Brillouin
zone} is constant and equals $2{\bar{q}}_{\alpha\nu} =
2q^0_{\alpha\nu}$, where $q^0_{\alpha\nu}$ is the ground-state
limiting bare momentum given in Eqs. (\ref{qcanGSefa}) and
(\ref{qcanGS}).

This confirms that the width of the $\alpha\nu$ pseudofermion
momentum domain remains unchanged, as in the case of the momentum
bands of non-interacting particles. This result is consistent with
the expressions given in Eq. (\ref{FL}) for the rapidity and
rapidity-momentum functions, respectively, of the excited states.
When expressed in terms of the pseudofermion momentum ${\bar{q}}$
such functions have the same form both for the ground state and
excited states. This introduces the requirement that the width of
the momentum domain where these functions are defined must be the
same for the ground state and the excited states. Thus we have
just confirmed that this requirement is fulfilled through the
cancelling associated with the value of the momentum functional
(\ref{Qcan1ql}).

On the other hand, the bare-momentum width of the $c0$
pseudoparticles is independent of the pseudoparticle occupancy
configurations and reads $2q_{c0}^0=2\pi/a$. In Appendix C we find
that the transformation (\ref{sumint0}) leaves this width
invariant and shifts the bare-momentum values $q=0$ and $q=\pm
q_{c0}^0$ by the same amount as follows,

\begin{equation}
0\rightarrow {Q_{c0} (0)\over L} \, ; \hspace{1cm} \pm
q_{c0}^0\rightarrow \pm q_{c0}^0 + {Q_{c0} (0)\over L} \, ,
\label{pfconqc0}
\end{equation}
where $Q_{c0} (0)/L$ is the $q=0$ value of the momentum functional
$Q_{c0} (q)/L$ given in Eq. (\ref{Qcan1j}). Thus, for the $c0$
band the shift is the same for these three bare-momentum values.
As a result the width of the corresponding pseudofermion momentum
domain remains unchanged and is given by $2q_{c0}^0=2\pi/a$. Note
that the value of such a shift is a functional which depends on
the actual occupancy configurations of the excited states
pseudoparticle bare-momentum distribution function deviations.

In general, a $\alpha\nu$ pseudoparticle is different from the
corresponding $\alpha\nu$ pseudofermion. Below we find that for
$\alpha\nu$ pseudoparticle branches such that $\alpha\nu\neq c0$
and $\alpha\nu\neq s1$, the only exception is for bare-momentum
values $q$, such that $q\rightarrow 0$ in the particular case when
these pseudoparticles are involved in the bare-momentum
distribution function deviations of zero-momentum excited states.
According to the above results concerning the $\nu>0$ branches, as
$q\rightarrow \pm q_{\alpha\nu}$ the excited-state $\alpha\nu$
pseudoparticle limiting bare-momentum values (\ref{q0Dqag}) map
onto the same momentum values $\bar{q}=\pm q^0_{\alpha\nu}$
independently of the value of the small deviations $\pm \Delta
q_{\alpha\nu}$ on the right-hand side of Eq. (\ref{q0Dqag}). Like
the corresponding $\alpha\nu$ pseudoparticles \cite{II}, as
$\bar{q}\rightarrow \pm {\bar{q}}_{\alpha\nu}=\pm q^0_{\alpha\nu}$
the $\alpha\nu$ pseudofermion and the rotated $\alpha\nu$
pseudofermion become the same quantum object. In that limit, the
energy $\epsilon_{c\nu}(\bar{q}) = 2\mu\,\nu +
\epsilon^0_{c\nu}(\bar{q})$ (and $\epsilon_{s\nu}(\bar{q}) =
2\mu_0 H\,\nu + \epsilon^0_{s\nu}(\bar{q})$)  of the composite
$c\nu$ pseudofermion (and $s\nu$ pseudofermion) is additive in the
$-1/2$ Yang holon energy $2\mu$ (and $-1/2$ HL spinon energy
$2\mu_0 H$) because $\epsilon^0_{c\nu}(\bar{q})\rightarrow 0$ (and
$\epsilon^0_{s\nu}(\bar{q})\rightarrow 0$) as $\bar{q}\rightarrow
\pm {\bar{q}}_{c\nu}=\pm q^0_{c\nu}$ (and $\bar{q}\rightarrow \pm
{\bar{q}}_{s\nu}=\pm q^0_{s\nu}$). It follows that in that limit
the $\nu$ $-1/2$ holons ($\alpha =c$) or $-1/2$ spinons ($\alpha
=s$) contained in the composite $\alpha\nu$ pseudofermion acquire
the same localized character as the $-1/2$ Yang holons ($\alpha
=c$) or $-1/2$ HL spinons ($\alpha =s$) \cite{II}, respectively.

%%%%%%%%%%%%%%%%%%%%%%%%%%%%%%%%%%%%%%%%%%%%%%%%%%%%%%%%%%%%%%%%
\subsection{TRANSFORMATION LAWS OF {\it FERMI-POINT} PSEUDOPARTICLES}

Let us next consider excitations corresponding to J-CPHS ground
states whose {\it Fermi points} are defined in Eqs.
(\ref{HSqiFan})-(\ref{DqiFan}). Our goal is to find out how the
{\it Fermi bare momentum} $q_{Fc\nu,\,\iota}$ given in these
equations transforms under the pseudoparticle - pseudofermion
transformation (\ref{sumint0}). In this case, such a
transformation can formally be written as,

\begin{equation}
q_{F\alpha\nu,\,\iota} =  \iota\,q^0_{F\alpha\nu} + \Delta
q_{F\alpha\nu,\,\iota} \, \rightarrow \,
{\bar{q}}_{F\alpha\nu,\,\iota} = \iota\,q^0_{F\alpha\nu} + \Delta
{\bar{q}}_{F\alpha\nu,\,\iota} \, , \label{qFtrans}
\end{equation}
where according to Eq. (\ref{DqiFan}), $\Delta
q_{F\alpha\nu,\,\iota} = \iota\,[2\pi/ L]\,\Delta
N_{\alpha\nu,\,\iota}$. In Appendix C, we find that the deviation
$\Delta {\bar{q}}_{F\alpha\nu,\,\iota}$ on the right-hand side of
Eq. (\ref{qFtrans}) is given by,

\begin{equation}
\Delta {\bar{q}}_{F\alpha\nu,\,\iota} = \iota\,{2\pi\over
L}\,\Delta N^f_{\alpha\nu,\,\iota} \, ; \hspace{1cm} \Delta
N^f_{\alpha\nu,\,\iota} = {\Delta N^f_{\alpha\nu}\over 2} +
\iota\,\Delta J^f_{\alpha\nu} \, . \label{DqFpfF}
\end{equation}
Here

\begin{equation}
{\Delta N^f_{\alpha\nu}\over 2} =
\sum_{\alpha'=c,s}\,\sum_{\nu'=\delta_{\alpha',s}}^{\infty}\,
{\xi}^0_{\alpha\nu,\,\alpha'\nu'}\,{\Delta N_{\alpha'\nu'}\over 2}
\, ; \hspace{1cm} \Delta J^f_{\alpha\nu} =
\sum_{\alpha'=c,s}\,\sum_{\nu'=\delta_{\alpha',s}}^{\infty}\,
{\xi}^1_{\alpha\nu,\,\alpha'\nu'}\,\Delta J_{\alpha'\nu'} \, ,
\label{NfJf}
\end{equation}
and

\begin{equation}
\xi^j_{\alpha\nu,\,\alpha'\nu'} =
\delta_{\alpha,\,\alpha'}\,\delta_{\nu,\,\nu'} + \sum_{\iota'=\pm
1}(\iota')^j\,\Phi_{\alpha\nu,\,\alpha'\nu'}^f
(q^0_{F\alpha\nu},\iota'\,q^0_{F\alpha'\nu'}) \, ; \hspace{0.5cm}
j = 0,\,1 \, , \label{xi}
\end{equation}
where $\Phi_{\alpha\nu,\,\alpha'\nu'}^f (\bar{q},\bar{q}')$ is the
two-pseudofermion phase shift given in Eq. (\ref{Phi-barPhipf}).

The pseudoparticle - pseudofermion transformation (\ref{sumint0}),
refers to a subspace spanned by states of finite energy $\omega$.
Let us show that in the specific limit of small energy
$\omega\rightarrow 0$ that unitary transformation is directly
related to the primary-field conformal dimensions of the 1D
Hubbard model two-component conformal field theory
\cite{Frahm,Carmelo91,Carmelo92}. We consider the particular case
of excited states with finite occupancies in the $c0$ and $s1$
pseudoparticle bands only. For electronic densities and spin
densities such that $0<n<1/a$ and $0<m<n$, respectively, this
corresponds to the low-energy Hilbert subspace considered in the
studies of Refs. \cite{Carmelo91,Carmelo92}. For simplicity, in
this specific case we use the notation of these references such
that $c\equiv c,0$ and $s\equiv s,1$. The indices $\alpha\nu$ of
all quantities are replaced by the index $\alpha$ such that
$\alpha =c,\,s$. In the present low-energy Hilbert subspace the
above general expressions simplify to,

\begin{equation}
\iota\,q^0_{F\alpha} + \iota\,{2\pi\over L}\Delta
N_{\alpha,\,\iota}\rightarrow \iota\,q^0_{F\alpha} +
\iota\,{2\pi\over L}\Delta N^f_{\alpha,\,\iota} \, ,
\label{qFtransGSSS}
\end{equation}
where

\begin{equation}
\Delta N_{\alpha,\,\iota} = {\Delta N_{\alpha}\over 2} +
\iota\,\Delta J_{\alpha} \, ; \hspace{1cm} \Delta
N^f_{\alpha,\,\iota} = {\Delta N^f_{\alpha}\over 2} +
\iota\,\Delta J^f_{\alpha} \, , \label{DNDNfGSSS}
\end{equation}
and

\begin{equation}
{\Delta N^f_{\alpha}\over 2} = \sum_{\alpha'=c,s}\,
{\xi}^0_{\alpha,\,\alpha'}\,{\Delta N_{\alpha'}\over 2} \, ;
\hspace{1cm} \Delta J^f_{\alpha} = \sum_{\alpha'=c,s}\,
{\xi}^1_{\alpha,\,\alpha'}\,\Delta J_{\alpha'} \,
.\label{DNfDJfGSSS}
\end{equation}
Here

\begin{equation}
\xi^j_{\alpha,\,\alpha'}= \delta_{\alpha,\,\alpha'} +
\sum_{\iota'=\pm 1}(\iota')^j\,\Phi_{\alpha,\,\alpha'}^f
(q^0_{F\alpha},\iota'\,q^0_{F\alpha'}) \, . \label{xiGSSS}
\end{equation}

Let us combine Eqs. (\ref{DNDNfGSSS}) and (\ref{DNfDJfGSSS}) and
rewrite the number $\Delta N^f_{\alpha,\,\iota}$ as,

\begin{equation}
\Delta N^f_{\alpha,\,\iota} = \Bigl[\sum_{\alpha'=c,s}\,
{\xi}^0_{\alpha,\,\alpha'}\,{\Delta N_{\alpha'}\over 2} +
\iota\,\sum_{\alpha'=c,s}\, {\xi}^1_{\alpha,\,\alpha'}\,\Delta
J_{\alpha'}\Bigr] \, . \label{Gcd}
\end{equation}

Based on the form (\ref{xiGSSS}) of the parameters
$\xi^j_{\alpha,\,\alpha'}$ one can by manipulation of the integral
equations given in Appendix B, define these parameters in terms of
related integral equations. Importantly, comparison of these
latter equations and associated expressions (\ref{xiGSSS}) and
(\ref{Gcd}) with the results of Refs. \cite{Carmelo91,Carmelo92}
reveals that the number $\Delta N^f_{\alpha,\,\iota}$ given in Eq.
(\ref{Gcd}) is such that,

\begin{equation}
[\Delta N^f_{\alpha,\,\iota}]^2 = 2\,\Delta^{\iota}_{\alpha} =
\Bigl[\sum_{\alpha'=c,s}\, \xi^0_{\alpha,\,\alpha'}\,{\Delta
N_{\alpha'}\over 2} + \iota\,\sum_{\alpha'=c,s}\,
\xi^1_{\alpha,\,\alpha'}\,\Delta J_{\alpha'}\Bigr]^2 \, ,
\label{cd}
\end{equation}
where the quantity $\Delta^{\iota}_{\alpha}$ is the $\alpha$
conformal dimension of the primary fields \cite{Belavin,Frahm}
associated with the $\alpha =c$ and $\alpha =s$ excitation
branches. Moreover, the parameters $\xi^1_{\alpha,\,\alpha'}$ and
$\xi^0_{\alpha,\,\alpha'}$ are entries of the transpose of the
dressed charge matrix and of the inverse of the transpose of the
dressed charge matrix, respectively, of the low-energy
two-component 1D Hubbard model conformal field theory
\cite{Frahm,Carmelo91,Carmelo92}.

These results confirm that for the low-energy subspace defined
above and for electronic densities and spin densities such that
$0<n<1/a$ and $0<m<n$, respectively, the conformal invariance of
the 1D Hubbard model is directly related to the pseudoparticle -
pseudofermion unitary transformation (\ref{sumint0}). For these
values of the electronic and spin densities the low-energy excited
states are described by pseudoparticle bare-momentum occupancy
configurations such that there are both pseudoparticle and
pseudoparticle holes in the $c\equiv c,0$ and $s\equiv s,1$
pseudoparticle bands. However, the  pseudoparticle - pseudofermion
unitary transformation is more general and refers to all values of
energy associated with few-electron excitations.

The above results provide interesting information about the
transformation laws of $c0$ pseudoparticles (and $s1$
pseudoparticles) of bare momentum values at the {\it Fermi
points}, $\,q=\iota\,2k_F+\iota\,{2\pi\over L}\Delta
N_{c0,\,\iota}$ (and $q=\iota\,k_{F\downarrow}+\iota\,{2\pi\over
L}\Delta N_{s1,\,\iota}$) where $\iota =\pm 1$. Under the
pseudoparticle - pseudofermion transformation (\ref{sumint0}),
these pseudoparticles are mapped onto $c0$ pseudofermions (and
$s1$ pseudofermions) with momentum values at the {\it Fermi
points} $\,\bar{q}=\iota\,2k_F+\iota\,{2\pi\over
L}\sqrt{2\,\Delta^{\iota}_{c}}$ (and
$\bar{q}=\iota\,k_{F\downarrow}+\iota\,{2\pi\over
L}\sqrt{2\,\Delta^{\iota}_{s}}$) where $\Delta^{\iota}_{c}$ (and
$\Delta^{\iota}_{s}$) is the conformal dimension of the $c0$ (and
$s$) excitation branch primary field
\cite{Frahm,Carmelo91,Carmelo92}. We then conclude that these
conformal dimensions are such that the quantities
$\sqrt{2\,\Delta^{\iota}_{\alpha}}$ are nothing but deviations in
the values of $c0$ ($\alpha =c$) and $s1$ ($\alpha =s$)
pseudofermion momentum {\it Fermi points} resulting from
low-energy excitations. (We note that the positive and negative
value of the root $\sqrt{2\,\Delta^{\iota}_{\alpha}}$ refers to
pseudofermion creation and annihilation, respectively.)
Conformal-field theory can be used to evaluate expressions for
low-energy few-electron spectral functions
\cite{Frahm,Carmelo97pp}. We recall that our general pseudofermion
description provides the momentum deviation values resulting from
finite-energy excitations for all values of pseudofermion momentum
and for all pseudofermion branches. Then it is not unexpected that
the general pseudofermion description introduced in this paper can
be used in the evaluation of few-electron spectral function
expressions for all values of energy \cite{spectral0,V,spectral}.

%%%%%%%%%%%%%%%%%%%%%%%%%%%%%%%%%%%%%%%%%%%%%%%%%%%%%%%%%%%%%%%%
\subsection{INVARIANCE UNDER THE PSEUDOPARTICLE - PSEUDOFERMION
UNITARY TRANSFORMATION}

According to the results of Appendix A, there are no $c\nu$
pseudoparticles and $s\nu$ pseudoparticles belonging to $\nu>0$
and $\nu>1$ branches, respectively, in the ground state. This is
consistent with the {\it Fermi values} of Eq. (\ref{q0Fcs}), such
that $q_{F\alpha\nu}^0=0$ for these $\alpha\nu$ pseudoparticle
branches. In Appendix C, it is shown that the momentum functional
$Q_{\alpha\nu} (0)/L$ given in Eq. (\ref{Qcan1j}) vanishes at
$q_{F\alpha\nu}^0=0$ for these $\alpha\nu$ pseudoparticle
branches,

\begin{equation}
{Q_{\alpha\nu} (0)\over L} =
\sum_{\alpha'=c,s}\,\sum_{\nu'=1-\delta_{\alpha',\,c}}^{\infty}\,
\int_{-q_{\alpha'\nu'}}^{q_{\alpha'\nu'}} dq\,
\Phi_{\alpha\nu,\,\alpha'\nu'}(0,q ') \, \Delta N_{\alpha'\nu'}(q)
= 0\, , \label{QanFq=0}
\end{equation}
provided that the excited states are partial J-CPHS ground states
and the bare-momentum distribution function deviations on the
right-hand side of Eq. (\ref{QanFq=0}) are for the $c0$ and $s1$
pseudoparticle branches such that,

\begin{equation}
\Delta N_{c0} (q) = \Delta N_{c0} (-q) \, ; \hspace{0.5cm} \Delta
N_{s1} (q) = \Delta N_{s1} (-q) \, . \label{DNcDNs}
\end{equation}
Such excited states have vanishing momentum. We recall that
partial J-CPHS ground states are states that have J-CPHS
ground-state occupancy configurations for all the $\alpha\nu$
pseudoparticle branches other than the $c0$ and $s1$
pseudoparticle branches. It follows that for the former branches
the vanishing pseudoparticle {\it Fermi points},

\begin{equation}
q_{F\alpha\nu,\,\iota}=\iota\, q_{F\alpha\nu}^0 + \Delta
q_{F\alpha\nu,\,\iota} = 0 + \iota\,{2\pi\over L}\,\Delta
N_{\alpha\nu,\,\iota} \, , \label{qaniCPHSGS}
\end{equation}
remain invariant under the pseudoparticle - pseudofermion
transformation (\ref{sumint0}), or

\begin{equation}
q_{F\alpha\nu,\,\iota} \rightarrow {\bar{q}}_{F\alpha\nu,\,\iota}
= q_{F\alpha\nu,\,\iota} \, . \label{qaniCPHSGStrans}
\end{equation}

This means that a $\alpha\nu$ pseudoparticle belonging to a branch
other than the $c0$ and $s1$ branches, carrying bare momentum
$q\rightarrow 0$, and involved in excitation processes associated
with zero-momentum excited states whose $c0$ and $s1$
bare-momentum distribution function deviations obey Eq.
(\ref{DNcDNs}) remains invariant under the pseudoparticle -
pseudofermion transformation (\ref{sumint0}). It follows that in
this limiting case the $q\rightarrow 0$ $\alpha\nu$ pseudoparticle
is the same quantum object as the corresponding
$\bar{q}\rightarrow 0$ $\alpha\nu$ pseudofermion and then,

\begin{equation}
q \,\, \rightarrow \,\, \bar{q}=q \, ; \hspace{0.5cm} {\rm as}
\hspace{0.5cm} q\rightarrow 0 \, . \label{q00}
\end{equation}

This invariance occurring as $q\rightarrow 0$ for $\alpha\nu$
pseudoparticles belonging to branches such that $\alpha\nu\neq c0$
and $\alpha\nu\neq s1$, is associated with the free,
non-interacting, and delocalized character of these quantum
objects. Indeed, in this limit the residual interactions vanish
and these pseudoparticles are free, non interacting, and
delocalized quantum objects provided that the excitation processes
they are involved in correspond to excited states of zero
momentum. On the other hand, according to the discussions
presented in Sec. III, as $q$ approaches the limiting
bare-momentum values (\ref{q0Dqag}), $q\rightarrow\pm
q_{\alpha\nu}$, the $\alpha\nu$ pseudoparticles belonging to
branches other than the $c0$ pseudoparticle branch become
non-interacting and localized. Such a $q\rightarrow\pm
q_{\alpha\nu}$ non-interacting and localized behavior is
associated with another invariance: As $q\rightarrow\pm
q_{\alpha\nu}$ these $\alpha\nu$ pseudoparticles become invariant
under the electron - rotated-electron unitary transformation
\cite{I,II}. Thus, as the bare momentum $q$ approaches the
limiting values $\pm q_{\alpha\nu}$ the $\alpha\nu$ pseudoparticle
and the rotated $\alpha\nu$ pseudoparticle become the same quantum
object. In contrast, according to Eq. (\ref{pfconqan}) in that
limit, these $\alpha\nu$ pseudoparticles are not invariant under
the pseudoparticle - pseudofermion unitary transformation. Their
non-interacting and localized behavior is achieved by a
cancellation of the term $\iota\,\Delta q_{\alpha\nu}$ appearing
in the limiting bare-momentum expression on the right-hand side of
Eq. (\ref{q0Dqag}) by the momentum functional $Q_{\alpha\nu}
(\iota\,q_{\alpha\nu})/L$ of Eq. (\ref{Qcan1ql}). Moreover, for
other bare momentum values $q$ such that $q\neq 0$ and $q\neq \pm
q_{\alpha\nu}$ the $\alpha\nu$ pseudoparticle is different both
from the $\alpha\nu$ pseudofermion and rotated $\alpha\nu$
pseudoparticle.

We thus conclude that as $q\rightarrow 0$ a $\alpha\nu$
pseudoparticle involved in deviations associated with
zero-momentum excited states becomes free, non-interacting,
delocalized, and the same quantum object as the corresponding
$\alpha\nu$ pseudofermion. As the limiting values are approached
and thus $q\rightarrow \pm q_{\alpha\nu}$, the $\alpha\nu$
pseudoparticle becomes non-interacting, localized, and the same
quantum object as the corresponding rotated $\alpha\nu$
pseudoparticle. In this latter case, the energy of this composite
object becomes additive in the energy of its $\nu$ $-1/2$ holons
($\alpha =c$) or $\nu$ $-1/2$ spinons ($\alpha =s$). Note that one
of the effects of increasing the bare-momentum absolute value from
$\vert\,q\vert =0$ to $\vert\,q\vert = q_{\alpha\nu}$ is to
enhance the localization character of the $\alpha\nu$
pseudoparticles. If a pseudoparticle has bare-momentum $q$ and
$q+\Delta q$ for an initial ground state and an excited state,
respectively, then the limiting free, non-interacting, and
delocalized character and non-interacting and localized character
are achieved when $Q_{\alpha\nu} (q)/L = 0$ and $Q_{\alpha\nu}
(q)/L = -\Delta q$, respectively, where $Q_{\alpha\nu} (q)/L$ is
the momentum functional (\ref{Qcan1j}). For other values of that
momentum functional the pseudoparticle is interacting. The
pseudoparticles are never invariant under both the pseudoparticle
- pseudofermion and pseudoparticle - rotated-pseudoparticle
unitary transformations. This is because invariance under the
pseudoparticle - pseudofermion unitary transformation is
associated with a free, non-interacting, and delocalized
character, whereas invariance under the pseudoparticle -
rotated-pseudoparticle unitary transformation corresponds to a
non-interacting and localized character. Therefore the
impossibility of simultaneous invariance under these two
transformations is related to Heisenberg's uncertainty relation.

Another interesting example of invariance under the pseudoparticle
- pseudofermion transformation (\ref{sumint0}) concerns the
bare-momentum difference $(q-q')$ when $q$ and $q'$ differ by a
finite multiple of $2\pi/L$. This case is of physical importance
because $(q-q')$ corresponds to the momentum of a pseudoparticle -
pseudoparticle hole excitation in the $c0$ or $s1$ band. This type
of process is associated with the excited states B and plays a key
role in the few-electron spectral-weight distribution \cite{V}. In
this case, we find in Appendix C that in the thermodynamic limit,
$(q-q')$ is indeed invariant under the pseudoparticle -
pseudofermion transformation and thus,

\begin{equation}
(q-q') \rightarrow ({\bar{q}}-{\bar{q}}') = (q-q') \, .
\label{phqq'}
\end{equation}

%%%%%%%%%%%%%%%%%%%%%%%%%%%%%%%%%%%%%%%%%%%%%%%%%%%%%%%%%%%%%%%%
\section{CONCLUDING REMARKS}

In this paper we introduced a pseudofermion operational
description for the 1D Hubbard model. We found that in the
thermodynamic limit the wave function of excited states belonging
to the Hilbert subspace of the ground-state normal-ordered 1D
Hubbard model associated with few-electron excitations factorizes
for all values of $U/t$. This factorization results from the
non-interacting character of the pseudofermions whose occupancy
configurations describe these excited states. While the
pseudoparticle description studied in Refs. \cite{I,II,IIIb}
corresponds to the whole Hilbert space, the pseudofermions are
defined in the Hilbert subspace associated with few-electron
excitations. In such a subspace, the pseudofermions are related to
the pseudoparticles by a unitary transformation. We classified the
statistics of the latter quantum objects according to a
generalized Pauli principle \cite{Haldane91}.

Our study included the introduction of the pseudoparticle -
pseudofermion unitary transformation and of an operator algebra
for both the pseudoparticles and pseudofermions. Such a
transformation removes the residual interactions of the
$\alpha\nu$ pseudoparticles, which are mapped onto the
non-interacting $\alpha\nu$ pseudofermions. The $\alpha\nu$
pseudoparticle residual interactions are cancelled by the momentum
functional $Q_{\alpha,\,\nu}(q)/L$ of Eq. (\ref{Qcan1j}). The
information recorded in the pseudoparticle interactions is
transferred over to the momentum two-pseudofermion phase shifts of
that functional. These phase shifts control the few-electron
spectral properties through the same unconventional momentum
functional.

We introduced creation and annihilation operators for both the
pseudoparticles and pseudofermions and derived the anticommutation
relations of these pseudofermion operators. We also studied and
discussed the transformation laws of the pseudoparticles under the
pseudoparticle - pseudofermion transformation. This study included
the discussion of the physics behind both these transformation
laws and the invariance under the pseudoparticle - pseudofermion
and pseudoparticle - rotated-pseudoparticle unitary
transformations of the pseudoparticles for some specific
bare-momentum values. Invariance under the former (and the latter)
transformation is associated with a free, non-interacting, and
delocalized character (and non-interacting and localized
character) for these quantum objects. Thus as a result of
Heisenberg's uncertainty relation, these two invariances never
occur simultaneously. We also find that under the pseudoparticle -
pseudofermion unitary transformation, the $c0$ and $s1$
pseudoparticles of bare momentum at the {\it Fermi points} are
mapped onto corresponding pseudofermions whose momentum {\it Fermi
points} expressions are related to the conformal dimensions of the
two-component conformal-field theory primary fields
\cite{Belavin,Frahm}. It follows that in the limit of low energy
the conformal invariance of the 1D Hubbard model is related to the
general pseudoparticle - pseudofermion unitary transformation.

The pseudofermion algebra introduced in this paper is used
elsewhere in the evaluation of finite-energy few-electron
spectral-function expressions \cite{V,spectral}. Fortunately, as a
consequence of the above wave function factorization, the
few-electron spectral functions can be written as a convolution of
pseudofermion, $-1/2$ Yang holon, and/or $-1/2$ HL spinon spectral
functions for {\it all} values of energy and on-site repulsion $U$
\cite{V}.

%%%%%%%%%%%%%%%%%%%%%%%%%%%%%%%%%%%%%%%%%%%%%%%%%%%%%%%%%%%%%%%%%%%%%%%%%%
\begin{acknowledgments}
I thank Karlo Penc for many useful and illuminating discussions
concerning the issues studied in this paper. I also thank Daniel
Bozi, Ralph Claessen, Francisco (Paco) Guinea, Katrina Hibberd,
Lu\'{\i}s Miguel Martelo, Pedro Sacramento, and Jo\~ao Lopes dos
Santos for stimulating discussions.
\end{acknowledgments}
%%%%%%%%%%%%%%%%%%%%%%%%%%%%%%%%%%%%%%%%%%%%%%%%%%%%%%%%%%%%%%%%%%%%%%%%%%
\appendix

%%%%%%%%%%%%%%%%%%%%%%%%%%%%%%%%%%%%%%%%%%%%%%%%%%%%%%%%%%%%%%%%
\section{PSEUDOPARTICLE REPRESENTATION, EFFECTIVE PSEUDOPARTICLE LATTICE, AND
GROUND-STATE QUANTITIES}

In this Appendix, we summarize some aspects of the pseudoparticle,
description which are needed for the studies of this paper. This
includes a brief summary of the basic pseudoparticle properties,
introduction to the pseudoparticle bare momentum \cite{I}, local
pseudoparticle concept, and effective lattices \cite{IIIb}. In
addition, we provide the ground-state pseudoparticle bare-momentum
distribution functions, as well as other ground-state quantities.

According to the results of Ref. \cite{IIIb}, the $\alpha\nu$
pseudoparticle bare-momentum description obtained naturally from
the Bethe-ansatz solution in Refs.
\cite{I,Carmelo91,Carmelo92,Carmelo97} is related by Fourier
transform to a local $\alpha\nu$ pseudoparticle description in
terms of spatial coordinates of an effective $\alpha\nu$ lattice.
These concepts are needed and useful for both the operational
pseudoparticle and pseudofermion representations introduced in
this paper.

The $\alpha\nu$ pseudoparticles carry bare momentum $q$. This is
the continuum bare momentum associated with the discrete
bare-momentum values $q_j$ such that,

\begin{equation}
q_{j+1}-q_j={2\pi\over L} \, . \label{differ0}
\end{equation}
These discrete values read \cite{I},

\begin{equation}
q_j = {2\pi\over L}\, I_j^{\alpha\nu} \,  , \label{qj}
\end{equation}
where the $I_j^{\alpha\nu}$ numbers are integers or half-odd
integers \cite{I,II}. The index number $j$ can have the values
$j=1,2,...,N^*_{\alpha\nu}$. Here the number $N^*_{\alpha\nu}$
equals the number of discrete bare-momentum values in the
$\alpha\nu$ pseudoparticle band and is such that,

\begin{equation}
N^*_{\alpha\nu}=N_{\alpha\nu}+ N^h_{\alpha\nu} \, , \label{N*}
\end{equation}
where $N_{\alpha\nu}$ is the number of $\alpha\nu$ pseudoparticles
and $N^h_{\alpha\nu}$ is the number of $\alpha\nu$ pseudoparticle
holes. The latter number is given by,

\begin{equation}
N^h_{c0} = N_a - N_{c0} \, ; \hspace{1cm} N^h_{\alpha\nu} =
2\,S_{\alpha} + 2\sum_{\nu'=\nu +1}^{\infty} (\nu' -\nu)
N_{\alpha\nu'} = L_{\alpha} + 2\sum_{\nu'=\nu +1}^{\infty} (\nu'
-\nu) N_{\alpha\nu'} \, ; \hspace{0.5cm} \alpha =c,s \, ,
\hspace{0.5cm} \nu >0 \, , \label{Nhag}
\end{equation}
where $S_c$ and $S_s$ are the values of $\eta$-spin and spin,
respectively, which read

\begin{equation}
S_c = {1\over 2}[N_a -N_{c0}] - \sum_{\nu
=1}^{\infty}\nu\,N_{c\nu} \, ; \hspace{1cm} S_s = {1\over 2}N_{c0}
- \sum_{\nu =1}^{\infty}\nu\,N_{s\nu} \, . \label{ScsNnu}
\end{equation}
Combination of Eqs. (\ref{Nhag}) and (\ref{ScsNnu}) leads to,

\begin{equation}
N^h_{c\nu} = N_a -N_{c0} - \sum_{\nu'=1}^{\infty} \Bigl(\nu + \nu'
- \vert\nu - \nu'\vert\Bigl) N_{c\nu'} \, ; \hspace{0.75cm}
N^h_{s\nu} = N_{c0} - \sum_{\nu'=1}^{\infty} \Bigl(\nu + \nu' -
\vert\nu - \nu'\vert\Bigl) N_{s\nu'} \, ; \hspace{0.5cm} \nu
>0 \, . \label{Nhcsn}
\end{equation}
For $\alpha\nu\neq c0$ (and $\alpha\nu =c0$) the numbers
$I_j^{\alpha\nu}$ on the right-hand side of Eq. (\ref{qj}) are
integers (half-odd integers), if $ N^*_{\alpha\nu}$ (and $N_a/2 -
\sum_{\alpha =c,s}\sum_{\nu = 1}^{\infty}N_{\alpha\nu}$) is odd
(even). This non-perturbative shake-up effect is related to the so
called orthogonality catastrophe \cite{Anderson}. It reveals that
the values of the available bare-momentum values $q_j$ of each
$\alpha\nu$ pseudoparticle band might be different for different
CPHS ensemble subspaces.

The holons, spinons, and pseudoparticles are related to the
electrons through the rotated electrons, as discussed in Sec. III
and in Ref. \cite{I,IIIb}. The unitary transformation which maps
electrons onto rotated electrons is defined in Sec. II. Since the
momentum operator is invariant under such a transformation, also
the lattice occupied by rotated electrons has the same lattice
constant $a$ and length $L$ as the original electronic lattice. In
Ref. \cite{IIIb} it was shown that there is an {\it effective
$\alpha\nu$ pseudoparticle lattice} for each $\alpha\nu$
pseudoparticle branch.

In contrast to the case of a non-interacting band, the value of
the number of available discrete bare-momentum values
$N^*_{\alpha\nu}$ defined in Eqs. (\ref{N*}), (\ref{Nhag}), and
(\ref{Nhcsn}) might be different for different CPHS ensemble
subspaces. This value was identified in Ref. \cite{IIIb} with the
number of sites of the corresponding effective $\alpha\nu$
pseudoparticle lattice. This lattice has the same length
$L=N^*_{\alpha\nu}\,a_{\alpha\nu}$ as the original real-space
lattice, where

\begin{equation}
a_{\alpha\nu} = a\,{N_a\over N^*_{\alpha\nu}} =  {L\over
N^*_{\alpha\nu}} \, , \label{aan}
\end{equation}
is the effective $\alpha\nu$ lattice constant. Since
$N^*_{\alpha\nu}$ is different for different CPHS ensemble
subspaces, it follows from Eq. (\ref{aan}) that the value of the
corresponding effective $\alpha\nu$ lattice constant
$a_{\alpha\nu}$ also changes. The spatial coordinates of the
effective $\alpha\nu$ pseudoparticle lattice are $x_j
=a_{\alpha\nu}\,j$ where $j=1,2,3,...,N^*_{\alpha\nu}$. The value
of $N^*_{\alpha\nu}$ defined in Eqs. (\ref{N*})-(\ref{Nhcsn})
determines the corresponding limiting bare-momentum values of the
$\alpha\nu$ pseudoparticle {\it Brillouin zone}.

In the case of the $c0$ pseudoparticle band, the discrete
bare-momentum values $q_j$ belong to the following range,

\begin{equation}
q_{c0}^{-} \leq q_j \leq q_{c0}^{+} \, , \label{rangeqjc}
\end{equation}
Here,

\begin{equation}
q_{c0}^{+} = - q_{c0}^{-}= {\pi\over a}\bigl[1-{1\over N_a}\bigr]
\, , \label{qcev}
\end{equation}
for ${\tilde{N}}_{c0}$ even and

\begin{equation}
q_{c0}^{+} = {\pi\over a} \, ; \hspace{1cm} q_{c0}^{-} = -
{\pi\over a} \bigl[1-{2\over N_a}\bigr] \, , \label{qcodd}
\end{equation}
for ${\tilde{N}}_{c0}$ odd. Note that since $N^*_{c0}=N_a$, the
lattice constant of the effective $c0$ lattice equals the lattice
constant $a$ of the original electronic lattice, that is
$a_{c0}=a$. Thus, for the $c0$ band the lattice constant $a_{c0}$
has the same value for all CPHS ensemble subspaces.

For the $\alpha\nu$ pseudoparticle bands such that $\nu>0$, the
discrete bare-momentum values $q_j$ are distributed symmetrically
relative to zero, and are such that $\vert q_j \vert \leq
q_{\alpha\nu}$. The two band momenta $\pm q_{\alpha\nu}$ are the
limiting bare-momentum values associated with the limits of the
$\alpha\nu$ pseudoparticle {\it Brillouin zone}. The limiting
bare-momentum $q_{\alpha\nu}$ reads,

\begin{equation}
q_{\alpha\nu} = {\pi\over a_{\alpha\nu}}[1-{1\over
N^*_{\alpha\nu}}] \, , \label{qag}
\end{equation}
where the lattice constant $a_{\alpha\nu}\geq a$ is given in Eq.
(\ref{aan}).

We close this Appendix by providing some useful ground-state
quantities. Following the results of Refs. \cite{II,Carmelo97},
the ground-state pseudoparticle bare-momentum distribution
functions and $-1/2$ Yang holon and $-1/2$ HL spinon numbers read,

\begin{equation}
N^{0}_{c0} (q) = \Theta\Bigl(q_{Fc0}^{+} - q \Bigl) \, ,
\hspace{0.5cm} 0\leq q\leq q_{c0}^{+} \, ; \hspace{1cm} N^{0}_{c0}
(q) = \Theta\Bigl(q - q_{Fc0}^{-}\Bigl) \, , \hspace{0.5cm}
q_{c0}^{-}\leq q\leq 0 \, ; \label{Nc0}
\end{equation}

\begin{equation}
N^{0}_{s1} (q) = \Theta\Bigl(q_{Fs1} - q \Bigl) \, ,
\hspace{0.5cm} 0\leq q\leq q_{s1} \, ; \hspace{1cm} N^{0}_{s1} (q)
= \Theta\Bigl(q + q_{Fs1}\Bigl) \, , \hspace{0.5cm} -q_{s1}\leq
q\leq 0 \, ; \label{Ns10}
\end{equation}

\begin{equation}
N^{0}_{\alpha\nu}(q) = L^0_{\alpha,\,-1/2} = 0 \, ; \hspace{0.5cm}
-q_{\alpha\nu}\leq q\leq q_{\alpha\nu} \, ; \hspace{0.5cm} \alpha
= c,\,s  \, ; \hspace{0.5cm} \nu \geq 1 + \delta_{\alpha ,\,s} \,
. \label{Ncnsn0}
\end{equation}
Except for $1/L$ corrections, the {\it Fermi values} on the
right-hand side of Eq. (\ref{Nc0}) are given by $q^{\pm}_{Fc0} =
\pm q^0_{Fc0}$, where the ground-state values $q^0_{Fs1}$ and
$q^0_{Fc0}$ are given in Eq. (\ref{q0Fcs}). Here we have ignored
terms of order of $1/L$ and used the expressions $k_{F\sigma}=\pi
n_{\sigma}$ and $2k_F=\pi n$. The expressions of the {\it Fermi
values} including $1/L$ contributions are provided in Appendix C
of Ref. \cite{I}.

For the ground-state CPHS ensemble subspace the expression of the
number $N^*_{\alpha\nu}$ given in Eqs. (\ref{N*}), (\ref{Nhag}),
and (\ref{Nhcsn}) simplifies. Let us denote by
$N^{0,*}_{\alpha\nu}$ the ground-state value of the numbers
$N^*_{\alpha\nu}$. These numbers read,

\begin{equation}
N^{0,*}_{c\nu}=(N_a -N) \, ; \hspace{1cm}
N^{0,*}_{s1}=N_{\uparrow} \, ; \hspace{1cm}
N^{0,*}_{s\nu}=(N_{\uparrow} -N_{\downarrow}) \, , \hspace{0.3cm}
\nu > 1 \, , \label{N*csnu}
\end{equation}
whereas $N^{0,*}_{c0}=N^*_{c0}$ is given by $N^*_{c0}=N_a$ for the
whole Hilbert space.

In addition to the $-1/2$ Yang holon and $-1/2$ HL spinon numbers
already given in Eq. (\ref{Ncnsn0}), the ground state belongs to a
CPHS ensemble subspace with the following values for the
pseudoparticle, $\pm 1/2$ holon, and $\pm 1/2$ spinon numbers,

\begin{equation}
M^0_{c,\,-1/2} = 0 \, ; \hspace{0.5cm} M^0_{c,\,+1/2} =
L^0_{c,\,+1/2} = N_a - N \, ; \hspace{0.5cm} N^0_c = N \, ;
\hspace{0.5cm} N^0_{c\nu} = 0 \, , \label{NcGS}
\end{equation}
in the charge sector and

\begin{equation}
M^0_{s,\,-1/2} = N^0_{s1} = N_{\downarrow} \, ; \hspace{0.5cm}
M^0_{s,\,+1/2} = N_{\uparrow} \, ; \hspace{0.5cm} N^0_{s\nu} = 0
\, , \hspace{0.25cm} \nu\geq 2 ; \hspace{0.5cm} L^0_{s,\,+1/2} =
N_{\uparrow}-N_{\downarrow} \, , \label{NsGS}
\end{equation}
in the spin sector.

Finally, note that for $\nu>0$ the number (\ref{N*}) can be
expressed in terms of the ground-state values (\ref{N*csnu}) as
follows,

\begin{equation}
N^*_{\alpha\nu} = N^{0,*}_{\alpha\nu} + \Delta L_{\alpha} +
2\sum_{\nu'=\nu +1}^{\infty} (\nu' -\nu) N_{\alpha\nu'} \, ;
\hspace{0.5cm} \alpha =c,s \, , \hspace{0.5cm} \nu >0 \, .
\label{N*DN}
\end{equation}
Moreover, for $\alpha\nu=s1$ the pseudofermion hole number defined
in Eqs. (\ref{Nhag}) and (\ref{Nhcsn}) can be written as,

\begin{equation}
N^h_{s1} = N_{c0} - 2\sum_{\nu'=1}^{\infty} N_{s\nu'} \, .
\label{Nhs1}
\end{equation}

%%%%%%%%%%%%%%%%%%%%%%%%%%%%%%%%%%%%%%%%%%%%%%%%%%%%%%%%%%%%%%%%%%%
\section{THE TWO-PSEUDOFERMION PHASE SHIFTS $\bar{\Phi}_{\alpha\nu,\,\alpha'\nu'}
\left(r, r'\right)$}

Here we provide the set of integral equations which define the
two-pseudofermion phase shifts $\bar{\Phi
}_{\alpha\nu,\,\alpha'\nu'}\left(r,r'\right)$ on the right-hand
side of Eqs. (\ref{tilPcc})-(\ref{tilPanan}), (\ref{Phi-barPhi}),
and (\ref{Phi-barPhipf}).

Let us start by introducing the following {\it Fermi surface}
parameters,

\begin{equation}
r^0_c = {4t\,\sin Q\over U} \, ; \hspace{1cm} r^0_s = {4t\,B\over
U} \, , \label{r0cs}
\end{equation}
where the parameters $Q$ and $B$ are defined in Eq. (\ref{QB}).

In order to derive the integral equations which define the
two-pseudofermion phase shifts $\bar{\Phi
}_{\alpha\nu,\,\alpha'\nu'}\left(r,r'\right)$, we first use in
Eqs. (\ref{Tapco1})-(\ref{Tapco3}) the bare-momentum distribution
functions of the general form (\ref{N0DNq}) as well as the
rapidity functional expressions given in Eq. (\ref{FL}) and
rapidity-momentum functional expression provided in Eq.
(\ref{FL}). Such functionals are written in terms of corresponding
ground-state rapidity and rapidity-momentum functions whose
argument is the momentum functional. The expression of this
functional is given in Eq. (\ref{barqan}) with the momentum
$Q_{\alpha\nu}(q)/L$ provided in Eq. (\ref{Qcan1j}). Expansion of
the obtained equations up to first order in the bare-momentum
distribution function deviations on the right-hand side of Eqs.
(\ref{DNq}) and (\ref{Qcan1j}) leads to expression
(\ref{Phi-barPhi}) with the two-pseudofermion phase shift
$\bar{\Phi }_{\alpha\nu,\,\alpha'\nu'} (r ,\,r')$ uniquely defined
by the integral equations given below. A first group of
two-pseudofermion phase shifts obey integral equations by their
own. These equations read,

\begin{equation}
\bar{\Phi }_{s1,\,c0}\left(r,r'\right) = -{1\over{\pi}}{\rm
arc}{\rm tan}(r-r') + \int_{-r^0_s}^{r^0_s}
dr''\,G(r,r'')\,{\bar{\Phi }}_{s1,\,c0}\left(r'',r'\right) \, ,
\label{Phis1c}
\end{equation}

\begin{equation}
\bar{\Phi }_{s1,\,c\nu}\left(r,r'\right) =  -
{1\over{\pi^2}}\int_{-r^0_c}^{r^0_c} dr''{{\rm arc}{\rm tan}
\Bigl({r''-r'\over\nu}\Bigr)\over{1+(r-r'')^2}} +
\int_{-r^0_s}^{r^0_s} dr''\,G(r,r'')\,{\bar{\Phi
}}_{s1,\,c\nu}\left(r'',r'\right) \, , \label{Phis1cn}
\end{equation}
and

\begin{eqnarray}
\bar{\Phi }_{s1,\,s\nu}\left(r,r'\right) & = & {\delta_{1
,\,\nu}\over\pi}\,\arctan\Bigl({r-r'\over 2}\Bigl) + {(1-\delta_{1
,\,\nu})\over\pi}\Bigl\{ \arctan\Bigl({r-r'\over \nu-1}\Bigl) +
\arctan\Bigl({r-r'\over
\nu+1}\Bigl)\Bigr\} \nonumber \\
& - &  {1\over{\pi^2}}\int_{-r^0_c}^{r^0_c} dr''{{\rm arc}{\rm
tan} \Bigl({r''-r'\over\nu}\Bigr)\over{1+(r-r'')^2}} +
\int_{-r^0_s}^{r^0_s} dr''\,G(r,r'')\,{\bar{\Phi
}}_{s1,\,s1}\left(r'',r'\right) \, . \label{Phis1sn}
\end{eqnarray}
Here the kernel $G(r,r')$ is given by,

\begin{equation}
G(r,r') = - {1\over{2\pi}}\left[{1\over{1+((r-r')/2)^2}}\right]
\left[1 - {1\over 2}
\left(t(r)+t(r')+{{l(r)-l(r')}\over{r-r'}}\right)\right] \, ,
\label{G}
\end{equation}
where

\begin{equation}
t(r) = {1\over{\pi}}\left[{\rm arc}{\rm tan}(r + r^0_c) - {\rm
arc}{\rm tan}(r -r^0_c)\right] \, , \label{t}
\end{equation}
and

\begin{equation}
l(r) = {1\over{\pi}}\left[ \ln (1+(r + r^0_c)^2) - \ln (1+(r
-r^0_c)^2)\right] \, . \label{l}
\end{equation}

The kernel defined in Eqs. (\ref{G})-(\ref{l}) was first
introduced in Ref. \cite{Carmelo92} within the low-energy
two-component $c\equiv c0$ and $s\equiv s1$ pseudoparticle theory
studied in that reference.

A second group of two-pseudofermion phase shifts are expressed in
terms of the basic functions given in Eqs.
(\ref{Phis1c})-(\ref{Phis1sn}) as follows,

\begin{equation}
\bar{\Phi }_{c0,\,c0}\left(r,r'\right) =
{1\over{\pi}}\int_{-r^0_s}^{r^0_s} dr''{\bar{\Phi
}_{s1,\,c0}\left(r'',r'\right) \over {1+(r-r'')^2}} \, ,
\label{Phicc}
\end{equation}

\begin{equation}
\bar{\Phi }_{c0,\,c\nu}\left(r,r'\right) = -{1\over{\pi}}{\rm
arc}{\rm tan}\Bigl({r-r'\over \nu}\Bigr) +
{1\over{\pi}}\int_{-r^0_s}^{r^0_s} dr''{\bar{\Phi
}_{s1,\,c\nu}\left(r'',r'\right) \over {1+(r-r'')^2}} \, ,
\label{Phiccn}
\end{equation}
and

\begin{equation}
\bar{\Phi }_{c0,\,s\nu}\left(r,r'\right) = -{1\over{\pi}}{\rm
arc}{\rm tan}\Bigl({r-r'\over \nu}\Bigr) + {1\over{\pi}}
\int_{-r^0_s}^{r^0_s} dr''{\bar{\Phi
}_{s1,\,s\nu}\left(r'',r'\right) \over {1+(r-r'')^2}} \, .
\label{Phicsn}
\end{equation}

Finally, the remaining two-pseudofermion phase shifts can be
expressed either in terms of the functions
(\ref{Phicc})-(\ref{Phicsn}) only,

\begin{equation}
{\bar{\Phi }}_{c\nu,\,c0}\left(r,r'\right) = {1\over{\pi}}{\rm
arc}{\rm tan}\Bigl({r-r'\over {\nu}}\Bigr) -
{1\over{\pi}}\int_{-r^0_c}^{r^0_c} dr''{{\bar{\Phi
}}_{c0,\,c0}\left(r'',r'\right) \over {\nu[1+({r-r''\over
{\nu}})^2]}} \, , \label{Phicnc}
\end{equation}

\begin{equation}
\bar{\Phi }_{c\nu,\,c\nu'}\left(r,r'\right) =
{1\over{2\pi}}\Theta_{\nu,\,\nu'}(r-r') -
{1\over{\pi}}\int_{-r^0_c}^{r^0_c} dr''{\bar{\Phi
}_{c0,\,c\nu'}\left(r'',r'\right) \over
{\nu[1+({r-r''\over\nu})^2]}} \, , \label{Phicncn}
\end{equation}
and

\begin{equation}
\bar{\Phi }_{c\nu,\,s\nu'}\left(r,r'\right) = -
{1\over{\pi}}\int_{-r^0_c}^{r^0_c} dr''{\bar{\Phi
}_{c0,\,s\nu'}\left(r'',r'\right) \over
{\nu[1+({r-r''\over\nu})^2]}} \, , \label{Phicnsn}
\end{equation}
or both in terms of the basic functions
(\ref{Phis1c})-(\ref{Phis1sn}) and of the phase shifts
(\ref{Phicc})-(\ref{Phicsn}),

\begin{equation}
{\bar{\Phi }}_{s\nu ,\,c0}\left(r,r'\right) = - {{\rm arc}{\rm
tan}\Bigl({r-r'\over {\nu}}\Bigr)\over \pi} +
{1\over{\pi}}\int_{-r^0_c}^{r^0_c} dr''{{\bar{\Phi
}}_{c0,\,c0}\left(r'',r'\right) \over {\nu[1+({r-r''\over
\nu})^2]}} - \int_{-r^0_s}^{r^0_s} dr''{\bar{\Phi
}}_{s1,\,c0}\left(r'',r'\right)
{\Theta^{[1]}_{\nu,\,1}(r-r'')\over{2\pi}} \, ; \hspace{0.5cm} \nu
> 1 \, , \label{Phisnc}
\end{equation}

\begin{equation}
{\bar{\Phi }}_{s\nu ,\,c\nu'}\left(r,r'\right) =
{1\over{\pi}}\int_{-r^0_c}^{r^0_c} dr''{{\bar{\Phi
}}_{c0,\,c\nu'}\left(r'',r'\right) \over {\nu[1+({r-r''\over
\nu})^2]}} - \int_{-r^0_s}^{r^0_s} dr''{\bar{\Phi
}}_{s1,\,c\nu'}\left(r'',r'\right)
{\Theta^{[1]}_{\nu,\,1}(r-r'')\over {2\pi}} \, ; \hspace{0.5cm}
\nu > 1 \, , \label{Phisncn}
\end{equation}
and

\begin{equation}
{\bar{\Phi }}_{s\nu ,\,s\nu'}\left(r,r'\right) =
{\Theta_{\nu,\,\nu'}(r-r')\over{2\pi}} +
{1\over{\pi}}\int_{-r^0_c}^{r^0_c} dr''{{\bar{\Phi
}}_{c0,\,s\nu'}\left(r'',r'\right) \over {\nu[1+({r-r''\over
\nu})^2]}} - \int_{-r^0_s}^{r^0_s} dr''{\bar{\Phi
}}_{s1,\,s\nu'}\left(r'',r'\right)
{\Theta^{[1]}_{\nu,\,1}(r-r'')\over{2\pi}} \, ; \hspace{0.5cm} \nu
> 1 \, . \label{Phisnsn}
\end{equation}
In the above two-pseudofermion phase shift expressions the
functions $\Theta_{\nu,\,\nu'}(x)$ and
$\Theta^{[1]}_{\nu,\,\nu'}(x)$ read,

\begin{eqnarray}
\Theta_{\nu,\,\nu'}(x) & = & \delta_{\nu
,\,\nu'}\Bigl\{2\arctan\Bigl({x\over 2\nu}\Bigl) + \sum_{l=1}^{\nu
-1}4\arctan\Bigl({x\over 2l}\Bigl)\Bigr\} + (1-\delta_{\nu
,\,\nu'})\Bigl\{ 2\arctan\Bigl({x\over \vert\,\nu-\nu'\vert}\Bigl)
\nonumber \\
& + &  2\arctan\Bigl({x\over \nu+\nu'}\Bigl) +
\sum_{l=1}^{{\nu+\nu'-\vert\,\nu-\nu'\vert\over 2}
-1}4\arctan\Bigl({x\over \vert\, \nu-\nu'\vert +2l}\Bigl)\Bigr\}
\, , \label{Theta}
\end{eqnarray}
and

\begin{eqnarray}
\Theta^{[1]}_{\nu,\,\nu'}(x) & = & {d\Theta_{\nu,\,\nu'}(x)\over
dx} = \delta_{\nu ,\nu'}\Bigl\{{1\over \nu[1+({x\over 2\nu})^2]}+
\sum_{l=1}^{\nu -1}{2\over l[1+({x\over 2l})^2]}\Bigr\} +
(1-\delta_{\nu ,\nu'})\Bigl\{ {2\over |\nu-\nu'|[1+({x\over
|\nu-\nu'|})^2]} \nonumber \\
& + & {2\over (\nu+\nu')[1+({x\over \nu+\nu'})^2]} +
\sum_{l=1}^{{\nu+\nu'-|\nu-\nu'|\over 2} -1}{4\over
(|\nu-\nu'|+2l)[1+({x\over |\nu-\nu'|+2l})^2]}\Bigr\} \, ,
\label{The1}
\end{eqnarray}
respectively. Note that the latter function is the $x$ derivative
of the function defined in Eq. (\ref{Theta}).

In spite of the different notation and except for simplifications
introduced here as a result of some integrations performed
analytically, the integral equations
(\ref{tilPcc})-(\ref{tilPanan}) are equivalent to the system of
coupled integral equations (B30)-(B40) of Ref. \cite{Carmelo97}.

The bare-momentum two-pseudofermion phase shifts are defined by
Eq. (\ref{Phi-barPhi}), in terms of the above phase shifts
$\bar{\Phi }_{\alpha\nu,\,\alpha'\nu'} (r ,\,r')$ associated with
the integral equations (\ref{Phis1c})-(\ref{Phisncn}). In
applications of the pseudofermion description to the study of
few-electron spectral properties the phase shifts
$\Phi_{s1,\,s1}(q,\,q')$, $\Phi_{s1,\,c0}(q\,q')$,
$\Phi_{c0,\,c0}(q,\,q')$, $\Phi_{c0,\,s1}(q,\,q')$,
$\Phi_{c0,\,c1}(q,\,q')$, and $\Phi_{s1,\,c1}(q,\,q')$ with $q$ at
the {\it Fermi points} play a major role \cite{V}. By manipulation
of the above integral equations, we find that in the limits of
zero spin density $m\rightarrow 0$ and $U/t\rightarrow 0$, these
two-pseudofermion phase shifts with $q$ at the {\it Fermi points}
and the second bare-momentum denoted by $q$ are given by,

\begin{eqnarray}
\Phi_{s1,\,s1}(\iota\,k_F,\,q) & = & {\iota\over 2\sqrt{2}} \, ,
\hspace{0.5cm} q \neq \iota\,k_F \, ; \nonumber \\
& = & \iota\Bigl[{3\over 2\sqrt{2}}-1\Bigr] \, , \hspace{0.5cm} q
= \iota\,k_F \, , \hspace{0.5cm} \iota = \pm 1 \, , \label{Phiss}
\end{eqnarray}

\begin{eqnarray}
\Phi_{s1,\,c0}(\iota\,k_F,\,q) & = & {1\over
2\sqrt{2}}\Bigl\{-\iota\, \Theta (2k_F-\vert q\vert) + {\rm sgn}
(q)\, \Theta (\vert q\vert -2k_F)\Bigl\} \, ,
\hspace{0.5cm} q \neq\iota\, 2k_F \, ; \nonumber \\
& = & - {\iota\over 2\sqrt{2}} \, , \hspace{0.5cm} q = \iota\,
2k_F \, , \hspace{0.5cm} \iota = \pm 1 \, , \label{Phisc0}
\end{eqnarray}

\begin{eqnarray}
\Phi_{c0,\,c0}(\iota\,2k_F,\,q) & = & {1\over
2\sqrt{2}}\Bigl\{-\iota\, \Theta (2k_F-\vert q\vert) + {\rm sgn}
(q)\, \Theta (\vert q\vert -2k_F)\Bigl\} \, ,
\hspace{0.5cm} q \neq\iota\, 2k_F \, ; \nonumber \\
& = & \iota\Bigl[{3\over 2\sqrt{2}}-1\Bigr] \, , \hspace{0.5cm} q
= \iota \, 2k_F \, , \hspace{0.5cm} \iota = \pm 1 \, ,
\label{PhiccU0}
\end{eqnarray}

\begin{eqnarray}
\Phi_{c0,\,s1}(\iota\,2k_F,\,q) & = & -{\iota\over 2\sqrt{2}} \, ,
\hspace{0.5cm} q \neq \iota\, k_F \, ; \nonumber \\
& = & {{\rm sgn} (q) \over 2\sqrt{2}} \, , \hspace{0.5cm} q =
\iota\, k_F \, , \hspace{0.5cm} \iota = \pm 1 \, , \label{PhicsU0}
\end{eqnarray}

\begin{eqnarray}
\Phi_{c0,\,c1}(\iota\,2k_F,\,q) & = & {{\rm sgn} (q) \over
\sqrt{2}} \, ,
\hspace{0.5cm} q \neq 0 \, ; \nonumber \\
& = & -{\iota\over \sqrt{2}} \, , \hspace{0.5cm} q = 0 \, ,
\hspace{0.5cm} \iota = \pm 1 \, , \label{PhictU0}
\end{eqnarray}
and

\begin{equation}
\Phi_{s1,\,c1}(\iota\,k_F,\,q) = 0 \, , \hspace{0.5cm} \iota = \pm
1 \, , \label{PhistU0}
\end{equation}
respectively, where $\Theta (x) = 1$ for $x\geq 0$, and $\Theta
(x) = 0$ for $x< 0$.

For $m\rightarrow 0$ and $U/t\rightarrow\infty$ the phase-shift
expression (\ref{Phiss}) for $\Phi_{s1,\,s1}(\iota\,k_F,\,q)$
remains valid, whereas the two-pseudofermion phase shifts
$\Phi_{s1,\,c0}(q,\,q')$, $\Phi_{c0,\,c0}(q,\,q')$,
$\Phi_{s1,\,c1}(q,\,q')$, $\Phi_{c0,\,s1}(q,\,q')$, and
$\Phi_{c0,\,c1}(q,\,q')$ with $q$ at the {\it Fermi points} and
the second bare-momentum denoted by $q$ read,

\begin{equation}
\Phi_{s1,\,c0}(\iota\,k_F,\,q) = - {\iota\over 2\sqrt{2}} \, ,
\hspace{0.5cm} \iota = \pm 1 \, , \label{Phisc}
\end{equation}

\begin{equation}
\Phi_{c0,\,c0}(\iota\,2k_F,\,q) = \Phi_{s1,\,c1}(\iota\,k_F,\,q) =
0 \, , \hspace{0.5cm} \iota = \pm 1 \, , \label{PhiccstUinf}
\end{equation}

\begin{equation}
\Phi_{c0,\,s1}(\iota\,2k_F,\,q) = {q\over 4k_F} \, ,
\hspace{0.5cm} \vert\,q\vert\leq k_F \, , \hspace{0.5cm} \iota =
\pm 1 \, , \label{PhicsUinf}
\end{equation}
and

\begin{equation}
\Phi_{c0,\,c1}(\iota\,2k_F,\,q) = {q\over 2[\pi -2k_F]} \, ,
\hspace{0.5cm} \vert\,q\vert\leq \pi -2k_F \, , \hspace{0.5cm}
\iota = \pm 1 \, . \label{PhictUinf}
\end{equation}

%%%%%%%%%%%%%%%%%%%%%%%%%%%%%%%%%%%%%%%%%%%%%%%%%%%%%%%%%%%%%%%%%%%%%%%%%%
\section{TRANSFORMATION LAWS UNDER THE PSEUDOPARTICLE - PSEUDOFERMION SUBSPACE UNITARY ROTATION}

Here we study the momentum functional $Q_{\alpha\nu} (q)/L$ given
in Eq. (\ref{Qcan1j}) for specific values of $q$. In some cases we
consider general excited states obeying relations (\ref{DNqzero}),
whereas in other cases we consider particular cases of such
general excited states.

We start by confirming that the second term on the right-hand side
of Eq. (\ref{differ}) is of $[1/L]^2$ order. Let us consider a
more general situation and also confirm the validity of Eq.
(\ref{phqq'}) and show that $(q-q')$ is invariant under the
pseudoparticle - pseudofermion transformation when
$q=q'+N_{ph}[2\pi/L]$ and $N_{ph}=\pm 1,\,\pm 2,...$ is a finite
integer number. Equation (\ref{differ}) corresponds to the
particular case when $N_{ph}=1$. Thus we want to show that the
quantity $[Q_{\alpha\nu} (q+N_{ph}{2\pi\over L})-Q_{\alpha\nu}
(q)]/L$ involving the functional (\ref{Qcan1j}) is of $(1/L)^2$
order. By expressing this quantity in terms of the derivative
$\partial Q_{\alpha\nu} (q)/\partial q$ one finds,

\begin{equation}
{Q_{\alpha\nu} (q+N_{ph}{2\pi\over L})-Q_{\alpha\nu} (q)\over L} =
{2\pi\over L^2}\,N_{ph}\,{\partial Q_{\alpha\nu} (q)\over\partial
q} \, . \label{DQexpansion}
\end{equation}
Analysis of the form of the derivative $\partial Q_{\alpha\nu}
(q)/\partial q$ reveals that it is of $[1/L]^0$ order. This is the
confirmation that the momentum contribution (\ref{DQexpansion}) is
of second order in $1/L$. Thus, since within the present large-$L$
pseudofermion description $[1/L]^2$ bare-momentum and momentum
contributions vanish, $(q-q')$ is indeed invariant under the
pseudoparticle - pseudofermion transformation.

The remaining of this Appendix is complementary to the studies of
Sec. VI about the transformation laws of the pseudoparticle -
pseudofermion subspace unitary rotation. Next we consider that
$q=\pm q_{\alpha\nu}$ for $\nu>0$ pseudoparticle branches. Our
goal is the study of the momentum functional,

\begin{equation}
{Q_{\alpha\nu} (\iota\, q_{\alpha\nu})\over L} =
\sum_{\alpha'=c,s}\,\sum_{\nu'=1-\delta_{\alpha',\,c}}^{\infty}\,
\int_{-q_{\alpha'\nu'}}^{q_{\alpha'\nu'}} dq'\,
\Phi_{\alpha\nu,\,\alpha'\nu'}(\iota\, q_{\alpha\nu},q ') \,
\Delta N_{\alpha'\nu'}(q') \, ; \hspace{0.5cm} \alpha = c,\,s \, ;
\hspace{0.3cm} \nu > 0 \, ; \hspace{0.3cm} \iota = \pm 1 \, ,
\label{Qcan1qlApp}
\end{equation}
and confirm that it is given by expression (\ref{Qcan1ql}) where
the deviation $\Delta q_{c\nu} $ is provided in Eq.
(\ref{Dq0Dqag}). It is assumed that the bare momentum $q'$ belongs
to the domain $q'\in (-q_{\alpha'\nu'},\,+q_{\alpha'\nu'})$ and
can be such that $q'\rightarrow \pm q_{\alpha'\nu'}$ but $q'\neq
q_{\alpha'\nu'}$.

First we note that according to Eq. (\ref{Rqan}), the
two-pseudofermion phase shift $\Phi_{\alpha\nu,\,\alpha'\nu'}(\pm
q_{\alpha\nu},q ')$ expressed in terms of the pseudoparticle bare
momentum on the right-hand side of Eq. (\ref{Qcan1qlApp}) is such
that,

\begin{equation}
\Phi_{\alpha\nu,\,\alpha'\nu'}(\pm q_{\alpha\nu},q ') = \bar{\Phi
}_{\alpha\nu,\,\alpha'\nu'} \left(\pm\infty,
{\Lambda^{0}_{\alpha\nu}(q')\over u}\right) \, ,
\label{Phi-barPhiINF}
\end{equation}
where the two-pseudofermion phase shift $\bar{\Phi
}_{\alpha\nu,\,\alpha'\nu'}\left(r,r'\right)$ is defined in
Appendix B. In order to achieve this result we used that
$\Lambda^{0}_{\alpha\nu} (\pm q_{\alpha\nu})=\pm\infty$, as given
in Eq. (\ref{Rqan}).

By manipulation of the integral equations given in Appendix B, we
find that all the two-pseudofermion phase shifts of form
(\ref{Phi-barPhiINF}) vanish except the following ones,

\begin{equation}
\Phi_{\alpha\nu,\,c0}(\pm q_{\alpha\nu},q') =
\pm{[\delta_{\alpha,\,c} - \delta_{\alpha,\,s}]\over 2} \, ;
\hspace{0.5cm} \Phi_{\alpha\nu,\,\alpha\nu'}(\pm q_{\alpha\nu},q')
= \mp{\delta_{\nu,\,\nu'}\over 2} \pm {\nu + \nu' -\vert\,\nu -
\nu'\vert\over 2} \, ; \hspace{0.3cm} \alpha = c,\,s \, ;
\hspace{0.2cm} \nu,\,\nu'
> 0 \, . \label{PhiINFanan}
\end{equation}

Use of Eq. (\ref{PhiINFanan}) on the right-hand side of Eq.
(\ref{Qcan1qlApp}) leads to,

\begin{eqnarray}
{Q_{\alpha\nu} (\pm q_{\alpha\nu})\over L} & = & \mp {1\over
2}\Bigr[\int_{-q_{\alpha\nu}}^{q_{\alpha\nu}} dq' \Delta
N_{\alpha\nu} (q') -[\delta_{\alpha,\,c} -
\delta_{\alpha,\,s}]\int_{q_{c0}^-}^{q_{c0}^+} dq' \Delta N_{c0}
(q') - \sum_{\nu'=1}^{\infty} \Bigl(\nu + \nu' - \vert\nu -
\nu'\vert\Bigl)\int_{-q_{\alpha\nu'}}^{q_{\alpha\nu'}}
dq' \Delta N_{\alpha\nu'} (q')\Bigl] \, ; \nonumber \\
& = & \mp {\pi\over L}\Bigr[\Delta N_{\alpha\nu}
-[\delta_{\alpha,\,c} - \delta_{\alpha,\,s}]\Delta N_{c0} -
\sum_{\nu'=1}^{\infty} \Bigl(\nu + \nu' - \vert\nu -
\nu'\vert\Bigl) \Delta N_{c\nu'}\Bigl] = \mp \Delta q_{\alpha\nu}
\, ; \hspace{0.5cm} \alpha = c,\,s \, ; \hspace{0.3cm} \nu > 0 \,
, \label{Qcan1qlFIN}
\end{eqnarray}
where we also used expression (\ref{Dq0Dqag}). Finally, note that
expression (\ref{Qcan1qlFIN}) is equivalent to Eq.
(\ref{Qcan1ql}).

Let us now consider the quantity,

\begin{equation}
{Q_{c0} (q_{c0}^{\pm})\over L} =
\sum_{\alpha'=c,s}\,\sum_{\nu'=1-\delta_{\alpha',\,c}}^{\infty}\,
\int_{-q_{\alpha'\nu'}}^{q_{\alpha'\nu'}} dq'\,
\Phi_{c0,\,\alpha'\nu'}(q_{c0}^{\pm},q ') \, \Delta
N_{\alpha'\nu'}(q')  \, , \label{Qc0App}
\end{equation}
and confirm that it obeys relation (\ref{pfconqc0}). This is
easily confirmed by noting that according to Eq. (\ref{Rqan}) the
two-pseudofermion phase shifts
$\Phi_{c0,\,\alpha'\nu'}(q_{c0}^{\pm},q ')$ on the right-hand side
of Eq. (\ref{Qcan1qlApp}) are such that,

\begin{equation}
\Phi_{c\nu,\,\alpha'\nu'}(q_{c0}^{\pm},q ') =
\Phi_{c\nu,\,\alpha'\nu'}(0,q ') = \bar{\Phi }_{c0,\,\alpha'\nu'}
\left(0, {\Lambda^{0}_{c0}(q')\over u}\right) \, ,
\label{Phic0App}
\end{equation}
where the two-pseudofermion phase shifts $\bar{\Phi
}_{c0,\,\alpha'\nu'}\left(r,r'\right)$ are defined in Appendix B.
The result (\ref{Phic0App}) implies the validity of Eq.
(\ref{pfconqc0}).

Our next task involves evaluation of the momentum functional
$Q_{\alpha\nu} (q)/L$ given in Eq. (\ref{Qcan1j}) for
$q=\iota\,q_{F\alpha\nu}$ in the particular case when the excited
states associated with the deviations $\Delta N_{\alpha'\nu'}(q')$
on the right-hand side of that equation are J-CPHS ground states.
The {\it Fermi bare momentum} values that limit the compact
pseudoparticle bare-momentum occupancy configurations of these
states are given in Eqs. (\ref{HSqiFcs}) and (\ref{DqiFan}). While
the above expressions derived in this Appendix refer to general
excited states whose deviations obey relations (\ref{DNqzero}),
the following results are valid only for the particular case when
these excited states are J-CPHS ground-states. Our goal is to
arrive to expressions (\ref{DqFpfF})-(\ref{xi}). This implies
evaluation of the following quantity,

\begin{equation}
\Delta {\bar{q}}_{F\alpha\nu,\,\iota} = \iota\,{2\pi\over L}\Delta
N_{\alpha\nu,\,\iota} + {Q_{\alpha\nu}
(\iota\,q_{F\alpha\nu}^0)\over L} \, , \label{Dq0QanFSApp}
\end{equation}
where

\begin{equation}
{Q_{\alpha\nu} (\iota\,q_{F\alpha\nu}^0)\over L} =
\sum_{\alpha'=c,s}\,\sum_{\nu'=1-\delta_{\alpha',\,c}}^{\infty}\,
\int_{-q_{\alpha'\nu'}}^{q_{\alpha'\nu'}} dq'\,
\Phi_{\alpha\nu,\,\alpha'\nu'}(\iota\,q_{F\alpha\nu}^0,q ') \,
\Delta N_{\alpha'\nu'}(q') \, , \label{QanFSApp}
\end{equation}
and the bare-momentum distribution function deviations $\Delta
N_{\alpha'\nu'}(q')$ correspond to final J-CPHS ground states.

The phase shifts $\Phi_{\alpha\nu,\,\alpha'\nu'}(q,q ')$ and
$\bar{\Phi }_{\alpha\nu,\,\alpha'\nu'}\left(r,r'\right)$ have the
following property,

\begin{equation}
\Phi_{\alpha\nu,\,\alpha'\nu'}(q,q ') = -
\Phi_{\alpha\nu,\,\alpha'\nu'}(-q,-q ') \hspace{1cm}
\bar{\Phi}_{\alpha\nu,\,\alpha'\nu'}\left(r,r'\right)=-\bar{\Phi
}_{\alpha\nu,\,\alpha'\nu'}\left(-r,-r'\right) \, . \label{antiPP}
\end{equation}
This symmetry is found by analysis of the integral equations given
in Appendix B, which define the two-pseudofermion phase shifts
$\bar{\Phi }_{\alpha\nu,\,\alpha'\nu'}\left(r,r'\right)$
associated with the phase shifts
$\Phi_{\alpha\nu,\,\alpha'\nu'}(q,q ')$ through Eq.
(\ref{Phi-barPhi}). Once the ground-state rapidity functions that
appear on the right-hand side of Eq. (\ref{Phi-barPhi}) in the
argument of the two-pseudofermion phase shifts $\bar{\Phi
}_{\alpha\nu,\,\alpha'\nu'}\left(r,r'\right)$ are odd functions of
the bare momentum, the second relation given in Eq. (\ref{antiPP})
implies the validity of the first relation of the same equation.

Since the pseudoparticle bare-momentum distribution function
deviations of the J-CPHS ground-states include creation or
annihilation of pseudoparticles in the vicinity of the {\it Fermi
points} only, we can replace $q'$ by ${\rm sgn}
(q')\,q_{F\alpha'\nu'}^0$ in the argument of the phase shift
$\Phi_{\alpha\nu,\,\alpha'\nu'}(\iota\,q_{F\alpha\nu}^0,q')$ on
the right-hand side of Eq. (\ref{QanFSApp}). This leads to,

\begin{equation}
\Delta {\bar{q}}_{F\alpha\nu,\,\iota} = \iota\,{2\pi\over L}\Delta
N_{\alpha\nu,\,\iota} +
\sum_{\alpha'=c,\,s}\,\sum_{\nu'=1-\delta_{\alpha',\,c}}^{\infty}\,
\int_{-q_{\alpha'\nu'}}^{q_{\alpha'\nu'}} dq'\, \Delta
N_{\alpha'\nu'}(q')\,
\Phi_{\alpha\nu,\,\alpha'\nu'}(\iota\,q_{F\alpha\nu}^0,{\rm sgn}
(q')\,q_{F\alpha'\nu'}^0) \, . \label{DqFanPF}
\end{equation}
By performing the $q'$ integrations and using Eq. (\ref{antiPP})
we arrive to,

\begin{equation}
\Delta {\bar{q}}_{F\alpha\nu,\,\iota} = {2\pi\over L}
\sum_{\alpha'=c,\,s}\,\sum_{\nu'=1-\delta_{\alpha',\,c}}^{\infty}\,\sum_{\iota''=\pm
1}\,\Bigl[\,\iota''\,\delta_{\alpha ,\,\alpha'}\,\delta_{\nu
,\,\nu'}\,\delta_{\iota ,\,\iota''} +
\iota\,\Phi_{\alpha\nu,\,\alpha'\nu'}(q_{F\alpha\nu}^0,\iota\,\iota''\,
q_{F\alpha'\nu'}^0)\Bigl] [{\Delta N_{\alpha'\nu'}\over 2}
+\iota''\,\Delta J_{\alpha'\nu'}] \, . \label{DqFanPF2}
\end{equation}
After performing the $\iota''$ summation and using again Eq.
(\ref{antiPP}) we find,

\begin{eqnarray}
\Delta {\bar{q}}_{F\alpha\nu,\,\iota} & = & \iota\,{2\pi\over L}
\sum_{\alpha'=c,\,s}\,\sum_{\nu'=1-\delta_{\alpha',\,c}}^{\infty}\,\Bigl[\,\delta_{\alpha
,\,\alpha'}\,\delta_{\nu ,\,\nu'} + \sum_{\iota'=\pm 1}
\Phi_{\alpha\nu,\,\alpha'\nu'}(q_{F\alpha\nu}^0,\iota'
\,q_{F\alpha'\nu'}^0)\Bigl]{\Delta N_{\alpha'\nu'}\over 2} \nonumber \\
& + & {2\pi\over L}
\sum_{\alpha'=c,\,s}\,\sum_{\nu'=1-\delta_{\alpha',\,c}}^{\infty}\,\Bigl[\,\delta_{\alpha
,\,\alpha'}\,\delta_{\nu ,\,\nu'} + \sum_{\iota'=\pm 1}\,\iota'\,
\Phi_{\alpha\nu,\,\alpha'\nu'}(q_{F\alpha\nu}^0,\iota'
\,q_{F\alpha'\nu'}^0)\Bigl]\Delta
J_{\alpha'\nu'} \, . \label{DqFanPF3}
\end{eqnarray}
Note that since $\Delta {\bar{q}}_{F\alpha\nu,\,\iota}
=\iota\,{2\pi\over L}\Delta N^f_{\alpha\nu,\,\iota}$ this result
is equivalent to Eqs. (\ref{DqFpfF})-(\ref{xi}).

Finally, let us confirm the validity of Eq. (\ref{QanFq=0}) and
show that the momentum functional $Q_{\alpha\nu} (0)/L$ given in
Eq. (\ref{Qcan1j}) vanishes for $\alpha\nu$ pseudoparticle
branches such that $\alpha ,\nu\neq c0$ and $\alpha ,\nu\neq s1$
provided that the deviations $\Delta N_{c0}(q)$ and $\Delta
N_{s1}(q)$ on the right-hand side of that equation obey Eq.
(\ref{DNcDNs}). In this case, the excited states are partial
J-CPHS ground states. These are states that have J-CPHS
ground-state occupancy configurations for $\alpha\nu$
pseudoparticles such that $\alpha ,\nu\neq c0$ and $\alpha
,\nu\neq s1$.

There are no $\alpha\nu$ pseudoparticles belonging to branches
other than $c0$ and $s1$ in the initial ground state. It follows
that excited states obeying relations (\ref{DNqzero}) and having
J-CPHS ground-state occupancy configurations for $\alpha\nu$
pseudoparticles such that $\alpha ,\nu\neq c0$ and $\alpha
,\nu\neq s1$ have only a vanishing density of these quantum
objects. Such J-CPHS ground state occupancy configurations
correspond to compact bare-momentum domains of vanishing width
centered at $q=q_{F\alpha\nu}^0=0$. The generation from the ground
state of these excited states includes creation of $\alpha\nu$
pseudoparticles such that $\alpha ,\nu\neq c0$ and $\alpha
,\nu\neq s1$ for bare momentum $q\rightarrow 0$. A simple and
useful example, is the creation of a single $\alpha\nu$
pseudoparticle at bare momentum $q=q_{F\alpha\nu}^0=0$. To start
with, we assume that the deviations $\Delta N_{c0} (q)$ and
$\Delta N_{s1} (q)$ are small but have arbitrary values, whereas
the deviations $\Delta N_{\alpha\nu}(q)$ of the remaining
excited-state occupied $\alpha\nu$ pseudoparticle branches
correspond to J-CPHS ground state occupancy configurations
associated with {\it Fermi points} of the form given in Eq.
(\ref{qaniCPHSGS}).

Use of the integral equations provided in Appendix B leads in the
case of such partial J-CPHS ground states to,

\begin{equation}
\Phi_{\alpha\nu,\,\alpha'\nu'}(0,\iota'\,0) = 0 \, ;
\hspace{0.5cm} \nu,\,\nu'>0 \hspace{0.2cm} {\rm for}
\hspace{0.2cm} \alpha,\,\alpha' =c \, ; \hspace{0.3cm}
\nu,\,\nu'>1 \hspace{0.2cm} {\rm for} \hspace{0.2cm}
\alpha,\,\alpha' =s \, . \label{limPhiaa}
\end{equation}
Use of Eq. (\ref{limPhiaa}) in expression (\ref{xi}) leads to,

\begin{equation}
{\xi}^j_{\alpha\nu,\,\alpha'\nu'}=
\delta_{\alpha,\,\alpha'}\,\delta_{\nu,\,\nu'} \, ; \hspace{0.5cm}
\nu,\,\nu'>0 \hspace{0.2cm} {\rm for} \hspace{0.2cm}
\alpha,\,\alpha' =c \, ; \hspace{0.3cm} \nu,\,\nu'>1
\hspace{0.2cm} {\rm for} \hspace{0.2cm} \alpha,\,\alpha' =s \, ;
\hspace{0.3cm} j = 0,\,1 \, . \label{xiq0}
\end{equation}
Moreover, it follows from Eq. (\ref{antiPP}) that,

\begin{equation}
\Phi_{\alpha\nu,\,c0}(0,q ') = - \Phi_{\alpha\nu,\,c0}(0,-q ') \,
; \hspace{1cm} \Phi_{\alpha\nu,\,s1}(0,q ') = -
\Phi_{\alpha\nu,\,s1}(0,-q ') \, ; \hspace{0.5cm}
\nu>\delta_{\alpha,\,s} \, . \label{limPhiacs}
\end{equation}

Let us next consider the same type of excited states but with
small deviations $\Delta N_{c0} (q)$ and $\Delta N_{s1} (q)$
obeying Eq. (\ref{DNcDNs}). These excited states have vanishing
momentum and their $c0$ and $s1$ current number deviations $\Delta
J_{c0}$ and $\Delta J_{s1}$, respectively, vanish, {\it i.e.}
$\Delta J_{c0}=\Delta J_{s1}=0$. In this case it follows from Eqs.
(\ref{limPhiaa})-(\ref{xiq0}) that,

\begin{eqnarray}
{Q_{\alpha\nu} (0)\over L} & = &
\sum_{\alpha'=c,s}\,\sum_{\nu'=1-\delta_{\alpha',\,c}}^{\infty}\,
\int_{-q_{\alpha'\nu'}}^{q_{\alpha'\nu'}} dq'\,
\Phi_{\alpha\nu,\,\alpha'\nu'}(0,q ') \, \Delta
N_{\alpha'\nu'}(q') \nonumber \\
& = & \int_{-q_{c0}}^{q_{c0}} dq'\, \Phi_{\alpha\nu,\,c \,0}(0,q
') \, \Delta N_{c0}(q') + \int_{-q_{s1}}^{q_{s1}} dq'\,
\Phi_{\alpha\nu,\,s1}(0,q
') \, \Delta N_{s1}(q') \nonumber \\
& + & {2\pi\over L}
\sum_{\alpha'=c,\,s}\,\sum_{\nu'=1+\delta_{\alpha',\,s}}^{\infty}\,\Bigl[\,\iota\,\sum_{\iota'=\pm
1} \Phi_{\alpha\nu,\,\alpha'\nu'}(0,\iota'\,0){\Delta
N_{\alpha'\nu'}\over 2} + \sum_{\iota'=\pm 1}\,\iota'\,
\Phi_{\alpha\nu,\,\alpha'\nu'}(0,\iota'\,0)\Delta
J_{\alpha'\nu'}\Bigl] = 0 \, . \label{QanFq=0App}
\end{eqnarray}
This result confirms the validity of Eq. (\ref{QanFq=0}).

%%%%%%%%%%%%%%%%%%%%%%%%%%%%%%%%%%%%%%%%%%%%%%%%%%%%%%%%%%%%%%%%%%%%%%%%%%

\end{document}